\documentclass[12pt]{article}

\usepackage{graphicx}
\usepackage{appendix}

\usepackage{color}
\usepackage{amsmath,amsfonts,amssymb,amsthm}
\usepackage{thmtools}
\usepackage{mathtools}
\usepackage{bbm}
\usepackage{comment}
\usepackage{enumitem}
\usepackage{epigraph}
\usepackage{float}
\usepackage{booktabs}
\usepackage{caption}
\usepackage{subcaption}
\usepackage{tikz}
\usepackage{mathtools}
\usepackage{todonotes}
\usepackage{xcolor}
\usepackage[mathlines]{lineno}
\usepackage{nicefrac}
\usepackage{colonequals}
\usepackage[onehalfspacing]{setspace}

\usetikzlibrary{positioning}
\usetikzlibrary{matrix}

\usepackage{hyperref}
\hypersetup{
	colorlinks=true,
	linkcolor=blue,
	filecolor=blue,
	urlcolor=blue,
	citecolor=blue,
}

\newcommand*\patchAmsMathEnvironmentForLineno[1]{%
	\expandafter\let\csname old#1\expandafter\endcsname\csname #1\endcsname
	\expandafter\let\csname oldend#1\expandafter\endcsname\csname end#1\endcsname
	\renewenvironment{#1}%
	{\linenomath\csname old#1\endcsname}%
	{\csname oldend#1\endcsname\endlinenomath}}%
\newcommand*\patchBothAmsMathEnvironmentsForLineno[1]{%
	\patchAmsMathEnvironmentForLineno{#1}%
	\patchAmsMathEnvironmentForLineno{#1*}}%
\AtBeginDocument{%
	\patchBothAmsMathEnvironmentsForLineno{equation}%
	\patchBothAmsMathEnvironmentsForLineno{align}%
	\patchBothAmsMathEnvironmentsForLineno{flalign}%
	\patchBothAmsMathEnvironmentsForLineno{alignat}%
	\patchBothAmsMathEnvironmentsForLineno{gather}%
	\patchBothAmsMathEnvironmentsForLineno{multline}%
}

\usepackage[backend=biber, maxbibnames=9, maxcitenames=4,uniquelist=false, citestyle=authoryear, bibstyle=authoryear, sorting= nyt, block =space, hyperref=true,  natbib=true, giveninits=true]{biblatex}
\defbibheading{head}{}
\usepackage{csquotes}

\DeclareNameAlias{sortname}{family-given}%
\renewbibmacro*{journal+issuetitle}{%
	\usebibmacro{journal}%
	\setunit*{\addspace}%
	\iffieldundef{series}
	{}
	{\newunit
		\printfield{series}%
		\setunit{\addspace}}%
	\printfield{volume}%
	\iffieldundef{number}
	{}
	{\mkbibparens{\printfield{number}}}%
	\setunit{\addcomma\space}%
	\printfield{eid}%
	\setunit{\addspace}%
	\usebibmacro{issue+date}%
	\setunit{\addcolon\space}%
	\usebibmacro{issue}%
	\newunit}

\DeclareMathOperator*{\argmax}{argmax}

\declaretheorem[style=definition,qed=\myqed,numberwithin=section]{definition}
\declaretheorem[style=definition,qed=\myqed,sibling=definition]{example}
\declaretheorem[style=definition,qed=\myqed,sibling=definition]{remark}
\declaretheorem[style=definition,qed=\myqed]{assumption}

\theoremstyle{plain}
\newtheorem{proposition}[definition]{Proposition}
\newtheorem{lemma}[definition]{Lemma}
\newtheorem{corollary}[definition]{Corollary}
\newtheorem{theorem}[definition]{Theorem}

\newcommand{\R}{\mathbb{R}}
\newcommand{\F}{\mathcal{F}}
\newcommand{\Prob}{\mathbb{P}}
\renewcommand{\d}{\mathrm{d}}
\newcommand{\dF}{\mathrm{d}F}
\newcommand{\LV}{\left\vert}
\newcommand{\RV}{\right\vert}
\newcommand{\Fto}{F\left( \theta_1^* \right)}
\newcommand{\Ftt}{F\left( \theta_2^* \right)}
\newcommand{\ubar}[1]{\text{\b{$#1$}}}
\newcommand{\eps}{\varepsilon}

\newcommand{\keywords}[1]{\textbf{Keywords:}\quad #1}
\newcommand{\jel}[1]{\textbf{JEL Classification:}\quad #1}
\newcommand{\myqed}{\hfill$\square$}

\newcounter{lastnote}

\title{\date{\today}}

\title{Cursed Equilibria and Knightian Uncertainty in a Trading Game}
\author{
	Jurek Preker\thanks{Center for Mathematical Economics, Bielefeld University, Postfach 100131, 33501 Bielefeld, Germany. jurek.preker@uni-bielefeld.de} \thanks{This work was supported by a fellowship of the German Academic Exchange Service (DAAD).
    The author would like to thank Gerrit Bauch, Marina Halac, Daniel Hauser, Dominik Karos, Frank Riedel, and Philipp Strack for helpful questions, comments, and discussions.}
}

\date{\today}

\begin{document}

\maketitle

\begin{abstract}
    We introduce a novel equilibrium concept that incorporates Knightian uncertainty into the cursed equilibrium \citep{eyster_rabin_2005}.
    This concept is then applied to a two-player game in which agents can engage in trade or refuse to do so.
    While the Bayesian Nash equilibrium predicts that trade never happens, players do trade in a cursed equilibrium.
    The inclusion of uncertainty enhances this effect for cursed and uncertainty averse players.
    This contrasts general findings that uncertainty reduces trade but is consistent with behavior that has been observed in experiments.
\end{abstract}


\jel{C72, D81, D82, D91}

\section{Introduction} \label{sec:Introduction}

The theoretical prediction of \emph{no-trade theorems} \citep{milgrom_stokey_1982, tirole_1982} is contrasted by many experiments in which subjects do engage in trade \citep{sonsino_erev_gilat_2001, rogers_palfrey_camerer_2009, carrillo_palfrey_2009, carrillo_palfrey_2011}.
One attempt to rationalize such behavior is to assume that players are \emph{cursed} in the sense of \citet{eyster_rabin_2005}: they fail to recognize the connection between their opponents' types and actions.\footnote{
The name stems from the \emph{winner's curse} in auctions (see \cite{kagel_levin_2002} for an overview) which can also be explained by cursed behavior.
}
While \citet{szembrot_2018} finds that the \emph{cursed equilibrium}, compared to competing concepts modeling bounded rationality, does a good job in explaining subjects' behavior in experiments, it has frequently been observed that many subjects deviate even further from the rational equilibrium than they would do under cursedness \citep{eyster_rabin_2005, christensen_2008, carrillo_palfrey_2011}.
In this paper, we introduce a new equilibrium concept that combines the cursed equilibrium with \emph{Knightian uncertainty}---that is, missing knowledge about probability distributions---and apply this \emph{cursed Knight-Nash equilibrium} to a trading game.
We will see that this new concept can theoretically explain the experimentally observed behavior that many subjects seem to be ``overcursed''.

The basic form of the investigated game is quite simple:
There are two players.
Both receive a private type; types are independent and identically distributed according to the uniform distribution on $[0,1]$.
Upon seeing their own type realizations, players can simultaneously offer to trade and exchange their number with the opponent. 
If both want to trade, types are switched, otherwise each player keeps their type. 
Afterwards both types are revealed and the person with the higher type wins.

Before continuing to read, the reader is invited to pause for a moment and contemplate this game.
How would you play it?
Would you offer to exchange numbers if your realized type is $\tfrac{1}{10}$, or $\tfrac{1}{2}$, or $\tfrac{9}{10}$?
And, if you did not know the distribution of types, (how) would that change your answer?

Anecdotal evidence suggests that when types are uniformly distributed, many (though not all) people suggest the strategy ``offer to trade if my type is below $\tfrac{1}{2}$, refuse trade if it is above $\tfrac{1}{2}$''.
After all, if one's own type is below $\tfrac{1}{2}$, the probability that the opponent's type exceeds the own one is greater than the probability that it is smaller, so exchanging types should on average be desirable.
Similarly, one should not offer to trade if one's type is above $\tfrac{1}{2}$, as in this case, the opponent's type is probably smaller than one's own.

This reasoning is as understandable as it is incomplete.\footnote{
Actually, the author followed exactly these arguments when facing the above posed problem as a student.
He only realized his mistake some years later, when he was a teaching assistant and prepared for the tutorial.
}
It ignores that there is not only one player who decides whether trade happens, but that \emph{both players have to consent} and that there is no trade if one of them prevents it.
A profit-maximizing, forward-looking agent should therefore condition their decision whether to offer to trade or not on the event that the opponent wishes to trade.
As this happens only if the opponent's type is low (otherwise they would just want to keep their type), a rational player will in effect never offer to trade.

Players who make the mistake described above are cursed:
They correctly identify the frequency with which other players employ certain actions, but fail to recognize that their opponents condition their behavior on their type, behaving as if every type profile of their opponents played these frequencies in a type-independent mixed strategy.
They then respond optimally to this \emph{average strategy}.
In the game described above, when some player offers to trade if and only if their type lies below $\tfrac{1}{2}$, a cursed opponent thinks that independently of their type, they mix between trading and not with probability $\tfrac{1}{2}$ each.
Given such a strategy, it is then indeed a best response to offer to trade if and only if the own type lies below $\tfrac{1}{2}$, such that it is a cursed equilibrium if both players play this strategy.

If agents face Knightian uncertainty and do not know the true type-generating distribution, they still observe the average strategy of the opponents, which is derived from this true, unknown probability measure.
The remaining uncertainty about the opponent's type is tackled by the cautious maxmin approach \citep{gilboa_schmeidler_1989}.
Our analysis shows that if players behave rationally, it does not make a difference whether there is uncertainty or not: 
trade never happens.
However, cursed players do trade under uncertainty, even more than in the case of known probability distributions.
This contrasts previous results which find that uncertainty aversion usually reduces trade \citep{dow_werlang_1992}, but might shed light on the observation that subjects seem to be ``overcursed'' in experiments.
As the underlying game admits only two possible levels of ex-post utility, our findings cannot be explained by players' attitudes towards quantifiable risk, but must be due to uncertainty.

So, why do cursed players trade more under uncertainty than when type distributions are known?
It is crucial to observe that in our game, a cursed player with maxmin preferences does not necessarily consider not trading a ``safe option'', as its benefit depends on the probability that the opponent is of a smaller type, something that is unknown.
If, however, such a player offers to trade, they can presumably profit from high-type opponents who also engage in trade, as well as from low-type opponents who refuse to trade.\footnote{
This is not how a rational or even cursed opponent actually plays the game---on the contrary, the opponent will engage in trade if and only if their type is low---, but as players are cursed, they do not see this connection and behave as if their opponents used a type-independent mixed strategy.
}
Uncertainty about the opponent's type still decreases the expected utility compared to the case with known probability distributions, both for the non-trader and for the trader.
Yet, this effect is stronger for the non-trader than for the trader, as the former must hope for the opponent to have a low type, while the latter can (seemingly) profit from high-type as well as low-type opponents.
In some strategy profiles, engaging in trade can---from the perspective of a cursed player---provide a full hedge against any uncertainty about the opponent's probability distribution, in the very sense of Raiffa's critique of the concept of uncertainty aversion \citep{raiffa_1961}.\footnote{
    In his comment on \citet{ellsberg_1961}, Raiffa pointed out that when betting on the outcome of an Ellsberg urn that contains red and black balls, one might first draw a ball without observing its color, then toss a fair coin, and eventually bet on ``red'' or ``black'' depending on whether the coin landed heads or tails.
    In this case, there is an objective probability of $\tfrac{1}{2}$ to match the correct color, so the bet on the Ellsberg urn should be equivalent to a bet on an urn that is known to contain red and black balls in equal proportion.
}

The rest of this work is organized as follows:
After Section \ref{sec:Literature} provides an overview over the relevant literature, Section \ref{sec:Several_Equilibrium_Concepts} reviews, defines, and discusses the necessary equilibrium concepts.
Section \ref{sec:The_Trading_Game_Without_Uncertainty} formally sets up the game we are going to analyze, and investigates rational and cursed behavior in the absence of uncertainty.
In Section \ref{sec:The_Trading_Game_With_Uncertainty}, we will see how these results change if uncertainty is incorporated into the model, and Section \ref{sec:Conclusion} concludes.
Proofs are delegated to Appendix \ref{app:Proofs}.
Appendices \ref{app:Ambiguous_Cursed_Equilibria} and \ref{app:Partial_Cursedness} contain complementary results about another way to incorporate Knightian uncertainty into the cursed equilibrium, and about partial cursedness.

\section{Literature} \label{sec:Literature}

In its primitive form without uncertainty, the game investigated in Sections \ref{sec:The_Trading_Game_Without_Uncertainty} and \ref{sec:The_Trading_Game_With_Uncertainty} is a basic textbook question, resembling for example Exercise 28.1 in \citet{osborne_rubinstein_1994}.
A related problem is the \emph{compromise game} from \citet{carrillo_palfrey_2009}. 
In their framework, players with private types can ``retreat'' or ``fight'', and a compromise is reached when both retreat. 
The same arguments as in our problem reveal that in every Bayesian Nash Equilibrium, the outcome of the game is that no compromise arises:
whenever one player would like to achieve a compromise, the other should not. 

The underlying reasoning resembles adverse selection problems such the \emph{Market for Lemons} \citep{akerlof_1970}, though these results depend on information asymmetries between players, which do not exist in our model.
In the context of multi-sided private information, \citet{milgrom_stokey_1982} and \citet{tirole_1982} establish that rational, risk-averse agents do not trade in equilibrium.

Despite these \emph{no-trade theorems}, experiments have found that subjects do engage in trade when they are faced with pertinent problems in the laboratory.
\citet{carrillo_palfrey_2011} investigate a trading game in which a buyer and a seller receive private information about a good with common valuation.
Although a no-trade theorem holds, trade is frequently observed in their experiment.
Analogous results hold true for related \emph{betting games} in which privately informed agents decide whether to engage in zero-sum betting or not.
The theoretical prediction by \citet{sebenius_geanakoplos_1983} that betting never occurs among rational players has been contradicted by many experiments \citep{sonsino_erev_gilat_2001, rogers_palfrey_camerer_2009, sovik_2009}.
Similarly, \citet{holt_sherman_1994} analyze a bidding game in which participants bid positive values despite the optimal bid being 0.
The authors remark that these subjects did not account for the information provided by the fact that trade occurs, calling them ``naive''.

This reasoning has been formalized by the \emph{cursed equilibrium} \citep{eyster_rabin_2005} which can rationalize most of the empirical observations described above.
A subsequent theoretical application of the cursed equilibrium is provided in \citet{eyster_rabin_vayanos_2019}.
The authors study a financial market in which trade never occurs under rational behavior, but agents do trade if they are cursed, a behavior consistent with our results in Subsection \ref{subsec:How_Cursedness_Leads_Players_To_Trade}.
Recently, \citet{cohen_li_2023} and \citet{fong_lin_palfrey_2023} have adapted the concept to analyze extensive form games.

While the cursed equilibrium has proven quite helpful in the theoretical and experimental analysis of boundedly rational behavior, especially compared to competing concepts \citep{carrillo_palfrey_2009, szembrot_2018}, there are some effects that cursedness alone cannot explain.
In fact, there is evidence that people ``tend to bid in excess of the levels predicted by even the fully cursed equilibrium'' \citep[p.~1638]{eyster_rabin_2005}.
Similar results have been made by \citet{christensen_2008} and \citet{carrillo_palfrey_2011}, so further phenomena seem to be present.
While we do not claim that these observations are necessarily driven by Knightian uncertainty, it is striking that in our model, the presence of uncertainty in combination with cursedness yields exactly these results:
players trade more (and therefore deviate more from rational behavior) than in the absence of uncertainty.

When analyzing Bayesian games with Knightian uncertainty, we use the framework of \emph{incomplete information games with imprecise probabilistic information} as defined in \citet{decerf_riedel_2020}.
The components of such a game are identical to the ones of ``normal'' Bayesian games, with the sole exception that the single probability measure is replaced by a set of measures.\footnote{
Alternative approaches to incorporate uncertainty into Bayesian games can for example be found in \citet{stauber_2011} and \citet{zhang_luo_ma_leung_2014}.
}
The natural solution concept for these games is the \emph{Knight-Nash equilibrium} of \citet{decerf_riedel_2020}, a variant of which we use for our analysis.
As announced above, we find that including uncertainty into the framework in addition to cursedness further enlarges the set of player types that are willing to engage in trade.
In view of the existing literature on trading under uncertainty, this is, at least at first glance, a highly unintuitive result.
A large strand of papers, following \citet{dow_werlang_1992}, features models in which uncertainty---and in particular aversion thereof---reduces or prohibits trade, see \citet{guidolin_rinald_2013} for a literature review.
\citet{marinacci_2015} states that under maxmin behavior, ``the larger is [...] the perceived model uncertainty [...], the larger [...] the set of prices that result in no trade'' \citep[p.~1064]{marinacci_2015}.
In contrast to that, we find that if players are cursed, there is always more trade under uncertainty than in its absence; and increasing the amount of uncertainty can increase the set of types under which players are willing to trade, so more uncertainty means more trade, not less.

\section{Several Equilibrium Concepts} \label{sec:Several_Equilibrium_Concepts}

This section reviews and establishes the concepts that we are going to use in our analysis.
Consider a Bayesian game $\Gamma = \left( N, \left(A_k\right)_{k\in N}, \left(\Theta_k\right)_{k\in N}, \left(u_k\right)_{k \in N}, \Prob \right)$ with finite player set $N$ and generic player $k$ (``they'').
Each  player has a finite action space $a_k$ and a (finite or infinite) type space $\Theta_k$.
Types are distributed according to distribution $\Prob$ on $\bigtimes_{k\in N} \Theta_k$.\footnote{
One might also assume that players' action spaces are infinite; in this case, sums are to be replaced by integrals.
}
We will, for $k \in N$, denote the set of other players by $-k$, as well as $A_{-k}=\bigtimes_{j\neq k} A_j$ and $\Theta_{-k}=\bigtimes_{j\neq k} \Theta_j$.
In this game, a \emph{mixed strategy} of player $k$ is a mapping $\sigma_k \colon \Theta_k \to \Delta\left(A_k\right)$ that specifies a probability distribution over actions for each type of player $k$.
We write $\sigma_k\left(a_k \mid \theta_k\right)$ for the probability that under the mixed strategy $\sigma_k$, player $k$ plays action $a_k$ if they are of type $\theta_k$.
Similarly, $\sigma_{-k}\colon \Theta_{-k} \to \bigtimes_{j\neq k}\Delta(A_j)$ is a strategy profile of $k$'s opponents.
The standard concept to solve such a game is the \emph{Bayesian Nash equilibrium} \citep{harsanyi_1967, harsanyi_1968a, harsanyi_1968b}.
\citet{eyster_rabin_2005} introduce their cursed equilibrium as a modification of the Bayesian Nash equilibrium that accounts for players' incapacity to correctly investigate the correlation between the other players' types and their strategies.
We slightly adjust their definition in order to allow for infinite type spaces, and adopt it to our model.

\begin{definition} \label{def:cursed_equilibrium}
    Let  $\Gamma = \left( N, \left(A_k\right)_{k\in N}, \left(\Theta_k\right)_{k\in N}, \left(u_k\right)_{k \in N}, \Prob \right)$ be a Bayesian game as specified above.
    For player $k$, the \emph{average strategy} of the other players, given their own type $\theta_k$ and probability distribution $\Prob$, is
    \begin{align}
        \bar{\sigma}_{-k}\left(a_{-k} \mid \theta_k, \Prob\right) \colonequals \int_{\Theta_{-k}} \sigma_{-k}\left(a_{-k} \mid \theta_{-k}\right) \; \d\Prob\left(\theta_{-k} \mid \theta_k \right). \label{eq:average_strategy}
    \end{align}
    For a strategy profile $\sigma_{-k}$ of $k$'s opponents, $\sigma_k$ is a \emph{(fully) cursed best response} against $\sigma_{-k}$ if for each $\theta_k \in \Theta_k$, and each $a_k^*$ such that $\sigma_k\left(a_k^* \mid \theta_k\right)>0$,
    \begin{align}
        a_k^* &\in \argmax_{a_k \in A_k} v_k\left(a_k\mid \theta_k, \sigma_{-k}\right), \text{ where} \nonumber \\
        v_k\left(a_k\mid \theta_k, \sigma_{-k}\right) &\colonequals \int_{\Theta_{-k}}
        \sum_{a_{-k} \in A_{-k}} \bar{\sigma}_{-k}\left(a_{-k}\mid \theta_k, \Prob\right) 
        u_k\left(a_k, a_{-k} ; \theta_k, \theta_{-k}\right) \d\Prob\left(\theta_{-k} \mid \theta_k\right)\label{eq:definition_v}
    \end{align}
    is the \emph{perceived interim expected utility} of action $a_k$, given the own type $\theta_k$ and the opponents' strategy profile $\sigma_{-k}$.
    A strategy profile $\sigma^*$ is a \emph{(fully) cursed equilibrium} if for each $k\in N$, $\sigma^*_k$ is a (fully) cursed best response against $\sigma^*_{-k}$.
\end{definition}

\noindent As \citet{eyster_rabin_2005} point out, a cursed player mistakenly believes that their opponents play the mixed strategy profile $\bar \sigma_{-k}$ independent of their types $\theta_{-k}$, when in fact they play the strategy profile $\sigma_{-k}$ which does depend on their type.
Such a player then optimally responds to this strategy, choosing their action $a_k$ as to maximize $v_k\left(a_k\mid \theta_k, \sigma_{-k}\right)$.
In contrast, in a (interim) Bayesian Nash equilibrium\footnote{
    Interim and ex-ante equilibria in Bayesian games are equivalent if the type distribution has full support.
    In this work, we solely focus on the interim interpretation of Bayesian games, as this is the ``language'' in which the cursed equilibrium has been formulated.
}, 
each player maximizes their actual \emph{interim expected utility} 
\begin{align}
    w_k\left(a_k\mid \theta_k, \sigma_{-k}\right) &\colonequals \int_{\Theta_{-k}} \sum_{a_{-k} \in A_{-k}} \sigma_{-k}\left(a_{-k} \mid \theta_{-k}\right) u_k\left(a_k, a_{-k} ; \theta_k, \theta_{-k}\right) \; \d\Prob\left(\theta_{-k}\mid\theta_k \right).\label{eq:definition_w}
\end{align}

\noindent \citet{eyster_rabin_2005} also define the concept of partial cursedness, in which players optimally respond to a convex combination of the opponent's actual strategy and the average one.
In this work, we will solely focus on the fully cursed equilibrium\footnote{
    Complementary results on partial equilibria can be found in Appendix \ref{app:Partial_Cursedness}.
}
and will therefore drop the word ``fully''.

We shall now go on by generalizing the model to games in which the probability distribution $\Prob$ is not known.
To this end, let $\Gamma = \left( N, \left(A_k\right)_{k\in N}, \left(\Theta_k\right)_{k\in N}, \left(u_k\right)_{k \in N}, \mathcal{P} \right)$ be an \emph{incomplete information game with imprecise probabilistic information}.
The components of such a game are identical to the ones of the Bayesian game as specified above, except for the fact that there is a nonempty \emph{ambiguity set} $\mathcal{P}$ of probability distributions over $\Theta$ instead of a single distribution $\Prob$.

\begin{definition} \label{def:knight_nash_equilibrium}
    Let $\Gamma = \left( N, \left(A_k\right)_{k\in N}, \left(\Theta_k\right)_{k\in N}, \left(u_k\right)_{k \in N}, \mathcal{P} \right)$ be an incomplete information game with imprecise probabilistic information.
    For a strategy profile $\sigma_{-k}$ of  player $k$'s opponents, $\sigma_k$ is a \emph{maxmin best response} against $\sigma_{-k}$ if for each $\theta_k \in \Theta_k$, and each $a_k^*$ such that $\sigma_k\left(a_k^* \mid \theta_k\right)>0$,
    \begin{align*}
        a_k^* &\in \argmax_{a_k \in A_k} \ubar w_k\left(a_k\mid \theta_k, \sigma_{-k}\right),\text{ where} \\
        \ubar w_k\left(a_k\mid \theta_k, \sigma_{-k}\right) &\colonequals \inf_{\Prob \in \mathcal{P}} \int_{\Theta_{-k}} \sum_{a_{-k} \in A_{-k}} \sigma_{-k}\left(a_{-k} \mid \theta_{-k}\right) u_i\left(a_k, a_{-k} ; \theta_k, \theta_{-k}\right) \d\Prob\left(\theta_{-k}\mid\theta_k\right)
    \end{align*}
    is the \emph{infimal interim expected utility} of action $a_k$, given the own type $\theta_k$ and the opponents' strategy profile $\sigma_{-k}$.
    A strategy profile $\sigma^*$ is a \emph{Knight-Nash equilibrium} if for each $k\in N$, $\sigma_k^*$ is a maxmin best response against $\sigma_{-k}^*$.\footnote{
        Definition \ref{def:knight_nash_equilibrium} resembles Definition 8 in \citet{decerf_riedel_2020}.
        Their definition requires optimality in the ex-ante stage, in which a player does not yet know their own type, and has to take expectations over their type as well.
        In this work, we focus on the interim stage.
        It is straightforward to show that in the framework of this paper, every Knight-Nash equilibrium in the sense of Definition \ref{def:knight_nash_equilibrium} is also an equilibrium in the sense of \citet{decerf_riedel_2020}.
    }
\end{definition}

\noindent Definition \ref{def:knight_nash_equilibrium} mimics the definition of a Bayesian Nash equilibrium.
The only difference is that instead of choosing optimal strategies with respect to a unique distribution, the infimal expected value over all distributions is maximized.
Hence, the Knight-Nash equilibrium takes the maxmin or Gilboa-Schmeidler approach to tackle uncertainty (cf. \cite{gilboa_schmeidler_1989}).

We now want to include Knightian uncertainty into the concept of cursed equilibria.
\citet{eyster_rabin_2005} do not base their equilibrium concepts on an axiomatic foundation, but they provide the following learning story that justifies the concept of fully cursed equilibrium:
Players who observe previous generations playing the game at hand ``(i) have correct a priori beliefs about a game's informational structure, (ii) observe the behavior of a large number of players previously playing the game---but not  players' final payoffs---and (iii) are entirely unsophisticated about other players' strategic incentives'' \citep[p. 1633]{eyster_rabin_2005}.
In our framework that features Knightian uncertainty, the ``correct a priori beliefs'' imply that the ambiguity set $\mathcal{P}$ is \emph{correct}, hence the \emph{correct probability distribution} $\Prob^*$ that generates the players' types is contained in $\mathcal{P}$.
The average strategy of other players, observed from previous generations playing the game, is the average according to $\Prob^*$.
Consequently, this average strategy is not subject to the maxmin optimization problem of player $k$, but it is---given player $k$'s type and the strategies of the opponents---a fixed object, $\bar{\sigma}_{-k}\left(\cdot \mid \theta_k, \Prob^*\right)$.\footnote{
An alternative approach in which this average strategy is also subject to uncertainty is outlined in Appendix \ref{app:Ambiguous_Cursed_Equilibria}.
}
\begin{definition} \label{def:cursed_knight_nash_equilibrium_given_P*}
    Let $\Gamma = \left( N, \left(A_k\right)_{k\in N}, \left(\Theta_k\right)_{k\in N}, \left(u_k\right)_{k \in N}, \mathcal{P} \right)$ be an incomplete information game with imprecise probabilistic information, and let $\Prob^* \in \mathcal{P}$ be the true type generating distribution.
    For a strategy profile $\sigma_{-k}$ of  player $k$'s opponents, $\sigma_k$ is a \emph{maxmin cursed best response (under $\Prob^*$)} against $\sigma_{-k}$ if for each $\theta_k \in \Theta_k$, and each $a_k^*$ such that $\sigma_k\left(a_k^* \mid \theta_k\right)>0$,
    \begin{align}
        a_k^* &\in \argmax_{a_k \in a_k} \ubar v_k^{\Prob^*}\left(a_k \mid \theta_k, \sigma_{-k}\right), \text{ where} \nonumber \\
        \ubar v_k^{\Prob^*}\left(a_k \mid \theta_k, \sigma_{-k}\right) &\colonequals \inf_{\Prob \in \mathcal{P}} \int_{\Theta_{-k}}
        \sum_{a_{-k} \in A_{-k}} \bar{\sigma}_{-k}\left(a_{-k} \mid \theta_k, \Prob^* \right) 
        u_i\left(a_k, a_{-k} ; \theta_k, \theta_{-k}\right) \d\Prob\left(\theta_{-k}\mid\theta_k\right).\label{eq:definition_vP*_uncertainty}
    \end{align}
    A strategy profile $\sigma^*$ is a \emph{cursed Knight-Nash equilibrium (given $\Prob^*$)} if for each $k\in N$, $\sigma^*_k$ is a maxmin cursed best response given $\Prob^*$ against $\sigma^*_{-k}$.
\end{definition}

\noindent We may think of a cursed player acting according to Definition \ref{def:cursed_knight_nash_equilibrium_given_P*} as follows:
They play the game repeatedly, knowing that types are drawn according to some probability distribution in $\mathcal{P}$.
The player is only aware of $\mathcal{P}$, having no idea about the exact distribution $\Prob^*$.
Alternatively, they might have a correct estimate about the true distribution $\Prob^*$, but, as they are not sure and want to be cautious, base their decision on a distribution band $\mathcal{P}$ that surrounds $\Prob^*$.
Our player observes that whenever they are of type $\theta_k$, their opponents play action profile $a_{-k}^1$ in, say, 70\% of the cases, while in the remaining 30\%, the opponents play $a_{-k}^2$.
Using the law of large numbers, our player (correctly) deduces that the opponents' average strategy, given their own type $\theta_k$, is $\bar\sigma_{-k}\left(a_{-k}^1 \mid\theta_k, \Prob^*\right) = 0.7 = 1-\bar\sigma_{-k}\left(a_{-k}^2 \mid \theta_k, \Prob^*\right)$.
As they are cursed, they ignore that the other agents' actions might (and typically will) depend on their types, instead assuming that all types of the other players play the same mixed action profile $0.7a_{-k}^1+0.3a_{-k}^2$.
In the presence of Knightian uncertainty, our player does not know with which probability measure to calculate their expected utilities.
Therefore, they calculate their (cursed) expected utility for each $\Prob \in \mathcal{P}$ and each of their own actions $a_k \in a_k$ and, being cautious, chooses $a_k$ to be the maxminimizer of this expected utility.
Note that in order to deduce the average strategy given $\Prob^*$, our player need not know $\Prob^*$.
Moreover, as they are ignorant about the other players' strategic incentives, knowing $\bar{\sigma}_{-k}\left(\cdot \mid \theta_k, \Prob^*\right)$ tells them nothing about $\Prob^*$.\footnote{
Even if they were more sophisticated and could deduce the other players' strategies, not only their average strategies, from observing their play, this would not necessarily allow the player to pin down $\Prob^*$, as there are typically many $\Prob \in \mathcal{P}$ that, together with a strategy profile of the opponent, induce the same distribution over actions.
}

If $\mathcal{P}=\{\Prob^*\}$ is a singleton, Definition \ref{def:cursed_knight_nash_equilibrium_given_P*} collapses to the cursed equilibrium (Definition \ref{def:cursed_equilibrium}).
In particular, the cursed Knight-Nash equilibrium relates to the cursed equilibrium in the same way the Knight-Nash equilibrium relates to the Bayesian Nash equilibrium.

Finally, let us also define equilibria in which only some players are cursed, while others act rationally, given the ambiguity set $\mathcal{P}$.

\begin{definition} \label{def:cursed_unscursed_nash_equilibrium}
    Let $\Gamma = \left( N, \left(A_k\right)_{k\in N}, \left(\Theta_k\right)_{k\in N}, \left(u_k\right)_{k \in N}, \mathcal{P} \right)$ be an incomplete information game with imprecise probabilistic information, and let $\Prob^* \in \mathcal{P}$ be the true type generating distribution.
    Moreover, let $N=N_1\sqcup N_2$ be a partition of $N$.
    A strategy profile $\sigma^*$ is an \emph{$(N_1, N_2)$-cursed-uncursed Knight-Nash equilibrium under $\Prob^*$} if for all $k\in N_1$, $\sigma^*_k$ is a maxmin cursed best response under $\Prob^*$ against $\sigma^*_{-k}$; and for all $j\in N_2$, $\sigma^*_j$ is a maxmin best response against $\sigma^*_{-j}$.
\end{definition}

\section{The Trading Game Without Uncertainty}
\label{sec:The_Trading_Game_Without_Uncertainty}

In this work, we focus on a constant-sum trading game in which both players hold private information.
One might think of a sale, for example of an asset on the stock market.
Both players hold some private information about the advantageousness of the potential trade.
Trade only occurs if there is mutual agreement about it.
At some point after the trade, it turns out who has profited from the trade. 
For instance, when the deal is about selling an asset, then the buyer profits if it evolves positively, and they do not profit if otherwise.
The person who profits from trade is happy, the other person will wish that they had not engaged in trade.
This regret is modeled by assigning this person a reward in case that there is no agreement on trade.
The private information that the players have is modeled by their types.

\subsection{The Game}
\label{subsec:The_Game}

We set up the game as prescribed in the Introduction, with the sole generalization that types are not necessarily uniformly distributed, but follow some distribution $\Prob_F$ induced by a strictly increasing and continuous cumulative distribution function $F$ on $[0,1]$.
More formally, the player set is $N=\{1,2\}$ with generic player $k$ (``they''), and both players' action sets are $A_1=A_2=\left\{nt, tr\right\}$.
The type spaces are $\Theta_1=\Theta_2=[0,1]$, the distribution on $\Theta=[0,1]^2$ is $\Prob_F^{\otimes 2}$, and a strategy for player $k$ is a mapping $\sigma_k \colon [0,1] \mapsto \Delta\left(\{nt, tr\}\right)$ from private types to probability distributions over actions.
As types are independent across players, $\bar \sigma_{k} \left(a_{k} \mid \theta_k, \Prob_{F}^{\otimes 2}\right)=\bar \sigma_{k} \left(a_{k} \mid \theta_k', \Prob_{F}^{\otimes 2}\right)$ for $k\in\{1,2\}$ and all $\theta_k, \theta_k'\in[0,1]$, so we will abbreviate this notation to $\bar \sigma_{k} \left(a_{k} \mid F\right)$.
Player 1's utility function is 
\begin{align} \label{eq:utility_function_1}
    u_1\left( a_1,a_2;\theta_1,\theta_2 \right) = 
    \begin{cases}
        1 &\text{ if } \theta_1 > \theta_2 \text{ and } a_k=nt \text{ for some } k \in \{1,2\}, \\
        1 &\text { if } \theta_1 < \theta_2 \text{ and } a_1=a_2=tr, \\
        0 &\text{ else.}
    \end{cases}
\end{align}
Ties are irrelevant as distributions are independent and atomless. 
Player 2's utility is $u_2\left( a_1,a_2;\theta_1,\theta_2 \right)= 1-u_1\left( a_1,a_2;\theta_1,\theta_2 \right)$.

The distribution function $F$ might either be known to the players, in which case we are facing a Bayesian game, or it is unknown and players are only aware of a set $\F$ of distribution functions that contains $F$.
In the latter case, we arrive at an incomplete information game with imprecise probabilistic information.
Note that as $F$ is continuous and strictly increasing on $[0,1]$, it is invertible, and $F^{-1}$ is also a continuous and strictly increasing function on $[0,1]$.

The game defined above is a \emph{constant-sum game} and can be transformed into a \emph{zero-sum game} by subtracting $\tfrac{1}{2}$ from each player's utility.
Doing so, all essential results in this paper would remain unchanged.
We avoid such a normalization in order to simplify calculations.
Still, one can think of the game as a stylized version of any zero-sum trading game in the sense that it models the characteristics that were outlined at the beginning of the subsection.

We shall define an important class of strategies, namely those in which a player offers to trade if their type is low, and refuses to do so for higher types.

\begin{definition} \label{def:cut-off_strategy}
    For $k \in \{1,2\}$, a strategy $\sigma_k$ is called \emph{cut-off strategy} if there exists a threshold $\hat\theta \in [0,1]$ such that $\sigma_k\left(tr\mid \theta\right)=1$ for $\theta<\hat\theta$ and $\sigma_k\left(nt\mid \theta\right)=1$ for $\theta>\hat\theta$.
\end{definition}

\noindent If a threshold in the sense of Definition \ref{def:cut-off_strategy} exists, it must necessarily be unique.
The action of player $k$ at $\theta=\hat\theta$ does not matter as $\{\hat\theta\}$ is a nullset.
Without loss of generality, we might assume that they offer to trade for $\theta \leq \hat\theta$.
In this case, we denote the cut-off strategy with threshold $\hat\theta$ by $\sigma^{\hat\theta}$.
Sometimes, we want to stress that the strategy is employed by player $k$ and therefore write $\sigma_k^{\hat\theta}$.

The following lemma provides the interim expected utilities of actions $nt$ and $tr$ against any cut-off strategy of the opponent, both for a cursed as well as for a rational player.

\begin{lemma} \label{lem:v(nt)_and_v(tr)}
    Let player $-k$ play the cut-off strategy $\sigma_{-k}^{\hat\theta}$.
    Then, for any type $\theta_k$ of player $k$,
    \begin{align}
        w_k\left(nt \mid \theta_k, \sigma_{-k}^{\hat\theta} \right) &= v_k\left(nt \mid \theta_k, \sigma_{-k}^{\hat\theta} \right) = F(\theta_k), \label{eq:v(nt)} \\
        w_k\left(tr \mid \theta_k, \sigma_{-k}^{\hat\theta} \right) &= \LV F\left(\theta_k\right)-F(\hat\theta_2) \RV, \text{ and} \label{eq:w(tr)}\\
        v_k\left(tr \mid \theta_k, \sigma_{-k}^{\hat\theta} \right) &= F(\hat\theta_2)\left(1-F(\theta_k)\right)+(1-F(\hat\theta_2))F(\theta_k). \label{eq:v(tr)}
    \end{align}
\end{lemma}

\noindent Equation \eqref{eq:v(tr)} nicely illustrates the phenomenon of the cursed equilibrium:
Assume that player $k$ has private type $\theta_k$ and plays ``$tr$''.
In this case, their probability of winning---which is their interim expected utility---equals the probability that the opponent also plays ``tr'' (which is $F(\hat\theta_2)$) and has a a higher type ($1-F(\theta_k)$), plus the probability that the opponent plays ``$nt$'' ($1-F(\hat\theta_2)$) and has a lower type ($F(\theta_k)$).
These events are not stochastically independent such that the probability of their intersection cannot be calculated by multiplying the probabilities of the events.
However, the cursed player treats the events as independent and splits the probabilities accordingly, which leads them to arrive at \eqref{eq:v(tr)} instead of \eqref{eq:w(tr)}.

The rational Bayesian Nash equilibrium of our game is fairly simple:
both players almost surely never offer to exchange their types.

\begin{proposition}\label{prop:BNE_is_nt}
    A strategy profile $\left(\sigma_1, \sigma_2\right)$ is a Bayesian Nash equilibrium if and only if $\sigma_1\left(nt \mid \theta_1\right)=\sigma_2\left(nt \mid \theta_2\right)=1$ for $F$-almost all $\theta_1, \theta_2 \in [0,1]$.
\end{proposition}

\noindent The logic behind the proof of Proposition \ref{prop:BNE_is_nt} is simple, and has been observed numerous times before:
If player 1 sometimes offers to exchange types, player 2 wants to ``undercut'' them and only trade at types that are lower than the average type at which player 1 wants to trade.
In equilibrium, mutual behavior of this kind leads to trade being impossible.

\subsection{How Cursedness Leads Players To Trade}
\label{subsec:How_Cursedness_Leads_Players_To_Trade}

While trade is impossible in the rational equilibrium, it can be rationalized by cursedness.

\begin{proposition}\label{prop:cNE_is_median_cutoff}
    A strategy profile $(\sigma_1,\sigma_2)$ is a cursed equilbrium if only if it is a Bayesian Nash equilibrium or if both players play a cut-off strategy with threshold $F^{-1}\left(\tfrac{1}{2}\right)$.
\end{proposition}

\noindent Let us imagine the situation of a cursed player 1 whose opponent sometimes offers to trade, that is, $\bar\sigma_{2}\left(tr\mid F\right)>0$.
Player 1 correctly identifies the value of $\bar\sigma_{2}\left(tr\mid F\right)$.
However, they do not notice that player 2 offers to trade for some of their types, and declines to do so for other types.
Instead, player 1 thinks that player 2 plays $tr$ with probability $\bar\sigma_{2}\left(tr\mid F\right)$ and $nt$ with probability $1-\bar\sigma_{2}\left(tr\mid F\right)$, independently of their type.
Against such a type-independent mixed strategy, it would be optimal for player 1 to offer to trade if and only if the probability that $\theta_2>\theta_1$ exceeds the probability that the $\theta_2<\theta_1$.
This is the case if and only if $\theta_1$ lies below the median of the distribution $F$.
In particular, in this game, the play of a cursed player coincides with behavior under level-1 reasoning (cf. \cite{nagel_1995, stahl_wilson_1995}).

\section{The Trading Game With Uncertainty}
\label{sec:The_Trading_Game_With_Uncertainty}

In Section \ref{sec:The_Trading_Game_Without_Uncertainty}, we assumed the type distribution $F$ to be fixed. 
In this section, we weaken this assumption and consider $F$ to be subject of \emph{Knightian uncertainty}, arriving at the concept of an incomplete information game with imprecise probabilistic information.
We make the following assumption on the ambiguity set $\mathcal{P}$.

\begin{assumption} \label{ass:prior_distributions}
    Types are independent and identically distributed according to some strictly increasing, continuous cumulative distribution function $F^*$ on $[0,1]$.
    Let $F_l$ and $F_h$ be strictly increasing and continuous functions on $[0,1]$ that satisfy $F_l(\theta) \leq F^*(\theta)\leq F_h(\theta)$ for all $\theta \in [0,1]$, as well as
    \begin{align}
    \label{eq:symmetry_F}
        F_l\left({F^*}^{-1}(1-\theta)\right) = 1- F_h\left({F^*}^{-1}(\theta)\right).
    \end{align}
    The ambiguity set is $\mathcal{P}_{\F} = \left\{ \Prob_F^{\otimes 2} \colon F \in \F\right\}$, where $\F$ is the set of all monotonously increasing, continuous cumulative distribution functions $F$ on $[0,1]$ that satisfy $F_l(\theta)\leq F(\theta) \leq F_h(\theta)$ for all $\theta \in [0,1]$.
\end{assumption}

\noindent Given this assumption, we will, with an abuse of notation, speak about maxmin cursed best responses and cursed Knight-Nash equilibria \emph{under $F^*$} instead of \emph{under $\Prob_{F^*}^{\otimes 2}$}, and we will use $\ubar v_k^{F^*}$ as a shorthand for $\ubar v_k^{\Prob_{F^*}^{\otimes 2}}$. 
As in Section \ref{sec:The_Trading_Game_Without_Uncertainty}, we write $\bar \sigma_{k} \left(a_{k} \mid F^*\right)$ for the average strategy of player $k$ under $F^*$.

We can interpret Assumption \ref{ass:prior_distributions} as follows:
Types are distributed according to $F^*$.
While players have some estimate or guess that $F^*$ is the true distribution, they are not definitely sure about this.
Hence, in order to use a more robust approach that accounts for other distributions, they base their decision not on $F^*$ alone, but on the distribution band $\F$ that symmetrically encloses $F^*$ and consists of all distributions between some dominating distribution $F_l$ and some dominated distribution $F_h$.
As players do not have a second-order distribution over $\F$, other approaches towards uncertainty apart from playing maxmin strategies are not readily available, and as types are independent, knowing their own type does not give them any additional information about the opponent's type or its distribution.
If $F_l=F_h$, we are in the situation of Section \ref{sec:The_Trading_Game_Without_Uncertainty}.

\begin{remark}
\label{rmk:true_F_uniform}
    Without loss of generality, we can assume that $F^*$ is the uniform distribution on $[0,1]$---that is, $F^*(\theta)=\text{Id}(\theta)=\theta$ for all $\theta\in[0,1]$---and that the functions $F_l$ and $F_h$ satisfy
    \begin{align}
    \label{eq:symmetry_F_uniform}
        F_l(1-\theta)=1-F_h(\theta) \text{ for all } \theta \in [0,1].
    \end{align}
    Indeed, under Assumption \ref{ass:prior_distributions}, we can define
    \begin{align*}
        \mathcal{G} \colonequals \F \circ {F^*}^{-1} = \left\{ G \colon \text{there is } F \in \F \text{ such that }G(\theta) = F({F^*}^{-1}(\theta)) \text{ for all } \theta \in [0,1] \right\}.
    \end{align*}
    Then, Assumption \ref{ass:prior_distributions} is also satisfied if types are uniformly distributed and the ambiguity set is $\left\{ \Prob_{G}^{\otimes 2} \colon G \in \mathcal{G}\right\}$.
    In particular, for the upper and lower bounds of $\mathcal{G}$ (these are $F_h\circ{F^*}^{-1}$ and $F_h\circ{F^*}^{-1}$, respectively), Equation \eqref{eq:symmetry_F_uniform} holds.
    Identifying types with quantiles, we can always assume that $F^*=\text{Id}$ and that the upper and lower bounds of $\F$ satisfy \eqref{eq:symmetry_F_uniform}, as a situation in which some player is of type $\theta$ and faces uncertainty over the set $\F$ (which symmetrically encloses $F^*$) is behaviorally and mathematically equivalent to a situation in which this player is of type ${F^*}(\theta)$ and faces uncertainty over the set $\mathcal{G}$ (which symmetrically encloses $\text{Id}$).
\end{remark}

\noindent Let us give some examples and counterexamples for Assumption \ref{ass:prior_distributions}.

\begin{figure}
\begin{subfigure}[t]{0.5\textwidth}
    \centering
    \begin{tikzpicture}[scale=6]
            \draw[->] node[left] {0} (0,0) -- (1.1,0) node[right] {$\theta$};
        	\draw[->] (0,0) -- (0,1.1);
            \draw[scale=1,domain=0:0.5,variable=\t]  plot ({\t},{0.25*(\t)});
            \draw[scale=1,domain=0.5:1,variable=\t]  plot ({\t},{-0.75+1.75*(\t)});
            \draw[scale=1,domain=0:0.5,variable=\t]  plot ({\t},{1.75*(\t)});
            \draw[scale=1,domain=0.5:1,variable=\t]  plot ({\t},{0.75+0.25*(\t)});
            \draw[dotted] (1,1) -- (0,1) node[left] {$1$};
            \draw[dotted] (1,1) -- (1,0) node[below] {$1$};
            \draw[dotted] (2/7,0.5) -- (0,0.5) node[left] {$\tfrac{1}{2}$};
            \draw[dotted] (2/7,0.5) -- (2/7,0) node[below] {$F_{h, \kappa}^{-1}\left(\tfrac{1}{2}\right)$};
            \draw[dotted] (5/7,0.5) -- (2/7,0.5);
            \draw[dotted] (5/7,0.5) -- (5/7,0) node[below] {$F_{l, \kappa}^{-1}\left(\tfrac{1}{2}\right)$};
            \draw (0.3,0.7) node {$F_{h,\kappa}$};
            \draw (0.92,0.7) node {$F_{l,\kappa}$};
            \draw[scale=1, domain=0:1, variable=\t, dashed]  plot ({\t},{\t});
            \draw[dotted] (0.5,0.5) -- (0.5,0) node[below] {$\tfrac{1}{2}$};
        \end{tikzpicture}
    \caption{$F_{l,\kappa}$ and $F_{h,\kappa}$ for $\kappa=0.75$}
    \label{subfig:F_contamination}
\end{subfigure}%
\begin{subfigure}[t]{0.5\textwidth}
    \centering
    \begin{tikzpicture}[scale=6]
            \draw[->] node[left] {0} (0,0) -- (1.1,0) node[right] {$\theta$};
        	\draw[->] (0,0) -- (0,1.1);
            \draw[scale=1,domain=0:(2/3),variable=\t]  plot ({\t},{0.5*(\t)});
            \draw[scale=1,domain=(2/3):1,variable=\t]  plot ({\t},{2*(\t)+1-2});
            \draw[scale=1,domain=0:(1/3),variable=\t]  plot ({\t},{2*(\t)});
            \draw[scale=1,domain=(1/3):1,variable=\t]  plot ({\t},{0.5*(\t)+1-0.5});
            \draw[dotted] (1,1) -- (0,1) node[left] {$1$};
            \draw[dotted] (1,1) -- (1,0) node[below] {$1$};
            \draw[dotted] (0.25,0.5) -- (0,0.5) node[left] {$\tfrac{1}{2}$};
            \draw[dotted] (0.25,0.5) -- (0.25,0) node[below] {$F_{h, a}^{-1}\left(\tfrac{1}{2}\right)$};
            \draw[dotted] (0.75,0.5) -- (0.25,0.5);
            \draw[dotted] (0.75,0.5) -- (0.75,0) node[below] {$F_{l, a}^{-1}\left(\tfrac{1}{2}\right)$};
            \draw (0.25,0.7) node {$F_{h,a}$};
            \draw (0.92,0.7) node {$F_{l,a}$};
            \draw[scale=1, domain=0:1, variable=\t, dashed]  plot ({\t},{\t});
            \draw[dotted] (0.5,0.5) -- (0.5,0) node[below] {$\tfrac{1}{2}$};
        \end{tikzpicture}
    \caption{$F_{l,a}$ and $F_{h,a}$ for $a=2$}
    \label{subfig:F_triangle}
\end{subfigure}%
\vspace{.8cm}
\centering
\\
\begin{subfigure}[t]{0.5\textwidth}
    \centering
    \begin{tikzpicture}[scale=6]
            \draw[->] node[left] {0} (0,0) -- (1.1,0) node[right] {$\theta$};
        	\draw[->] (0,0) -- (0,1.1);
            \draw[scale=1,domain=0:0.8,variable=\t]  plot ({\t},{\t+0.2});
            \draw[scale=1,domain=0.8:1,variable=\t]  plot ({\t},{1});
            \draw[-] (0,0) -- (0.2,0);
            \draw[scale=1,domain=0.2:1,variable=\t]  plot ({\t},{\t-0.2});
            \draw[dotted] (1,1) -- (0,1) node[left] {$1$};
            \draw[dotted] (1,1) -- (1,0) node[below] {$1$};
            \draw[dotted] (0.3,0.5) -- (0,0.5) node[left] {$\tfrac{1}{2}$};
            \draw[dotted] (0.3,0.5) -- (0.3,0) node[below] {$F_{h, \eps}^{-1}\left(\tfrac{1}{2}\right)$};
            \draw[dotted] (0.7,0.5) -- (0.3,0.5);
            \draw[dotted] (0.7,0.5) -- (0.7,0) node[below] {$F_{l, \eps}^{-1}\left(\tfrac{1}{2}\right)$};
            \draw (0.3,0.6) node {$F_{h,\eps}$};
            \draw (0.9,0.6) node {$F_{l,\eps}$};
            \draw[scale=1, domain=0:1, variable=\t, dashed]  plot ({\t},{\t});
            \draw[dotted] (0.5,0.5) -- (0.5,0) node[below] {$\tfrac{1}{2}$};
        \end{tikzpicture}
    \caption{$F_{l,\eps}$ and $F_{h,\eps}$ for $\eps=0.2$}
    \label{subfig:F_eps_band}
\end{subfigure}%
\label{fig:different_quantifications_uncertainty}
\caption{Different Quantifications of Uncertainty}
\end{figure}
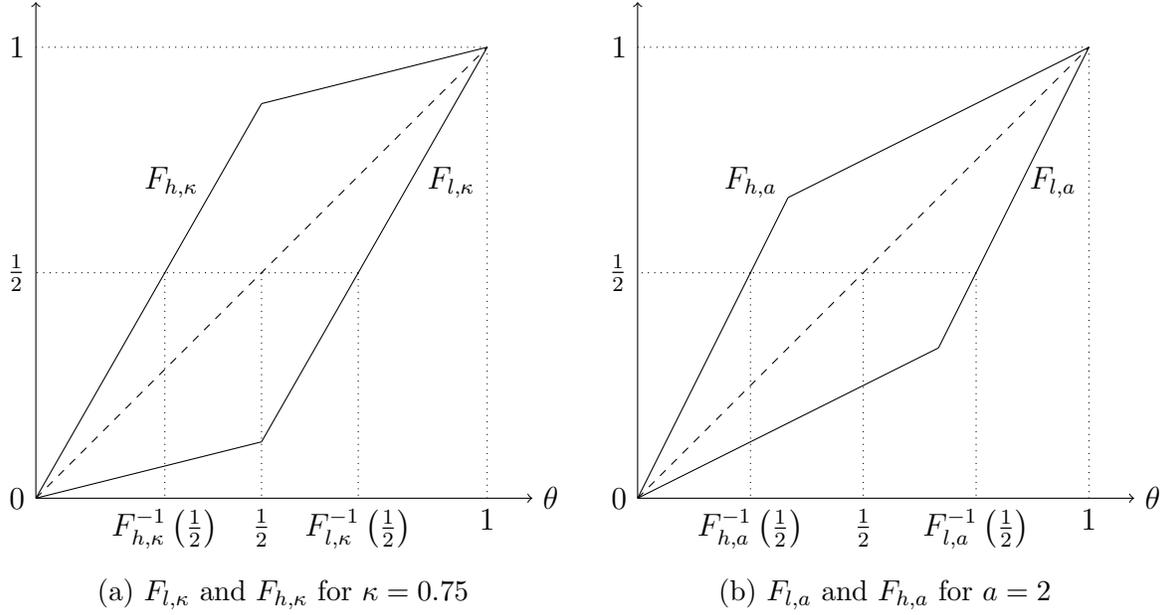
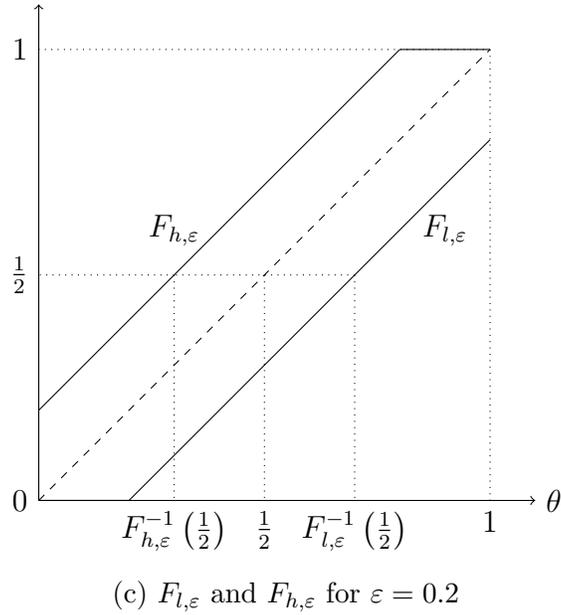

\begin{example}
\label{ex:parametrizations}
Following Remark \ref{rmk:true_F_uniform}, let the true distribution be the uniform distribution.
\begin{enumerate}
    \item\label{it:contamination} For $\kappa \in [0,1)$ and $\theta\in [0,1]$, let 
    \begin{align*}
        F_{l,\kappa}(\theta) &= \begin{cases}
            (1-\kappa)\theta &\text{ for } \theta \leq \tfrac{1}{2},\\
            -\kappa + (1+\kappa)\theta &\text{ for } \theta \geq \tfrac{1}{2},
        \end{cases} \\
        F_{h,\kappa}(\theta) &= \begin{cases}
            (1+\kappa)\theta &\text{ for } \theta \leq \tfrac{1}{2}, \\
            \kappa + (1-\kappa)\theta &\text{ for } \theta \geq \tfrac{1}{2},
        \end{cases}
    \end{align*}
    and let $\F_{\kappa}$ be the set of all continuous cumulative distribution functions $F$ on $[0,1]$ that satisfy $F_{l, \kappa}(\theta)\leq F(\theta) \leq F_{h, \kappa}(\theta)$ for all $\theta \in [0,1]$.
    It is immediate to verify that Assumptions \ref{ass:prior_distributions} is satisfied with $F^*=\text{Id}$, in particular, Equation \eqref{eq:symmetry_F_uniform} holds true.
    The distribution set $\F$ includes the so-called \emph{contamination of the uniform distribution on $[0,1]$}, that is, the set of all probability distributions on $[0,1]$ that are absolutely continuous with respect to the Lebesgue measure and whose Lebesgue density lies between $1-\kappa$ and $\kappa$ (cf. \cite{decerf_riedel_2020}).\footnote{
    Note that we allow for distributions that do not have a Lebesgue density.
    There are (pathologic) examples of cumulative distribution functions that are continuous but not absolutely continuous and therefore do not have density functions, for instance the Cantor distribution.
    }
    The limit case of $\kappa=0$ corresponds to the unambiguous uniform distribution.
    The distribution functions $F_{l,\kappa}$ and $F_{h,\kappa}$ are depicted in Subfigure \ref{subfig:F_contamination}.

     \item\label{it:triangles} For $a \in [1, \infty)$, let 
     \begin{align*}
        F_{l,a}(\theta) &= \begin{cases}
            \frac{1}{a}\theta &\text{for } \theta \leq \frac{a}{a+1}, \\
            a\theta+1-a &\text{for } \theta \geq \frac{a}{a+1},
        \end{cases} \\
        F_{h,a}(\theta) &= \begin{cases}
            a\theta &\text{for } \theta \leq \frac{1}{a+1}, \\
            \frac{1}{a}\theta+1-\frac{1}{a} &\text{for } \theta \geq \frac{1}{a+1}.
        \end{cases} 
    \end{align*}
    Let $\F_a$ be the set of all continuous distribution functions between $F_{l,a}$ and $F_{h,a}$, see Subfigure \ref{subfig:F_triangle}.
    Assumption \ref{ass:prior_distributions} is again satisfied.
    As in \ref{it:contamination}., the dominated and the dominant distribution form a triangle around the uniform distribution, 
    However, in this parametrization, the kinks do not lie at $\frac{1}{2}$, but parameter-dependently at $\frac{1}{a+1}$ and $\frac{a}{a+1}$.
    If $a$ approaches infinity, this allows to approximate the case of full uncertainty that includes every continuous distribution on $[0,1]$.
    If $a=1$, there is no uncertainty.
    
    \item\label{it:eps_band} For $\eps \in \left[0,\tfrac{1}{2}\right)$, let
    \begin{align*}
        F_{l, \eps}(\theta) &= \max\{\theta-\eps, 0\} \text{ for } \theta \in [0,1), \quad F_{l, \eps}(1)=1,\\
        F_{h, \eps}(\theta) &= \min\{\theta+\eps, 1\} \text{ for } \theta \in [0,1],
    \end{align*}
    and let $\F_{\eps}$ be defined as above, see Subfigure \ref{subfig:F_eps_band}.
    In this case, $\F_{\eps}$ contains all continuous distribution functions on $[0,1]$ that lie in an $\eps$-ball around the identity with respect to the supremum norm.
    There is no uncertainty in the limit case $\eps=0$.
    While Equation \eqref{eq:symmetry_F_uniform} is valid for all $\theta\in(0,1)$, $F_{l, \eps}$ and $F_{h, \eps}$ are not continuous everywhere, and piece-wise constant.
    Hence, Assumption \ref{ass:prior_distributions} is not satisfied.
    \qedhere
\end{enumerate}
\end{example}

\noindent Before starting with the analysis, we shall comment on the monotonicity assumptions in Assumption \ref{ass:prior_distributions}.

\begin{remark} \label{rmk:monotonicity_assumptions}
    On the one hand, Assumption \ref{ass:prior_distributions} prescribes $F^*$, $F_l$, and $F_h$ to be strictly increasing.
    This assumption guarantees the uniqueness of the equilibria that will be characterized in the following subsections, and it largely simplifies the analysis.
    One can check that if we allow these functions to be constant on intervals, all the equilibria found in these subsections remain equilibria.
    However, there might then also be further equilibria that differ from those characterized below on sets that are nullsets with respect to some of functions $F^*$, $F_l$, or $F_h$.
    In order to avoid cumbersome reformulations of the statements and keep the proofs as simple as possible, we make the simplifying assumption that the relevant functions are strictly monotonous.

    On the other hand, the assumption allows for $F \in \F \setminus \{F_l, F_h, F^*\}$ to be constant on intervals such that the set $\F$ is closed in $C\left([0,1]\right)$ with respect to the supremum norm.
    Including piece-wise constant functions in $\F$ will guarantee that all the relevant infima are attained by some function in $\F$ and can be replaced by minima.\footnote{
    As $\F$ is closed, but not necessarily compact, this does not directly follow from the assumptions, but has to be proven.
    }
    If we were to define $\F$ as the set of all \emph{strictly} increasing functions between $F_l$ and $F_h$, some of the infimal values that appear in the proofs would not be attained by some function in $\F$, but could only be approximated.
    However, all the statements could be proven along the same lines and would remain true.
\end{remark}

\subsection{Benchmark: Uncursed Players Do Not Trade Under Uncertainty}
\label{subsec:Benchmark_Uncursed_Players_Do_Not_Trade_Under_Uncertainty}

Let us start with analyzing the behavior of uncursed players.
That is, in this subsection, we are looking for Knight-Nash equilibria of the game (Definition \ref{def:knight_nash_equilibrium}).
It seems intuitive that in a trading game as the one investigated in this work, players will be willing to offer an exchange of types if their own type is low, and refuse to do so if it is high.
Thus, it seems plausible that both players will play cut-off strategies as defined in Definition \ref{def:cut-off_strategy}.
If this is assumed, nothing changes compared to the case without uncertainty:
The only equilibrium is where both never offer to trade.

\begin{proposition} \label{prop:KNE_is_nt}
    Under Assumption \ref{ass:prior_distributions}, the only Knight-Nash equilibrium in cut-off strategies is $\left(\sigma_1^0, \sigma_2^0\right)$.
\end{proposition}

\noindent Proposition \ref{prop:KNE_is_nt} comes at no surprise, as we already know that in the Bayesian Nash equilibrium without uncertainty, trade almost surely never happens (Proposition \ref{prop:BNE_is_nt}), and uncertainty usually further reduces the amount of trade in trading games \citep{dow_werlang_1992, marinacci_2015}.

If we do not restrict ourselves to cut-off strategies, there can be further equilibria, as illustrated by the following example.

\begin{example} \label{ex:uncursed_equilibrium_not_cut-off}
    Let Assumption \ref{ass:prior_distributions} be satisfied, and further assume that $F_l(\theta)< F_h(\theta)$ for all $\theta \in (0,1)$.
    Let $A = \left[\ubar a, \bar a \right] \subset (0,1)$ be a nondegenerate, closed interval such that $F_h \left(\ubar a\right) \leq F_l \left(\bar a \right)$.
    (For example, choose $\bar a \in \left(0, 1\right)$ arbitrary and set $\ubar a \colonequals F_h^{-1}\left(F_l\left(\bar a\right)\right)$.)
    Assume that player 1 plays the strategy $\sigma_1^A$ defined via 
    \begin{align*}
        \sigma_1^A\left(tr \mid \theta_1 \right)= \begin{cases}
            1 &\text{ if } \theta_1 \in A, \\
            0 &\text{ if } \theta_1 \notin A,
        \end{cases}
    \end{align*}
    that is, player 1 offers to trade if and only if their type lies in $A$.
    As $\ubar a >0$, this is not a cut-off strategy.
    We will show that $\sigma_2^{\left\{\ubar a\right\}}$, defined as the strategy in which player 2 offers to trade if and only if their type is $\theta_2 = \ubar a$,\footnote{
    This strategy is not to be confused with $\sigma_2^{\ubar a}$, the strategy in which player 2 trades if and only if their types lies in $\left[0, \ubar a\right]$.
    } 
    is a best response for player 2 against $\sigma_1^A$.
    As $\sigma_1^A$ is also a best response against $\sigma_2^{\left\{\ubar a\right\}}$---under any probability distribution in $\F$, there will almost surely never be trade anyway, so for player 1, every strategy is a best response---the strategy profile $\left(\sigma_1^A, \sigma_2^{\left\{\ubar a\right\}}\right)$ constitutes a Knight-Nash equilibrium in which both players do not play  cut-off strategies.
    For any type $\theta_2$, player 2's values of not offering to trade and of doing so are given by
    \begin{align}
        \ubar w_2 \left( nt \mid \theta_2, \sigma_1^A \right) &= \inf_{F \in \F} F\left(\theta_2\right) = F_l\left(\theta_2\right), \text{ and} \nonumber \\
         \ubar w_2 \left( tr \mid \theta_2, \sigma_1^A \right) &= \inf_{F \in \F} \Prob_F \left(a_1=tr, \theta_1 > \theta_2 \right) + \Prob_F \left(a_1=nt, \theta_1 < \theta_2 \right) \nonumber \\
        &= \inf_{F \in \F} \Prob_F \left(\theta_1 \in A, \theta_1 > \theta_2 \right) + \Prob_F \left(\theta_1 \notin A, \theta_1 < \theta_2 \right),\label{eq:uncursed_value(tr)_vs_sigma^A}
    \end{align}
    respectively.
    We will now estimate the value of \eqref{eq:uncursed_value(tr)_vs_sigma^A} against $F_l\left(\theta_2\right)$ for $\theta_2 < \ubar a$, $\theta_2 \in A$, and $\theta_2 > \bar a$ separately.

    For any $\theta_2 < \ubar a$, there exists a function $\tilde{F} \in \F$ (whose exact shape depends on $\theta_2$) that satisfies $\tilde{F}\left(\theta_2\right)= F_l\left(\theta_2\right)$ as well as $\tilde{F}(a) = F_h \left(\ubar a\right)= F_l\left(\bar a\right)$ for all $a \in A$.
    In this case, as $\tilde{F}$ is constant on $A$, $\Prob_{\tilde{F}}(A)=0$ such that
    \begin{align*}
        \ubar w_2 \left( tr \mid \theta_2, \sigma_1^A \right) = \inf_{F \in \F} \Prob_F(A) + F\left(\theta_2\right) \leq \Prob_{\tilde{F}}(A) + \tilde{F}\left(\theta_2\right)= F_l\left(\theta_2\right) = \ubar w_2 \left( nt \mid \theta_2, \sigma_1^A \right).
    \end{align*}
    If $\theta_2 \in A$, then
    \begin{align*}
        \ubar w_2 \left( tr \mid \theta_2, \sigma_1^A \right) = \inf_{F \in \F} F\left(\bar a\right) - F\left(\theta_2\right) + F\left(\ubar a\right).
    \end{align*}
    For $\theta_2=\ubar a$, this expression simplifies to $\inf_{F \in \F} F \left(\bar a\right)=F_l \left(\bar a\right)>F_l\left(\theta_2\right)$.
    For $\theta_2 \in \left(\ubar a, \bar a\right]$, there exists a function $\hat{F} \in \F$ with $\hat{F}\left(\ubar a\right)=F_l\left(\ubar a\right)$ and $\hat{F}\left(\theta_2\right) = \hat{F}\left(\bar a\right)$ such that
    \begin{align*}
        \ubar w_2 \left( tr \mid \theta_2, \sigma_1^A \right) \leq \hat{F}\left(\bar a\right) - \hat{F}\left(\theta_2\right) + \hat{F}\left(\ubar a\right) = F_l\left(\ubar a\right) \leq F_l\left(\theta_2\right) = \ubar w_2 \left( nt \mid \theta_2, \sigma_1^A \right).
    \end{align*}
    Finally, for $\theta_2 > \bar a$, it holds that
    \begin{align*}
        \ubar w_2 \left( tr \mid \theta_2, \sigma_1^A \right) &= \inf_{F \in \F} 0 + F\left(\theta_2\right) - F\left(\bar a\right) + F\left(\ubar a\right) \\
        &\leq F_l\left(\theta_2\right) - F_l\left(\bar a\right) + F_l\left(\ubar a\right) \leq F_l\left(\theta_2\right).
    \end{align*}
    
    \noindent We have thus shown that player 2 prefers to trade their type if $\theta_2 = \ubar a$, and (weakly) prefers not to do for $\theta_2 \neq \ubar a$.
    Hence, $\sigma_2^{\left\{\ubar a\right\}}$ is indeed a best response against $\sigma_1^A$.
\end{example}

\noindent Example \ref{ex:uncursed_equilibrium_not_cut-off} shows that there exists a Knight-Nash equilibrium in which both players do not play cut-off strategies.
The key aspect in the construction of the example is the existence of an interval $A$ that has zero probability under some probability measures in $\F$.
Even if we were to restrict $\F$ to the set of all strictly increasing cumulative distribution functions between $F_l$ and $F_h$, the calculations in Example \ref{ex:uncursed_equilibrium_not_cut-off} would go through, with the exception that $\tilde{F}$ and $\hat{F}$ would have to be approximated by sequences of functions in $\F$, which would not change the value of the infimum.
The crucial point is that the probability of $A$ is not bounded away from 0, unlike, for every fixed $\theta>0$, the probability of $[0, \theta]$.

The proof of Proposition \ref{prop:KNE_is_nt} and Example \ref{ex:uncursed_equilibrium_not_cut-off} do not use the symmetric structure of $\F$ as prescribed by Equation \eqref{eq:symmetry_F}.
Hence, the results from this subsection still hold if we assume that $F_l$ and $F_h$ are some arbitrary continuous and strictly increasing functions on $[0,1]$ with $F_l(\theta)\leq F_h(\theta)$ for all $\theta$, and $\F$ consists of all functions between these.

\subsection{How Uncertainty Leads Cursed Players To Trade Even More}
\label{subsec:How_Uncertainty_Leads_Cursed_Players_To_Trade_Even_More}

We shall now investigate how cursed players behave if they face uncertainty over the distribution of types, and find cursed Knight-Nash equilibria in the sense of Definition \ref{def:cursed_knight_nash_equilibrium_given_P*}.
One such equilibrium is that both players almost surely (under any probability distribution in $\F$, in particular $F^*$-almost surely) never offer to trade:
in this case, both players are indifferent between all their strategies as there will never be trade anyway.
We call such an equilibrium \emph{trivial}.
Note that such a strategy profile also constitutes a Bayesian Nash Equilibrium and a Knight-Nash equilibrium by Propositions \ref{prop:BNE_is_nt} and \ref{prop:KNE_is_nt}.
In the following, we look for \emph{non-trivial} equilibria in which at least one player sometimes offers to trade.
The following remark shows that whenever a non-trivial equilibrium exists, cursed players will prefer this over the trivial one.

\begin{remark} \label{rmk:cursed_player_prefers_non-trivial_equilibrium}
    Whenever a non-trivial cursed Knight-Nash equilibrium exists, it is weakly preferred over the trivial one:
    If, in such an equilibrium $\left(\sigma_1^*, \sigma_2^*\right)$, player $k$ decides to offer to trade at type $\theta_k$, they will only do so if
    \begin{align*}
        \ubar v_k^{F^*} \left(tr \mid \theta_k, \sigma_{-k}^*\right) \geq \ubar v_k^{F^*}  \left(nt \mid \theta_k, \sigma_{-k}^*\right) = \inf_{F \in \F} F(\theta_k) = F_l(\theta_k).
    \end{align*}
    Hence, in the non-trivial equilibrium, their perceived interim expected utility at type $\theta_k$ is at least $F_l(\theta_k)$ which is their perceived interim expected utility in any trivial equilibrium.
    (We will later see that there exist non-trivial equilibria in which this inequality is strict for some types.)
    It therefore seams reasonable that cursed players would rather play the non-trivial equilibrium $\left(\sigma_1^*, \sigma_2^*\right)$ than any trivial one, as they statewise weakly prefer the non-trivial equilibrium over the trivial one.
    This holds in particular as every strategy is a cursed best response against any strategy in which the opponent almost surely (under any distribution in $\F$) never offers to trade.
    Hence, cursed players do not see any danger in transitioning from a trivial equilibrium to a preferred non-trivial one.
\end{remark}

\noindent Following Example \ref{ex:uncursed_equilibrium_not_cut-off}, an uncursed Knight-Nash equilibrium need not be of cut-off type.
If players are cursed, this is no longer the case, and only cut-off strategies are played in equilibrium.
These cut-off must always lie between the median of the dominated and the median of the dominant distribution of $\F$.

\begin{proposition} \label{prop:cut-offs_ambiguity_in_middle_under_F*}
    Under Assumption \ref{ass:prior_distributions}, in every non-trivial cursed Knight-Nash equilibrium under $F^*$, both players play cut-off strategies with thresholds in $\left[ F_h^{-1}\left(\tfrac{1}{2}\right), F_l^{-1}\left(\tfrac{1}{2}\right) \right]$.
\end{proposition}

\noindent In the absence of uncertainty, Proposition \ref{prop:cNE_is_median_cutoff} has shown that a cursed player 1 is indifferent between exchanging their type and not doing so if and only if the probability that $\theta_1>\theta_2$ equals the probability that $\theta_1<\theta_2$, leading to a cursed equilibrium in which both players play a cut-off strategy with the median of the probability distribution.
Proposition \ref{prop:cut-offs_ambiguity_in_middle_under_F*} generalizes this result to the uncertain case: in any non-trivial cursed equilibrium,\footnote{
At this point, we do not yet know whether such an equilibrium actually exists.
We will see that it does.
}
player 1 must have a type $\theta_1$ at which they they do not know whether $\theta_1>\theta_2$ is more likely than $\theta_1<\theta_2$, or vice versa.

In order to be able to further classify the possible thresholds and their properties, we first assume that this set is a singleton.
In this case, there is no uncertainty about the median of the distribution of types: for all $F\in\F$, $F^{-1}\left(\tfrac{1}{2}\right)$ coincides, or equivalently, $F_h^{-1}\left(\tfrac{1}{2}\right)= F_l^{-1}\left(\tfrac{1}{2}\right)$.
There is then a unique fully cursed Knight-Nash equilibrium in which both players play the cut-off strategy with the type for which they know that it is equally likely that their opponent has a higher or a lower type.

\begin{corollary} \label{cor:fully_cursed_Knight_Nash_equilibrium_certain_median_under_F*}
    In addition to Assumption \ref{ass:prior_distributions}, let $F_h^{-1}\left(\tfrac{1}{2}\right)= {F^*}^{-1}\left(\frac{1}{2}\right)=F_l^{-1}\left(\tfrac{1}{2}\right)$. 
    Then, there is a cursed Knight-Nash equilibrium under $F^*$ in which both players play the cut-off strategy with threshold ${F^*}^{-1}\left(\tfrac{1}{2}\right)$, and this equilibrium is only non-trivial one.
\end{corollary}

\noindent As long as the median of the distribution is certain, the unique non-trivial cursed equilibrium consists of both players trading if and only if their type lies below ${F^*}^{-1}\left(\tfrac{1}{2}\right)$.
In particular, as $F^*$ is the true underlying distribution, the probability that they offer to trade is $\frac{1}{2}$.
If the median is uncertain, this is no longer the case.

To see this, imagine a cursed player 1 who has observed that player 2 trades in half of all games and who is themselves of type $\theta_1={F^*}^{-1}\left(\tfrac{1}{2}\right)$.
In case of a certain median, they would now be indifferent between playing $nt$ and $tr$.
If, however, the median is uncertain, their infimal expected utility from not offering to trade is
\begin{align*}
    \inf_{F \in \F} \Prob_F \left(\theta_{2}<{F^*}^{-1}\left(\tfrac{1}{2}\right)\right) = F_l\left({F^*}^{-1}\left(\tfrac{1}{2}\right)\right) < \tfrac{1}{2},
\end{align*}
while their perceived utility from offering to trade is
\begin{align*}
    &\Prob_{F^*}\left(a_{2}=nt\right) \Prob_F \left(\theta_{2}<\tfrac{1}{2}\right) + \Prob_{F^*}\left(a_{2}=tr\right) \Prob_F \left(\theta_{2}>\tfrac{1}{2}\right) \\
    = &\tfrac{1}{2} \Prob_F \left(\theta_{2}<\tfrac{1}{2}\right) + \tfrac{1}{2} \Prob_F \left(\theta_{2}>\tfrac{1}{2}\right) \\
    = &\tfrac{1}{2}
\end{align*}
\emph{for any possible distribution function} $F \in \F$.
Hence, while the value of not offering to trade is subject to minimization, the (cursed) value of a trade offer is not uncertain at all, even if the true type-generating distribution is unknown!
Therefore, a player who is of type ${F^*}^{-1}\left(\tfrac{1}{2}\right)$ and faces an opponent who trades in half of the cases will definitely want to trade.
In this game, the quantifiable risk about the opponent's action, as perceived by a cursed player, serves as a hedge against the fundamental uncertainty about their probability distribution.
Such an observation has already been made in the early days of decision-theoretic analysis of uncertainty, most prominently by \citet{raiffa_1961} in his critique against the seminal paper of \citet{ellsberg_1961}.

The reasoning above suggests that in a cursed equilibrium, thresholds must exceed ${F^*}^{-1}\left(\frac{1}{2}\right)$.
The following theorem shows that this is indeed true.

\begin{theorem} \label{thm:existence_chracterization_fully_cursed_KNE_under_F*}
    In addition to Assumption \ref{ass:prior_distributions}, let $F_h^{-1}\left(\tfrac{1}{2}\right)< F_l^{-1}\left(\tfrac{1}{2}\right)$.
    In any fully cursed Knight-Nash equilibrium under $F^*$, players 1 and 2 play cut-off strategies with thresholds greater than ${F^*}^{-1}\left(\frac{1}{2}\right)$.
    Moreover, there always exists a non-trivial cursed Knight-Nash equilibrium under $F^*$ in which both players play the cut-off strategy $\sigma^{\theta^*}$ with the threshold $\theta^*$ that satisfies
    \begin{align} \label{eq:symmetric_cursed_equilibrium_under_F*}
      F_l\left(\theta^*\right) = \left(1-2\theta^*\right) F_h\left(\theta^*\right) + \theta^*.
   \end{align}
   This equilibrium is unique within the class of symmetric strategy profiles.
\end{theorem}

\noindent While uncertainty usually reduces trade (cf. \cite{dow_werlang_1992}), in combination with cursedness, it might have the opposite effect.
To see why this is the case, recall that a player who does not trade their type profits from the opponent having a lower type than themselves.
Being uncertainty averse, they will therefore take the minimal probability of this event in order to evaluate their interim expected value of refusing to trade.
In particular, in this game, refusing to trade does not constitute a ``safe outside option'', but still yields an uncertain final payoff.
When engaging in trade, a cursed player can seemingly profit from the opponent having a high or a having a low type:
They profit from high-type opponents who engage in trade, and from low-type opponents who refuse to do so.\footnote{
As already remarked before, this is what a rational, or even cursed, opponent is actually \emph{not doing}.
However, cursed players do not realize this connection, behaving as if all types of opponents play the same mixed strategy.
}
While for a given own type and strategy of the opponent, one of these effects is typically stronger and determines which distribution in the uncertainty set minimizes the expected utility (In equilibrium, it is the first one, leading to $F_h$ being the minimizing distribution function in $\F$.), the other effect is still present and dampens the strength of the minimization.
Thus, while uncertainty decreases both the expected values of trading and of not trading, the decrease in the value of trading is mitigated by the fact that a cursed player who engages in trade seemingly profits from the opponent having a high and having a low type, while the player who refuses to trade only profits from low-type opponents.

The phenomenon that uncertainty reduces the willingness to trade even above the cursed equilibrium's level sheds light on the occasional observation that players deviate more from the standard rationality prediction than under full cursedness \citep{eyster_rabin_2005, christensen_2008}.
Our results suggest that one possible explanation for such behavior could lie in uncertainty about their opponents' types.

\begin{figure}
\begin{subfigure}[t]{0.5\textwidth}
    \centering
    \includegraphics[width=\textwidth]{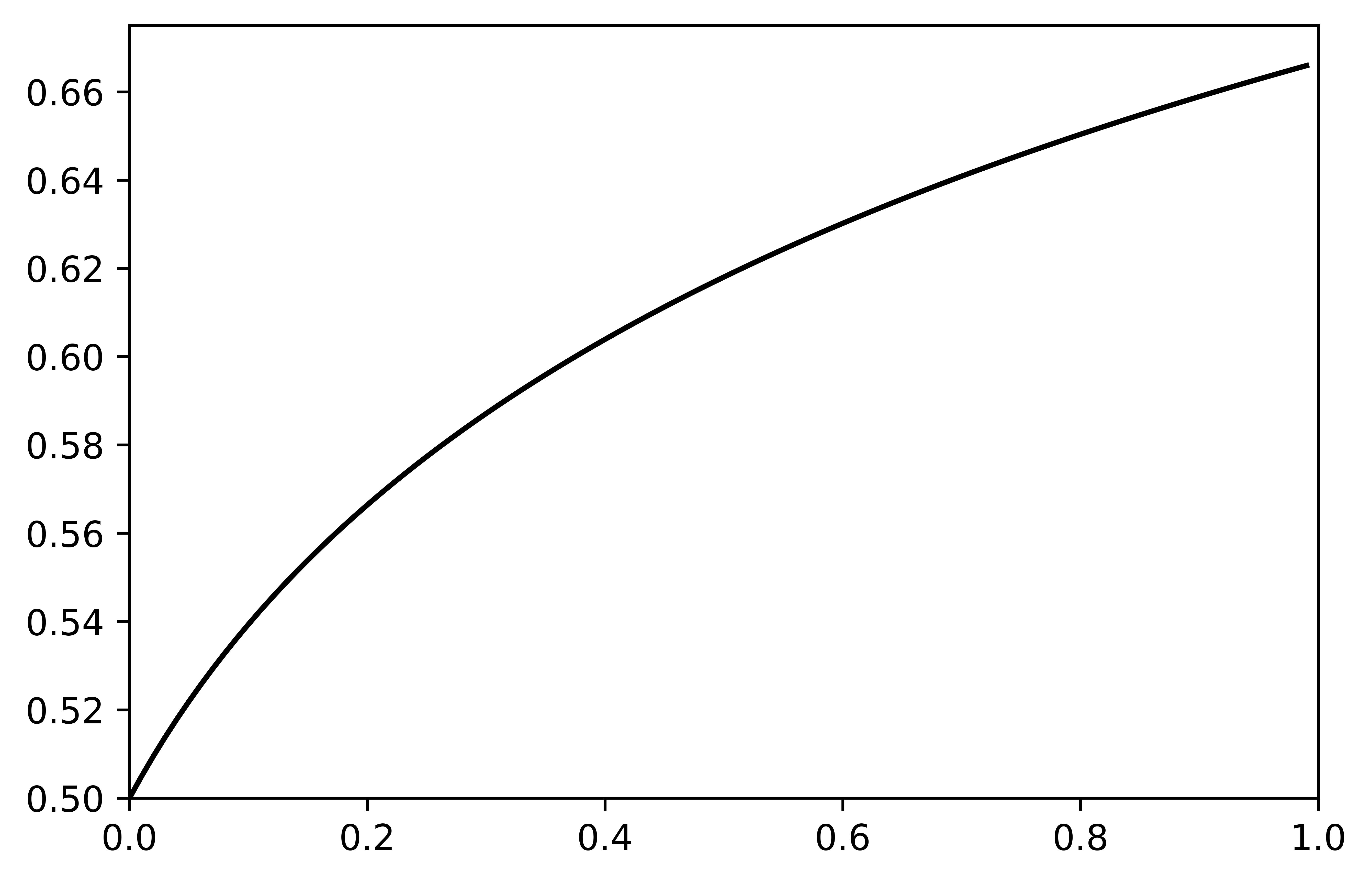}
    \caption{Threshold $\theta_{\kappa}^*$}
    \label{subfig:theta^*_kappa_F*}
\end{subfigure}%
\begin{subfigure}[t]{0.5\textwidth}
    \centering
    \includegraphics[width=\textwidth]{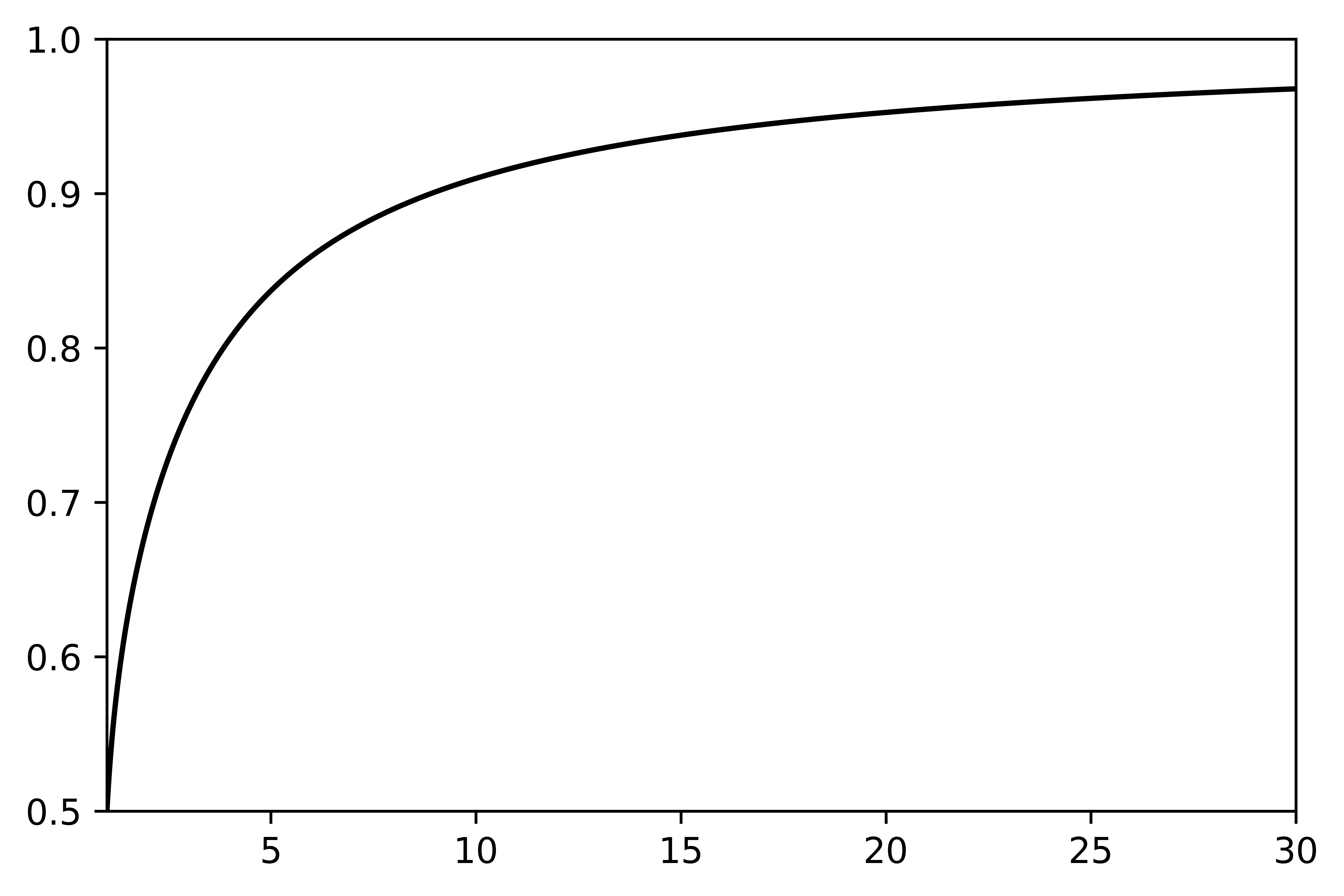}
    \caption{Threshold $\theta_{a}^*$}
    \label{subfig:theta^*a_F*}
\end{subfigure}%
\vspace{.8cm}
\centering
\\
\begin{subfigure}[t]{0.5\textwidth}
    \centering
    \includegraphics[width=\textwidth]{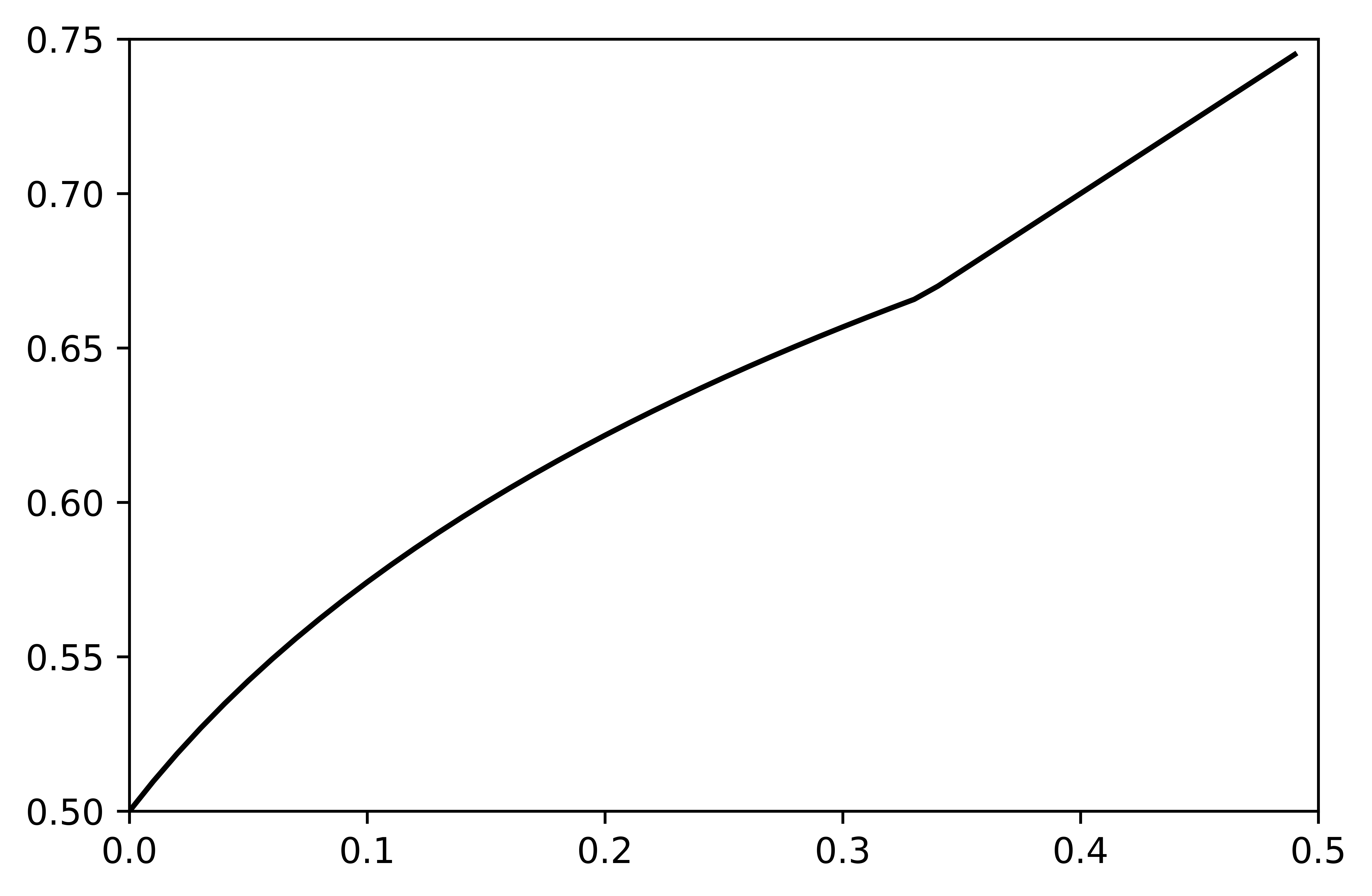}
    \caption{Threshold $\theta_{\eps}^*$}
    \label{subfig:theta^*_eps_F*}
\end{subfigure}%
\caption{Thresholds in Example \ref{ex:examples_revisited}}
\label{fig:thresholds_F*}
\end{figure}

\begin{example}
\label{ex:examples_revisited}
    For $\kappa \in (0,1)$, $a \in (1, \infty)$, and $\eps \in \left(0, \tfrac{1}{2}\right)$, let $\F_{\kappa}$, $\F_a$, and $\F_{\eps}$ be as in Example \ref{ex:parametrizations}, and let $F^*=\text{Id}$.
    We know from Theorem \ref{thm:existence_chracterization_fully_cursed_KNE_under_F*} that there exist symmetric cursed Knight-Nash equilibria with thresholds in $\left(\tfrac{1}{2},F_l^{-1}\left(\tfrac{1}{2}\right)\right)$ that can be found by solving Equation \eqref{eq:symmetric_cursed_equilibrium_under_F*}.\footnote{
    As already remarked in Example \ref{ex:parametrizations}, $\F_{\eps}$ does not satisfy Assumption \ref{ass:prior_distributions}.
    However, investigating the proofs shows that for all the statements, it is sufficient that $F_l$ and $F_h$ are continuous in $\left[F_h^{-1}\left(\tfrac{1}{2}\right), F_l^{-1}\left(\tfrac{1}{2}\right) \right]$; and that for any $\theta \in (0,1)$, either $F_l$ or $F_h$ are strictly increasing in a neighborhood around $\theta$, while the other function may be constant there.
    These assumptions are indeed satisfied by $\F_{\eps}$.
    See also Remark \ref{rmk:monotonicity_assumptions}.
    \label{fn:disclaimer_F_eps}
    }
    In the cases at hand, the thresholds in the symmetric equilibria are 
    \begin{align*}
        \theta_{\kappa}^* &= \frac{4\kappa-1 - \sqrt{8\kappa+1}}{4(\kappa-1)}, \\
        \theta_{a}^* &= \frac{-a^2-a+3+\sqrt{a^4+2a^3+3a^2-6a+1}}{4}, \text{ and } \\
        \theta_{\eps}^* &= \begin{cases}
            \frac{-2\eps+1+ \sqrt{4\eps^2+12\eps+1}}{4} &\text{ for } \eps \in \left(0,\tfrac{1}{3}\right), \\
            \frac{\eps+1}{2} &\text{ for } \eps \in \left[\tfrac{1}{3}, \tfrac{1}{2}\right).
        \end{cases}
    \end{align*} 
    Figure \ref{fig:thresholds_F*} depicts these thresholds as functions of their respective parameters.
    One can show that all these functions are strictly increasing and in their uncertainty parameters and that $\lim_{\kappa \to 1}\theta_{\kappa}^*=\tfrac{2}{3}$, $\lim_{a \to \infty}\theta_{a}^*=1$, and $\lim_{\kappa \to 1/2}\theta_{\eps}^*=\tfrac{3}{4}$.
\end{example}

\noindent Theorem \ref{thm:existence_chracterization_fully_cursed_KNE_under_F*} shows existence of equilibria as well as uniqueness within the class of symmetric strategy profiles, but makes no statements about uniqueness in general.
The following remark provides an example where multiple non-trivial cursed Knight-Nash equilibria exist.\footnote{
    The examples in Remarks \ref{rmk:multiple_equilibria} and \ref{rmk:no_monotonocity_types_uncertainty} are highly contrived and lack any economical interpretation.
    They just serve as counterexamples to the conjectures that the equilibrium is unique, or that the amount of types willing to trade in equilibrium is increasing in the amount of uncertainty.
    }

\begin{remark}
\label{rmk:multiple_equilibria}
    Following Theorem \ref{thm:existence_chracterization_fully_cursed_KNE_under_F*} and Lemma \ref{lem:v(nt)_and_v(tr)}, a strategy profile $\left(\sigma_1^{\theta_1^*}, \sigma_2^{\theta_2^*}\right)$ is a cursed Knight-Nash equilibrium given $\text{Id}$ if and only if $\theta_1^*, \theta_2^*\in\left[\tfrac{1}{2}, F_l^{-1}\left(\tfrac{1}{2}\right)\right]$ as well as 
    \begin{align}
    \label{eq:cursed_equilibrium_under_F*}
        \ubar v_1^{\text{Id}}\left(nt \mid \theta_1^*, \sigma_{2}^{\theta_2^*}\right) = \ubar v_1^{\text{Id}}\left(tr \mid \theta_1^*, \sigma_{2}^{\theta_2^*}\right) \quad \text{and} \quad \ubar v_2^{\text{Id}}\left(nt \mid \theta_2^*, \sigma_{1}^{\theta_1^*}\right) = \ubar v_2^{\text{Id}}\left(tr \mid \theta_2^*, \sigma_{1}^{\theta_1^*}\right).
    \end{align}
    While Theorem \ref{thm:existence_chracterization_fully_cursed_KNE_under_F*} shows that there always exists exactly one symmetric solution of \eqref{eq:cursed_equilibrium_under_F*} in which $\theta_1^*=\theta_2^*$, it is silent about asymmetric solutions.
    One can show that such solutions do not exist in the parametrizations from Example \ref{ex:parametrizations}.
    
    However, let $F^*=\text{Id}$ and set $g(\theta) \colonequals -4 \left(\tfrac{1}{16} \sin\left(4\pi\theta\right)+\theta-\tfrac{1}{2}\right)^2+1$.
    For $\theta \geq \tfrac{1}{2}$, let $F_h(\theta) \colonequals \tfrac{4}{5}\left(\theta-\tfrac{1}{2}\right)^2+\tfrac{4}{5}$ and $F_l(\theta) \colonequals \left(1-2F_h(\theta)\right)g(\theta)+F_h(\theta)$. 
    For $\theta<\tfrac{1}{2}$, define $F_h(\theta) \colonequals 1-F_l(1-\theta)$ and $F_l(\theta) \colonequals 1-F_h(1-\theta)$ such that Assumption \ref{ass:prior_distributions} is satisfied.
    One can show that in this case, $F_l^{-1}\left(\tfrac{1}{2}\right) \approx 0.9101$ and that Equation \eqref{eq:cursed_equilibrium_under_F*} admits three solutions in $\left(\tfrac{1}{2}, F_l^{-1}\left(\tfrac{1}{2}\right)\right]$: $\left(\theta_1^*, \theta_2^*\right) \approx \left(0.6069, 0.8809\right)$, $\left(\theta_1^*, \theta_2^*\right) = \left(0.75, 0.75\right)$, and $\left(\theta_1^*, \theta_2^*\right) \approx \left(0.8809, 0.6069\right)$.
    Hence, there are three non-trivial fully cursed Knight-Nash equilibria: one symmetric equilibrium (the one whose existence has been proven in Theorem \ref{thm:existence_chracterization_fully_cursed_KNE_under_F*}) and two asymmetric ones.
\end{remark}

\noindent Another question raised by Example \ref{ex:examples_revisited} is whether the amount of trade is monotonous in the amount of uncertainty, as suggested by the patterns in Figure \ref{fig:thresholds_F*}.
Again, the answer is negative.

\begin{remark}
\label{rmk:no_monotonocity_types_uncertainty}
    Let $F_{l}$ and $F_{h}$ as well as $G_{l}$ and $G_{h}$ be two pairs of distribution functions surrounding the same true distribution function $F^*$, and let $\F$ and $\mathcal{G}$ be the distribution sets in the sense of Assumption \ref{ass:prior_distributions}.
    Let there be more uncertainty under $\F$ than under $\mathcal{G}$, that is, $G_h(\theta)>F_h(\theta)$ and $G_l(\theta)<F_l(\theta)$ for all $\theta \in (0,1)$.
    If the preliminaries of Theorem \ref{thm:existence_chracterization_fully_cursed_KNE_under_F*} are met, there exist symmetric cursed Knight-Nash equilibria under $F^*$.
    In these equilibria, both players play the cut-off strategy with $\theta^*_{\F}$ (if the ambiguity set is $\F$) or $\theta^*_{\mathcal{G}}$ (if the ambiguity set is $\mathcal{G}$).
    In the parametrizations presented in Example \ref{ex:examples_revisited}, more uncertainty always induced higher thresholds, that is $\theta^*_{\mathcal{G}}>\theta_F^*$.
    However, this need not be the case, as the following example shows:

    Let $F^*=\text{Id}$ and for $\theta\geq \tfrac{1}{2}$, let $F_l(\theta)=\tfrac{11}{10}\theta-\tfrac{1}{10}, F_h(\theta)=\tfrac{9}{10}\theta+\tfrac{1}{10}, G_l(\theta)=\tfrac{23}{20}\theta-\tfrac{3}{20}, G_h(\theta)=\tfrac{17}{40}\sqrt[10]{2\theta-1}+\tfrac{23}{40}$.
    For $\theta<\tfrac{1}{2}$, let $F_{h}(\theta)=1-F_{l}(1-\theta)$ and $F_{l}(\theta)=1-F_{h}(1-\theta)$ as well as $G_{h}(\theta)=1-G_{l}(1-\theta)$ and $G_{l}(\theta)=1-G_{h}(1-\theta)$.
    Assumption \ref{ass:prior_distributions} is satisfied for the distribution sets $\F$ and $\mathcal{G}$.
    Moreover, $G_h(\theta)>F_h(\theta)$ and $G_l(\theta)<F_l(\theta)$ for all $\theta \in (0,1)$ such that under $\mathcal{G}$, there is more uncertainty than under $\mathcal{F}$.
    In both cases, there is a unique cursed Knight-Nash equilibrium, given by the cut-off strategies with threshold $\theta^*_{\F}\approx 0.5393$ for $\F$, and $\theta_{\mathcal{G}}^*\approx 0.5383$ for $\mathcal{G}$.
    Hence, the threshold under $\mathcal{G}$, where there is more uncertainty, is smaller.
\end{remark}

\noindent Remark \ref{rmk:no_monotonocity_types_uncertainty} shows that there is no monotonicity in the probability of a trade offer with respect to the true probability $F^*$.
However, in the unique symmetric cursed Knight-Nash equilibrium under $F^*$, the minimal probability (among all probability measures defined by the cumulative distribution functions in $\F$) that a player offers to trade is decreasing in the amount of uncertainty, while the maximal probability is increasing in it.
Hence, more uncertainty about the type distribution always leads to more uncertainty about trade happening or not, thus more uncertainty about the outcome of the game.

\begin{proposition} \label{prop:comparative_statics_max_min_prob_of_trade_under_F*}
    Let $\F$ and $\mathcal{G}$ be two uncertainty sets that satisfy Assumption \ref{ass:prior_distributions}, with the same true distribution $F^*$.
    Let $\theta^*_{\F}$ and $\theta^*_{\mathcal{G}}$ be the thresholds of the cut-off strategies played in the unique symmetric cursed Knight-Nash equilibrium.
    If $G_h(\theta)>F_h(\theta)$ and $G_l(\theta)<F_l(\theta)$ for all $\theta \in (0,1)$, then $G_l\left(\theta^*_{\mathcal{G}}\right)<F_l\left(\theta^*_{\mathcal{F}}\right)$ and $G_h\left(\theta^*_{\mathcal{G}}\right)>F_h\left(\theta^*_{\mathcal{F}}\right)$.
\end{proposition}

\subsection{When Cursedness and Rationality Come Together}

So far, we have considered situations in which both players are either uncursed (Subsection \ref{subsec:Benchmark_Uncursed_Players_Do_Not_Trade_Under_Uncertainty}), or cursed (Subsection \ref{subsec:How_Uncertainty_Leads_Cursed_Players_To_Trade_Even_More}).
We shall now investigate cases in which player 1 is cursed, while player 2 is not.
As before, there are always \emph{trivial} equilibrium in which both players almost surely always play $nt$.

\begin{theorem}\label{thm:existence_uniqueness_cursed_uncursed_KNE}
    Under Assumption \ref{ass:prior_distributions}, the unique non-trivial $\left(\{1\}, \{2\}\right)$-cursed-uncursed Knight-Nash equilibrium under $F^*$ in cut-off strategies\footnote{
    There can be further cursed-uncursed Knight-Nash equilibria outside the class of cut-off strategies, for instance the strategy profile $\left(\sigma_1^A, \sigma_2^{\left\{\ubar a\right\}}\right)$ from Example \ref{ex:uncursed_equilibrium_not_cut-off}.
    } 
    is $\left(\sigma_1^{F_l^{-1}\left(\frac{1}{2}\right)}, \sigma_2^{\left(F_l+F_h\right)^{-1}\left(\frac{1}{2}\right)}\right)$.
\end{theorem}

\noindent Theorem \ref{thm:existence_chracterization_fully_cursed_KNE_under_F*} shows that the presence of Knightian uncertainty increases the set of types for which a cursed player is willing to trade, compared to the situation without such uncertainty.
Theorem \ref{thm:existence_uniqueness_cursed_uncursed_KNE} shows that this effect is even stronger when the opponent is not cursed:
In this case, the cursed player plays a cut-off strategy with threshold $F_l^{-1}\left(\tfrac{1}{2}\right)$, which is not only greater than ${F^*}^{-1}\left(\tfrac{1}{2}\right)$, but also weakly (and, in most cases, strongly) exceeds the threshold from Theorem \ref{thm:existence_chracterization_fully_cursed_KNE_under_F*}.
This threshold is even monotone in the amount of uncertainty in the sense that if the set of possible distribution functions grows larger (and therefore the lower bound $F_l$ becomes smaller), the threshold $F_l^{-1}\left(\tfrac{1}{2}\right)$ increases.
The threshold for the uncursed player is not necessarily monotone in the amount of uncertainty:
If $\F$ admits more uncertainty than $\mathcal{G}$ (in the sense of Proposition \ref{prop:comparative_statics_max_min_prob_of_trade_under_F*}), $\left(F_{l}+F_{h}\right)^{-1}\left(\tfrac{1}{2}\right)$ can be greater than, equal, or less than $\left(G_{l}+G_{h}\right)^{-1}\left(\tfrac{1}{2}\right)$.

\section{Conclusion} \label{sec:Conclusion}

This paper introduces a new equilibrium concept that combines cursedness with Knightian uncertainty.
This concepts is then applied to investigate a trading game.
While the Bayesian Nash equilibrium predicts no trade, players do trade in the cursed equilibrium.
If the probability distribution of types is unknown, nothing changes in the uncursed set-up, if we focus on cut-off strategies.
However, if players face uncertainty, are pessimistic with respect to this uncertainty, and are cursed, they trade even more than without uncertainty.
This is in contrast with many previous results from the literature that find that uncertainty aversion reduces trade.

This work focuses on one particular game.
Despite its simple structure, the formal analysis is already quite involved.
Possible future research might generalize our results to a broader class of trading games.
For example, players' utility functions could be monotone functions of the difference between types, which would include our model (in which the utility is a step function in the difference between types) as a special case.
We expect that the basic results about Bayesian Nash and cursed equilibria (Subsection \ref{subsec:How_Cursedness_Leads_Players_To_Trade}) would still hold in a similar fashion, as well as the results about uncursed Knight-Nash equilibria (Subsection \ref{subsec:Benchmark_Uncursed_Players_Do_Not_Trade_Under_Uncertainty}).
It is less clear whether the findings about cursed Knight-Nash equilibria (Subsection \ref{subsec:How_Uncertainty_Leads_Cursed_Players_To_Trade_Even_More}) will persist, as many proofs heavily rely on the simple structure of our trading game.

A further possible line of research is to conduct an experiment based on this game, and to investigate how subjects actually behave in the laboratory, with and without Knightian uncertainty.

\appendix

\section{Proofs}
\label{app:Proofs}

\begin{proof}[Proof of Lemma \ref{lem:v(nt)_and_v(tr)}]
    Equation \eqref{eq:v(nt)} is clear:
    If a player does not offer to trade, their (cursed or uncursed) interim expected value does not depend on the opponent's action and is just given by $\Prob_F\left(\theta_{-k}<\theta_k\right)=F(\theta_k)$.

    For Equation \eqref{eq:w(tr)}, we have
    \begin{align*}
        w_k\left(tr \mid \theta_k, \sigma_{-k}^{\hat\theta} \right) &= \Prob_F\left(a_{-k}=tr, \theta_{-k}>\theta_k\right) + \Prob_F\left(a_{-k}=nt, \theta_{-k}<\theta_k\right) \\
        &= \Prob_F\left(\theta_k<\theta_{-k}\leq \hat\theta\right) + \Prob_F\left(\hat\theta < \theta_{-k} < \theta_k\right) \nonumber\\
        &= \LV F(\hat\theta_2)-F(\theta_k)\RV.\nonumber
    \end{align*}
    As the average strategy of the opponent is $\bar\sigma_{-k}^{\hat\theta}=F(\hat\theta_2)\cdot tr + \left(1-F(\hat\theta_2)\right)\cdot nt$, we obtain
    \begin{align*}
        v_k\left(tr \mid \theta_k, \sigma_{-k}^{\hat\theta} \right) &= \Prob_F\left(a_{-k}=tr\right)\Prob_F\left(\theta_{-k}>\theta_k\right) + \Prob_F\left(a_{-k}=nt\right)\Prob_F\left(\theta_{-k}<\theta_k\right) \\
        &= F(\hat\theta_2)\left(1-F(\theta_k)\right)+\left(1-F(\hat\theta_2)\right)F(\theta_k) \nonumber
    \end{align*}
    which shows Equation \eqref{eq:v(tr)}.
\end{proof}

\begin{proof}[Proof of Proposition \ref{prop:BNE_is_nt}]
    If, for $k \in \{1,2\}$, player $-k$ plays some strategy $\sigma_{-k}$ such that $\sigma_{-k}\left(nt \mid \theta_{-k}\right)=1$ for $F$-almost all $\theta_{-k}$, then 
    \begin{align*}
        w_k \left( tr \mid \theta_k, \sigma_{-k} \right) = w_k \left( nt \mid \theta_k, \sigma_{-k} \right) = F\left(\theta_k\right)
    \end{align*}
    for all $\theta_k \in [0,1]$.
    Hence, every strategy is a best response against $\sigma_{-k}$.
    It follows that any strategy profile in which both players $F$-almost surely play $nt$ is a Bayesian Nash equilibrium.

    Let us show that there cannot be more equilibria.
    For each player $k \in \{1,2\}$, define the cursed expected difference between not trading and trading at type $\theta_k$, given the opponent's strategy $\sigma_{-k}$, as
    \begin{align} \label{eq:definition_difference_d_k}
        d_k \left( \theta_k, \sigma_{-k} \right)
        \colonequals w_k\left(nt \mid \theta_k, \sigma_{-k}\right) - w_k\left(tr \mid \theta_k, \sigma_{-k}\right).
    \end{align}
    Let $T_k$ be the set in which player $k$ offers to trade with positive probability, that is, 
    \begin{align} \label{eq:definition_T_k}
        T_k \colonequals \left\{ \theta_k \in [0,1] \colon \sigma_k (tr \mid \theta_k) >0 \right\}.
    \end{align}
    By means of contradiction, assume that $T_k$ is a non-nullset for at least one player, without loss of generality $\Prob_F(T_2)>0$. 
    Then, with $d_1$ as defined in \eqref{eq:definition_difference_d_k},
    \begin{align*}
        d_1\left(\theta_1, \sigma_2\right) = \int_0^1 \sigma_2\left(tr \mid \theta_2 \right) \left(\mathbbm{1}_{\{\theta_1 > \theta_2\}} - \mathbbm{1}_{\{\theta_2 > \theta_1\}} \right) \d F \left(\theta_2\right),
    \end{align*}
    in particular
    \begin{align*}
        d_1\left(0, \sigma_2\right) = -\int_0^1 \sigma_2(tr \mid \theta_2) \d F\left(\theta_2\right) < 0 \quad \text{and} \quad
        d_1\left(\sup T_2, \sigma_2\right) = \int_0^1 \sigma_2(tr \mid \theta_2) \d F\left(\theta_2\right) > 0.
    \end{align*}
    By continuity, there exists $\varepsilon>0$ such that $d_1 \left(\theta_1, \sigma_2\right) <0$ for all $\theta<\varepsilon$ and $d_1 \left(\theta_1, \sigma_2 \right) >0$ for all $\theta_1 \geq \sup T_2 -\varepsilon$.
    Therefore, every strategy $\sigma_1$ that is a best response to $\sigma_2$ must satisfy $\sigma_1(tr \mid \theta_1)=1$ for $\theta<\varepsilon$ and $\sigma_1(tr \mid \theta_1)=0$ for $\theta_1 \geq \sup T_2 -\varepsilon$.
    Hence, $\Prob_F(T_1)>0$ and $\sup T_1 < \sup T_2$.
    In an equilibrium, $\sigma_2$ must be a best response against $\sigma_1$ as well.
    As $\Prob_F(T_1)>0$, repeating the same arguments with reversed roles yields $\sup T_2 < \sup T_1$, which is impossible.
\end{proof}

\begin{proof}[Proof of Proposition \ref{prop:cNE_is_median_cutoff}]
    First note that by the very same arguments as the ones from the first paragraph of the proof of Proposition \ref{prop:BNE_is_nt}, every Bayesian Nash equilibrium is also a cursed equilibrium.
    Now assume that player $-k$ plays a strategy $\sigma_{-k}$ such that $T_{-k}$ as defined in \eqref{eq:definition_T_k} is not a nullset.
    This is equivalent to $\bar\sigma_{-k}(tr\mid F)>0$.
    In this case, for all $\theta_k \in [0,1]$,
    \begin{align*}
        v_k \left(nt \mid \theta_k, \sigma_{-k} \right) &= F\left(\theta_k\right), \text{ while} \\
        v_k \left(tr \mid \theta_k, \sigma_{-k} \right)
        &= \bar\sigma_{-k}(tr\mid F) \left(1-F\left(\theta_k\right)\right) + \left(1-\bar\sigma_{-k}(tr\mid F)\right) F\left(\theta_k\right).
    \end{align*}
    As $\bar\sigma_{-k}(tr\mid F)>0$, it follows that $v_k \left(tr \mid \theta_k, \sigma_{-k} \right) \geq v_k \left(nt \mid \theta_k, \sigma_{-k} \right)$ if and only if $F\left(\theta_k\right) \leq \tfrac{1}{2}$.
    Hence, any cut-off strategy with threshold $F^{-1}\left(\tfrac{1}{2}\right)$ is a cursed best response against $\sigma_{-k}$, and any cursed best response is such a strategy.
    It follows that every strategy profile in which both players play a cut-off strategy with threshold $F^{-1}\left(\tfrac{1}{2}\right)$ is a cursed equilibrium, and that these, together with the Bayesian Nash equilibria, are all cursed equilibria.
\end{proof}

\begin{proof}[Proof of Proposition \ref{prop:KNE_is_nt}]
    Obviously, $\left(\sigma_1^0, \sigma_2^0\right)$ constitutes a Knight-Nash equilibrium:
    When the opponent never offers to trade---or only does so at type 0, which happens with probability 0 under any probability measure from $\mathcal{P}_{\F}$---, the own strategy does not matter, as there will not be a trade anyway.
    In particular, it is optimal to always play $nt$.

    Let us now assume that player 2 plays a cut-off strategy $\sigma_2^{\hat\theta}$ with $\hat\theta>0$.
    Then, following Lemma \ref{lem:v(nt)_and_v(tr)}, player 1's infimal interim expected utilities from playing ``$nt$'' or ``$tr$'', respectively, are
    \begin{align*}
        \ubar w_1 \left(nt \mid \theta_1, \sigma_2^{\hat\theta}\right)=  F_l(\theta_1) \qquad \text{and} \qquad \ubar w_1 \left(tr \mid \theta_1, \sigma_2^{\hat\theta}\right) = \inf_{F \in \F} \LV F(\theta_1)-F(\hat\theta_2) \RV,
    \end{align*} 
    respectively.
    The value of the second term depends on the location of $\hat\theta$ and $\theta_1$:
    If $F(\theta_1)\leq F_l(\hat\theta_2)$ for all $F \in \F$---that is, if $F_h(\theta_1)\leq F_l(\hat\theta_2)$---then there exists a function $G \in \F$ that satisfies $G(\theta_1)=F_h(\theta_1)$ and $G(\hat\theta_2)=F_l(\hat\theta_2)$, and $G$ minimizes $ \LV F(\theta_1)-F(\hat\theta_2) \RV$ over $\F$.
    Similarly, if $F_h(\hat\theta_2)\leq F_l(\theta_1)$, any function $G \in \F$ that satisfies $G(\hat\theta_2)=F_h(\hat\theta_2)$ and $G(\theta_1)=F_l(\theta_1)$ minimizes $ \LV F(\theta_1)-F(\hat\theta_2) \RV$; and if neither of the two is the case, there is a function that is constant between $\theta_1$ and $\hat\theta$, and any such function minimizes $ \LV F(\theta_1)-F(\hat\theta_2) \RV$.
    Putting all of this together, we see that the infimum is always attained and that
    \begin{align} \label{eq:w(tr)_uncertain_uncursed}
        \ubar w_1 \left(tr \mid \theta_1, \sigma_2^{\hat\theta}\right) = \min_{F \in \F} \LV F(\theta_1)-F(\hat\theta_2) \RV = \begin{cases}
            F_l(\hat\theta_2)-F_h(\theta_1) &\text{ for } \theta_1 \leq F_h^{-1}\left(F_l(\hat\theta_2)\right), \\
            0 &\text{ for } \theta_1 \in \left[ F_h^{-1}\left(F_l(\hat\theta_2)\right), F_l^{-1}\left(F_h(\hat\theta_2)\right) \right], \\
            F_l(\theta_1)-F_h(\hat\theta_2) &\text{ for } \theta_1 \geq F_l^{-1}\left(F_h(\hat\theta_2)\right).
        \end{cases}
    \end{align}
    It follows that $\ubar w_1 \left(tr \mid \theta_1, \sigma_2^{\hat\theta}\right) < \ubar w_1 \left(nt \mid \theta_1, \sigma_2^{\hat\theta}\right)$ for $\theta_1 \geq F_h^{-1}\left(F_l(\hat\theta_2)\right)$ such that for these $\theta_1$, any best reply requires to play $nt$.
    For $\theta_1 \leq  F_h^{-1}\left(F_l(\hat\theta_2)\right)$, $\ubar w_1 \left(nt \mid \theta_1, \sigma_2^{\hat\theta}\right)$ is strictly increasing, while $\ubar w_1 \left(tr \mid \theta_1, \sigma_2^{\hat\theta}\right)$ is decreasing in $\theta_1$.
    Moreover, 
    \begin{align*}
        \ubar w_1 \left(tr \mid 0, \sigma_2^{\hat\theta}\right) = F_l(\hat\theta_2)&>F_l(0)=\ubar w_1 \left(nt \mid 0, \sigma_2^{\hat\theta}\right) \text{ and} \\
        \ubar w_1 \left(tr \mid F_h^{-1}\left(F_l(\hat\theta_2)\right), \sigma_2^{\hat\theta}\right) = 0&< F_l \left(F_h^{-1}\left(F_l(\hat\theta_2)\right)\right) = \ubar w_1 \left(nt \mid F_h^{-1}\left(F_l(\hat\theta_2)\right), \sigma_2^{\hat\theta}\right)
    \end{align*}
    such that there must be a unique intersection at some $\theta_1^*$ with $0<\theta_1^*< F_h^{-1}\left(F_l(\hat\theta_2)\right) \leq \hat\theta$.
    Hence, the best response to the cut-off strategy $\sigma_2^{\hat\theta}$ is the cut-off strategy $\sigma_1^{\theta_1^*}$ with $\theta_1^*<\hat\theta$.
    But as players are symmetric, a Knight-Nash equilibrium requires that also $\hat\theta<\theta_1^*$, which is impossible.
\end{proof}

\noindent For the proof of Proposition \ref{prop:cut-offs_ambiguity_in_middle_under_F*}, we need the following Lemma.

\begin{lemma} \label{lem:value_of_minimum_under_F*}
    Let Assumption \ref{ass:prior_distributions} be satisfied.
    For $k \in \{1,2\}$ and any strategy $\sigma_{-k}$ of player $-k$,
    \begin{align*}
        \ubar v_k^{F^*} \left(nt \mid \theta_k, \sigma_{-k}\right) &= F_l\left(\theta_k\right), \text{ and} \\
        \ubar v_k^{F^*} \left(tr \mid \theta_k, \sigma_{-k}\right) &= 
        \begin{cases}
            \left( 1-2\bar\sigma_{-k} \left( tr \mid F^* \right) \right) F_l(\theta_1) + \bar\sigma_2 \left( tr \mid F^* \right) &\text{ if } \bar \sigma_{-k} \left( tr \mid F^* \right) \leq \tfrac{1}{2}, \\
            \left( 1-2\bar\sigma_{-k} \left( tr \mid F^* \right) \right) F_h(\theta_1) + \bar\sigma_2 \left( tr \mid F^* \right) &\text{ if } \bar \sigma_{-k} \left( tr \mid F^* \right) \geq \tfrac{1}{2}.
        \end{cases}
    \end{align*}
\end{lemma}

\begin{proof}
    Let $k=1$ without loss of generality.
    For any strategy $\sigma_2$ of player 2,
    \begin{align*}
        \ubar v_1^{F^*} \left( nt \mid \theta_1, \sigma_2 \right) = \inf_{F \in \F} \Prob_{F} \left(\theta_2 \leq \theta_1 \right) = F_l\left(\theta_1\right).
    \end{align*}
    If player 1 offers to trade, their cursed expected utility is given by
    \begin{align*}
        \ubar v_1^{F^*} \left( tr \mid \theta_1, \sigma_2 \right) &= \inf_{F \in \F} \int_0^1 \bar\sigma_2 \left( tr \mid F^* \right) \mathbbm{1}_{\{\theta_2>\theta_1\}} + \left(1-\bar\sigma_2 \left( tr \mid F^* \right)\right) \mathbbm{1}_{\{\theta_2<\theta_1\}} \dF (\theta_2) \\
        &= \inf_{F \in \F} \bar\sigma_2 \left( tr \mid F^* \right) \left(1-F(\theta_1)\right) + \left(1-\bar\sigma_2 \left( tr \mid F^* \right)\right) F(\theta_1) \\
        &= \inf_{F \in \F} \left( 1-2\bar\sigma_2 \left( tr \mid F^* \right) \right) F(\theta_1) + \bar\sigma_2 \left( tr \mid F^* \right).
    \end{align*}
    If $1-2\bar\sigma_2 \left( tr \mid F^* \right) \geq 0$, this infimum is attained by $F=F_l$, while if $1-2\bar\sigma_2 \left( tr \mid F^* \right) \leq 0$, it is attained by $F=F_h$.
\end{proof}

\noindent In order to prove Proposition \ref{prop:cut-offs_ambiguity_in_middle_under_F*}, we first establish that only cut-off strategies with thresholds in $(0,1)$ are played in equilibrium.

\begin{proof}[Proof that only cut-off strategies are played in equilibrium]
    Without loss of generality, let $\bar\sigma_2\left(tr \mid F^*\right)>0$.
    In order to establish that only cut-off strategies are played in equilibrium, it suffices to show that for any such strategy player 2 might employ, every maxmin fully cursed best response of player 1 is of the cut-off type, with some cut-off $\theta_1^* \in (0,1)$.
    Having shown this, we can reverse the roles of 1 and 2 and conclude that player 2 must play a cut-off strategy as well.
    
    Following Lemma \ref{lem:value_of_minimum_under_F*}, 
    \begin{align*}
        &\ubar v_1^{F^*} \left(nt \mid \theta_1, \sigma_2 \right) - \ubar v_1^{F^*} \left(tr \mid \theta_1, \sigma_2 \right) \\
        = 
        &\begin{cases}
            \left(2F_l(\theta_1)-1\right) \bar \sigma_{2} \left( tr \mid F^* \right) &\text{ if } \bar \sigma_{2} \left( tr \mid F^* \right) \leq \tfrac{1}{2},\\
            F_l\left(\theta_1\right) + \left(2\bar\sigma_2 \left(tr \mid F^* \right)-1 \right) F_h\left(\theta_1\right) - \bar\sigma_2 \left(tr \mid F^* \right) &\text{ if } \bar \sigma_{2} \left( tr \mid F^* \right) \geq \tfrac{1}{2}
        \end{cases}
    \end{align*}
    which is negative for $\theta_1=0$, positive for $\theta_1=1$, and strictly increasing in between.
    Hence, there must be a unique threshold $\theta_1^* \in (0,1)$ such that $\ubar v_1^{F^*} \left(tr \mid \theta_1, \sigma_2 \right)>\ubar v_1^{F^*} \left(nt \mid \theta_1, \sigma_2 \right)$ if and only if $\theta_1>\theta_1^*$.
    It follows that every best response of player 1 is of cut-off type with threshold $\theta_1^*$.
    Reverting the roles of the players, we see that in equilibrium, both players employ a cut-off strategy with tresholds in $(0,1)$.
\end{proof}

\noindent The following lemma is needed in order to localize these equilibria and therefore conclude the proof of Proposition \ref{prop:cut-offs_ambiguity_in_middle_under_F*}.

\begin{lemma} \label{lem:best_responses_under_F^*}
    Let Assumption \ref{ass:prior_distributions} be satisfied, and let player $-k$ play a cut-off strategy $\sigma_{-k}^{\hat\theta_{-k}}$ with threshold $\hat\theta_{-k}>0$.
    Then, there is a unique maxmin cursed best response under $F^*$ for player $k$.
    If player $-k$'s threshold satisfies $F^*\left(\hat\theta_{-k}\right)\leq \frac{1}{2}$, this best response is the cut-off strategy with threshold $F_l^{-1}\left(\tfrac{1}{2}\right)$.
    If $F^*\left(\hat\theta_{-k}\right)> \frac{1}{2}$, the best response is also a cut-off strategy, and the threshold lies in $\left[ F_h^{-1}\left(\tfrac{1}{2}\right), 1 \right)$.
\end{lemma}

\begin{proof}
    As usual, let $k=1$ and $-k=2$.
    When player 2 plays the cut-off strategy $\sigma_2^{\hat\theta_2}$ with threshold $\hat\theta_2 >0$, their average strategy under $F^*$ is given by $\bar\sigma_2^{\hat\theta_2}\left(tr \mid F^* \right) = F^*(\hat\theta_2) = 1- \bar\sigma_2^{\hat\theta_2}\left(nt \mid F^* \right)$.
    
    If $F^*(\hat\theta_2)\leq \frac{1}{2}$, then, by Lemma \ref{lem:value_of_minimum_under_F*}, $\ubar v_1^{F^*} \left(nt \mid \theta_1, \sigma_2^{\hat\theta_2}\right)=F_l\left(\theta_1\right)$, while 
    \begin{align*}
        \ubar v_1^{F^*} \left(tr \mid \theta_1, \sigma_2^{\hat\theta_2}\right) = F_l\left(\theta_1\right) + F^*(\hat\theta_2) - 2F_l\left(\theta_1\right) F^*(\hat\theta_2).
    \end{align*}
    It follows that $\ubar v_1^{F^*} \left(tr \mid \theta_1, \sigma_2^{\hat\theta_2}\right) > \ubar v_1^{F^*} \left(nt \mid \theta_1, \sigma_2^{\hat\theta_2}\right)$ if and only if $F_l\left(\theta_1\right)< \frac{1}{2}$.
    So indeed, the maxmin fully cursed best response against $\sigma_2^{\hat\theta_2}$ is the cut-off strategy with threshold $F_l^{-1}\left(\frac{1}{2}\right)$.

    Next, let $F^*(\hat\theta_2)>\tfrac{1}{2}$.
    We have already seen that the best response must be of cut-off type with threshold less than 1.
    As 
    \begin{align*}
        \ubar v_1^{F^*}\left(nt \mid  F_h^{-1}\left(\tfrac{1}{2}\right), \sigma_2^{\hat\theta_2}\right) = F_l\left(F_h^{-1}\left(\tfrac{1}{2}\right)\right) \leq \tfrac{1}{2} = \ubar v_1^{F^*}\left(tr \mid  F_h^{-1}\left(\tfrac{1}{2}\right), \sigma_2^{\hat\theta_2}\right), 
    \end{align*}
    the threshold for player 1 must be at least $F_h^{-1}\left(\tfrac{1}{2}\right)$.
\end{proof}

\begin{proof}[Proof of Proposition \ref{prop:cut-offs_ambiguity_in_middle_under_F*}]
    We have already shown that in equilibrium, only cut-off strategies with thresholds in $(0,1)$ are played.
    We now aim at localizing the thresholds of these equilibrium strategies.
    To this end, let $\left(\sigma_1^{\theta_1^*}, \sigma_2^{\theta_2^*}\right)$ be an equilibrium in which player $k$ plays the cut-off strategy with threshold $\theta_k^*$, for $k \in \{1,2\}$.

    Assume by means of contradiction that $\theta_1^* < F_h^{-1}\left(\tfrac{1}{2}\right)$ which implies $F^*\left(\theta_1^*\right)\leq \tfrac{1}{2}$.
    As $\sigma_1^{\theta_1^*}$ and $\sigma_2^{\theta_2^*}$ are mutual maxmin cursed best responses under $F^*$, $\theta_2^*=F_l^{-1}\left(\tfrac{1}{2}\right)$ by Lemma \ref{lem:best_responses_under_F^*}, in particular $F^*\left(\theta_2^*\right)\geq\frac{1}{2}$.
    Again by Lemma \ref{lem:best_responses_under_F^*}, $\theta_1^*\geq F_h^{-1}\left(\tfrac{1}{2}\right)$ which is impossible.
    Analogously, $\theta_2^* \geq F_h^{-1}\left(\tfrac{1}{2}\right)$.

    Assume next that, say, $\theta_2^*>F_l^{-1}\left(\tfrac{1}{2}\right)$ which implies that $F_h\left(\theta_2^*\right)>\tfrac{1}{2}$.
    In this case, $F^*\left(\theta_1^*\right)>\tfrac{1}{2}$ because otherwise, player 2's best response would be to play the cut-off strategy with threshold $F_l^{-1}\left(\tfrac{1}{2}\right)$, not the one with $\theta_2^*$.
    As player 2 is, given type $\theta_2^*$, indifferent between playing $nt$ and $tr$, Lemma \ref{lem:value_of_minimum_under_F*} yields
    \begin{align*}
        F_l\left(\theta_2^*\right) = \left(1-2F_h\left(\theta_2^*\right)\right) F^*\left(\theta_1^*\right) + F_h\left(\theta_2^*\right) < \left(1-2F_h\left(\theta_2^*\right)\right) \tfrac{1}{2} + F_h\left(\theta_2^*\right) = \tfrac{1}{2},
    \end{align*}
    where in the inequality, we used that $1-2F_h\left(\theta_2^*\right)<0$ and $F^*\left(\theta_1^*\right) > \tfrac{1}{2}$.
    The overall inequality contradicts our standing assumption that $\theta_2^*>F_l^{-1}\left(\tfrac{1}{2}\right)$.
\end{proof}

\begin{proof}[Proof of Corollary \ref{cor:fully_cursed_Knight_Nash_equilibrium_certain_median_under_F*}]
    As $\left[ F_h^{-1}\left(\tfrac{1}{2}\right), F_l^{-1}\left(\tfrac{1}{2}\right) \right]=\left\{{F^*}^{-1}\left(\tfrac{1}{2}\right)\right\}$ is a singleton, the proposed strategy profile is the only candidate for such an equilibrium by Proposition \ref{prop:cut-offs_ambiguity_in_middle_under_F*}.
    It directly follows from Lemma \ref{lem:best_responses_under_F^*} that both players' strategies are indeed mutual best responses to one another.
\end{proof}

\noindent The following Lemma is needed for the proof of Theorem \ref{thm:existence_chracterization_fully_cursed_KNE_under_F*}.

\begin{lemma} \label{lem:g_decreasing}
    Let $F_l$ and $F_h$ be continuous, strictly increasing cumulative distribution functions on $[0,1]$ that satisfy $F_l(\theta)\leq \theta\leq F_h(\theta)$ for all $\theta\in[0,1]$ as well as $F_h^{-1}\left(\tfrac{1}{2}\right)<F_l^{-1}\left(\tfrac{1}{2}\right)$ and Equation \eqref{eq:symmetry_F_uniform}.
    The continuous function $g \colon \left[\tfrac{1}{2}, F_l^{-1}\left(\tfrac{1}{2}\right)\right] \to \left[\tfrac{1}{2}, 1\right]$ defined via $g(\theta) \colonequals \frac{F_h(\theta)-F_l(\theta)}{2F_h(\theta)-1}$ is monotonously decreasing.    
\end{lemma}

\begin{proof}
    Continuity follows from $F_l$ and $F_h$ being continuous.
    It is obvious that for all $\theta$ in the relevant domain, the numerator of $g(\theta)$ is nonnegative, while the denominator is strictly positive.
    Let $\theta, \theta' \in \left[\tfrac{1}{2}, F_l^{-1}\left(\tfrac{1}{2}\right)\right]$ with $\theta \leq \theta'$.
    It then holds that
    \begin{align*}
        \frac{F_h\left(\theta\right)-F_l\left(\theta\right)}{F_h\left(\theta'\right)-F_l\left(\theta'\right)} \geq \frac{F_h\left(\theta\right)-F_l\left(\theta'\right)}{F_h\left(\theta'\right)-F_l\left(\theta'\right)} \geq \frac{F_h\left(\theta\right)-\tfrac{1}{2}}{F_h\left(\theta'\right)-\tfrac{1}{2}}
    \end{align*}
    where the first inequality follows from $F_l\left(\theta\right)\leq F_l\left(\theta'\right)$ and the second one from the fact that for any real numbers $a$ and $b$ with $a\leq b$, the mapping $x \mapsto \frac{a-x}{b-x}$ is monotonously decreasing on $(-\infty, b)$.
    Hence,
    \begin{align*}
        g\left(\theta\right) = \frac{1}{2}\frac{F_h\left(\theta\right)-F_l\left(\theta\right)}{F_h\left(\theta'\right)-F_l\left(\theta'\right)} \frac{F_h\left(\theta'\right)-F_l\left(\theta'\right)}{F_h\left(\theta\right)-\tfrac{1}{2}} \geq \frac{1}{2}\frac{F_h\left(\theta\right)-\tfrac{1}{2}}{F_h\left(\theta'\right)-\tfrac{1}{2}} \frac{F_h\left(\theta'\right)-F_l\left(\theta'\right)}{F_h\left(\theta\right)-\tfrac{1}{2}} = g\left(\theta'\right)
    \end{align*}
    which shows the claimed monononicity.
    Using Equation \eqref{eq:symmetry_F_uniform}, one finds that $g\left(\tfrac{1}{2}\right)=1$ and $g\left(F_l^{-1}\left(\tfrac{1}{2}\right)\right)=\tfrac{1}{2}$ which completes the proof.
\end{proof}

\begin{proof}[Proof of Theorem \ref{thm:existence_chracterization_fully_cursed_KNE_under_F*}]
   Following Remark \ref{rmk:true_F_uniform}, it is no loss of generality to assume that $F^*=\text{Id}$, and that $F_l$ and $F_h$ satisfy Equation \eqref{eq:symmetry_F_uniform}.
   We first show that in every cursed Knight-Nash equilibrium, players use cut-off strategies with thresholds above $\frac{1}{2}$.

   Observe that it directly follows from the assumptions that $F_l\left(\frac{1}{2}\right)<\frac{1}{2}<F_h\left(\frac{1}{2}\right)$ and that $F_h^{-1}\left(\frac{1}{2}\right)<\frac{1}{2}<F_l^{-1}\left(\frac{1}{2}\right)$.

   We know from Proposition \ref{prop:cut-offs_ambiguity_in_middle_under_F*} that in any such equilibrium, both players impose cut-off strategies with thresholds $\theta_1^*, \theta_2^* \in \left[F_h^{-1}\left(\tfrac{1}{2}\right), F_l^{-1}\left(\tfrac{1}{2}\right)\right]$.
   If $\theta_1^*\leq {F^*}^{-1}\left(\tfrac{1}{2}\right)=\tfrac{1}{2}$, Lemma \ref{lem:best_responses_under_F^*} implies that $\theta_2^*=F_l^{-1}\left(\tfrac{1}{2}\right)> \tfrac{1}{2}$.
   Hence, it is not possible that both $\theta_1^*$ and $\theta_2^*$ are less than or equal $\tfrac{1}{2}$.

   Assume next that $\theta_1^*\leq \tfrac{1}{2}$ and $\theta_2^* > \tfrac{1}{2}$.
   Again by Lemma \ref{lem:best_responses_under_F^*}, this is only possible if $\theta_2^*=F_l^{-1}\left(\tfrac{1}{2}\right)$.
   By Lemma \ref{lem:value_of_minimum_under_F*}, $\ubar v_1^{\text{Id}} \left(nt \mid \theta_1,  \sigma_2^{\theta_2^*}\right)$ is strictly increasing in $\theta_1$, and $\ubar v_1^{\text{Id}} \left(tr \mid \theta_1,  \sigma_2^{\theta_2^*}\right)$ is decreasing.
   As the two are equal for $\theta_1=\theta_1^*\leq\tfrac{1}{2}$, we must have that $\ubar v_1^{\text{Id}} \left(nt \mid \tfrac{1}{2}, \sigma_2^{\theta_2^*}\right) \geq \ubar v_1^{\text{Id}} \left(tr \mid \tfrac{1}{2}, \sigma_2^{\theta_2^*}\right)$.
   However,
   \begin{align}
       \ubar v_1^{\text{Id}} \left(tr \mid \tfrac{1}{2}, \sigma_2^{\theta_2^*}\right) &= \left(1-2F_l^{-1}\left(\tfrac{1}{2}\right)\right) F_h\left(\tfrac{1}{2}\right) + F_l^{-1}\left(\tfrac{1}{2}\right) \nonumber \\
       &= \left(1-2F_l^{-1}\left(\tfrac{1}{2}\right)\right) \left(1-F_l\left(\tfrac{1}{2}\right)\right) + F_l^{-1}\left(\tfrac{1}{2}\right) \label{eq:equality_F_h(1/2)=1-F_l(1/2)}\\
       &= \left(1-F_l^{-1}\left(\tfrac{1}{2}\right)\right) \left(1-2F_l\left(\tfrac{1}{2}\right)\right) + F_l\left(\tfrac{1}{2}\right) \nonumber \\
       &> F_l\left(\tfrac{1}{2}\right) \label{eq:inequality_equilbrium_above_1/2}\\
       &= \ubar v_1^{\text{Id}} \left(nt \mid \tfrac{1}{2}, \sigma_2^{\theta_2^*}\right), \nonumber
   \end{align}
   where we used Equation \eqref{eq:symmetry_F_uniform} in \eqref{eq:equality_F_h(1/2)=1-F_l(1/2)}, and $F_l\left(\tfrac{1}{2}\right)<\tfrac{1}{2}$ in \eqref{eq:inequality_equilbrium_above_1/2}.
   Hence, it is impossible that $\theta_1^*\leq \tfrac{1}{2}$ and, analogously, that $\theta_2^*\leq\tfrac{1}{2}$.

   We have thus shown that in every cursed Knight-Nash equilibrium under $F^*=\text{Id}$, both thresholds must exceed $\tfrac{1}{2}$.
   It remains to show that such an equilibrium actually exists.
   The continuous function $g$ defined in Lemma \ref{lem:g_decreasing} is monotonously decreasing on $\left[\tfrac{1}{2}, F_l^{-1}\left(\tfrac{1}{2}\right)\right]$, and $g\left(\tfrac{1}{2} \right)=1>\tfrac{1}{2}$ as well as $g\left(F_l^{-1}\left(\tfrac{1}{2}\right) \right) = \tfrac{1}{2}< F_l^{-1}\left(\tfrac{1}{2}\right)$, so there must be a unique $\theta^* \in \left(\tfrac{1}{2}, F_l^{-1}\left(\tfrac{1}{2}\right)\right)$ that satisfies $\theta^*=g\left(\theta^*\right)$.
   Rearranging terms, we see that for $k\in \{1,2\}$,
   \begin{align*}
       \ubar v_k^{\text{Id}}\left(nt \mid \theta^*, \sigma_{-k}^{\theta^*}\right) =F_l\left(\theta^*\right) = \left(1-2\theta^*\right) F_h\left(\theta^*\right) + \theta^* = \ubar v_k^{\text{Id}}\left(tr \mid \theta^*, \sigma_{-k}^{\theta^*}\right)
   \end{align*}
   which yields Equation \eqref{eq:symmetric_cursed_equilibrium_under_F*}.
   As $\theta^*>\frac{1}{2}$, $\ubar v_k^{\text{Id}}\left(tr \mid \theta_k, \sigma_{-k}^{\theta^*}\right)$ is strictly decreasing on $[0,1]$, while $\ubar v_k^{\text{Id}}\left(nt \mid \theta_k, \sigma_{-k}^{\theta^*}\right)$ is strictly increasing.
   It follows that for player $k$, $\sigma_k^{\theta^*}$ is a best reply against $\sigma_{-k}^{\theta^*}$, such that $\left(\sigma^{\theta^*}, \sigma^{\theta^*}\right)$ is indeed a symmetric non-trivial cursed Knight-Nash equilibrium.
   As any threshold used in such an equilibrium must satisfy \eqref{eq:symmetric_cursed_equilibrium_under_F*}, and as $\theta^*$ is the only number in $\left[\tfrac{1}{2}, F_l^{-1}\left(\tfrac{1}{2}\right)\right]$ that solves this equation, there cannot be any further symmetric non-trivial cursed Knight-Nash equilibria.    
\end{proof}

\begin{proof}[Proof of Proposition \ref{prop:comparative_statics_max_min_prob_of_trade_under_F*}]
    Remark that the symmetric equilibria $\left(\sigma_1^{\theta_{\mathcal{F}}^*}, \sigma_2^{\theta_{\mathcal{F}}^*}\right)$ and $\left(\sigma_1^{\theta_{\mathcal{G}}^*}, \sigma_2^{\theta_{\mathcal{G}}^*}\right)$ exist because of Theorem \ref{thm:existence_chracterization_fully_cursed_KNE_under_F*}.
    As before, we will assume without loss of generality that $F^*(\theta)=\theta$ for all $\theta$.
    We consider the two cases $\theta_{\mathcal{G}}^*\geq \theta_{\F}^*$ and $\theta_{\mathcal{G}}^*\leq \theta_{\F}^*$.
    
    In the first case, as $F_h$ and $G_h$ are increasing, $G_h\left(\theta^*_{\mathcal{G}}\right) \geq G_h\left(\theta^*_{\mathcal{F}}\right) > F_h\left(\theta^*_{\mathcal{F}}\right)$.
    Using that $1-2\theta^*_{\mathcal{G}}<0$ and $1-2F_h\left(\theta^*_{I}\right)<0$ as well as Equation \eqref{eq:symmetric_cursed_equilibrium_under_F*}, we obtain that
    \begin{align*}
        G_l\left(\theta^*_{\mathcal{G}}\right) &= \left(1-2\theta_{\mathcal{G}}^*\right) G_h\left(\theta^*_{\mathcal{G}}\right) + \theta_{\mathcal{G}}^* \\
        &< \left(1-2\theta_{\mathcal{G}}^*\right) F_h\left(\theta_{\F}^*\right) + \theta_{\mathcal{G}}^* \\
        &= \left(1-2F_h\left(\theta^*_{\F}\right)\right)\theta_{\mathcal{G}}^* + F_h\left(\theta^*_{\F}\right) \\
        &\leq \left(1-2F_h\left(\theta^*_{\F}\right)\right)\theta_{\F}^* + F_h\left(\theta^*_{\F}\right) = F_l\left(\theta_{\F}^*\right).
    \end{align*}
    In the second case, as $F_l$ and $G_l$ are increasing, $G_l\left(\theta^*_{\mathcal{G}}\right) \leq G_l\left(\theta^*_{\mathcal{F}}\right) < F_l\left(\theta^*_{\mathcal{F}}\right)$.
    Again using \eqref{eq:symmetric_cursed_equilibrium_under_F*}, we find that
    \begin{align*}
        G_h\left(\theta_{\mathcal{G}}^*\right)
        = \frac{1}{2} \frac{\theta_{\mathcal{G}}^*-G_l\left(\theta_{\mathcal{G}}^*\right)}{\theta_{\mathcal{G}}^*-\tfrac{1}{2}}
        \geq \frac{1}{2} \frac{\theta_{\F}^*-G_l\left(\theta_{\mathcal{G}}^*\right)}{\theta_{\F}^*-\tfrac{1}{2}}
        > \frac{1}{2} \frac{\theta_{\F}^*-F_l\left(\theta_{\F}^*\right)}{\theta_{\F}^*-\tfrac{1}{2}}
        = F_h\left(\theta_{\F}^*\right),
    \end{align*}
    where in the first inequality, we used that $\theta_{\F}^*>\theta_{\mathcal{G}}^*$ and that for any real numbers $a<b$, the map $x \mapsto \tfrac{x-a}{x-b}$ is strictly decreasing on $(b, \infty)$.
\end{proof}

\begin{proof}[Proof of Theorem \ref{thm:existence_uniqueness_cursed_uncursed_KNE}]
    Let us confirm that the described strategy profile indeed constitutes such an equilibrium.
    Following Equation \eqref{eq:w(tr)_uncertain_uncursed}, the infimal interim expected utilities of the uncursed player 2 against player 1's strategies are
    \begin{align*}
        \ubar w_2 \left( nt \mid \theta_2, \sigma_1^{F_l^{-1}\left(\tfrac{1}{2}\right)}\right) &= F_l(\theta_2), \text{ and} \\
        \ubar w_2 \left( tr \mid \theta_2, \sigma_1^{F_l^{-1}\left(\tfrac{1}{2}\right)}\right) &= 
        \begin{cases}
            \tfrac{1}{2}-F_h(\theta_2) &\text{ for } \theta_2 \leq F_h^{-1}\left(\tfrac{1}{2}\right), \\
            0 &\text{ for } \theta_2 \in \left[ F_h^{-1}\left(\tfrac{1}{2}\right), F_l^{-1}\left(F_h\left(F_l^{-1}\left(\tfrac{1}{2}\right)\right)\right) \right], \\
            F_l(\theta_2)-F_h\left(F_l^{-1}\left(\tfrac{1}{2}\right)\right) &\text{ for } \theta_2 \geq F_l^{-1}\left(F_h\left(F_l^{-1}\left(\tfrac{1}{2}\right)\right)\right).
        \end{cases}
    \end{align*}
    It is obvious that $\ubar w_2 \left( nt \mid \theta_2, \sigma_1^{F_l^{-1}\left(\tfrac{1}{2}\right)}\right) > \ubar w_2\left( tr \mid \theta_2, \sigma_1^{F_l^{-1}\left(\tfrac{1}{2}\right)}\right)$ for $\theta_2 \geq F_h^{-1}\left(\tfrac{1}{2}\right)$ and that the utility of trading is strictly decreasing on $\left[0, F_h^{-1}\left(\tfrac{1}{2}\right)\right)$, while that of not trading is strictly increasing.
    The unique intersection $\theta_2^*$ is where $F_l(\theta_2^*) = \tfrac{1}{2}-F_h(\theta_2^*)$, that is, $\theta_2^*=\left(F_l + F_h\right)^{-1}\left(\tfrac{1}{2}\right)$.
    Hence, the cut-off strategy with threshold $\left(F_l+F_h \right)^{-1}\left(\tfrac{1}{2}\right)$ is a maxmin best reply for player 2 against player 1 playing the cut-off strategy with threshold $F_l^{-1}\left(\tfrac{1}{2}\right)$.
    As $\left(F_l+F_h \right)^{-1}\left(\tfrac{1}{2}\right) < \tfrac{1}{2}$, Lemma \ref{lem:best_responses_under_F^*} yields that for player 1, $\sigma_1^{F_l^{-1}\left(\tfrac{1}{2}\right)}$ is indeed a maxmin cursed best response under $F^*$.

    Whenever player 2 plays a cut-off strategy with a threshold that is no more than ${F^*}^{-1}\left(\tfrac{1}{2}\right)$, player 1 responds to this with the strategy $\sigma_1^{F_l^{-1}\left(\tfrac{1}{2}\right)}$.
    Hence, in order to show uniqueness, it is sufficient to confirm that there cannot be a cursed-uncursed Knight-Nash equilibrium under $F^*$ in which player 2 plays a cut-off strategy with threshold $\theta_2 > {F^*}^{-1}\left(\tfrac{1}{2}\right)$.
    If this were the case, $F_l(\theta_1)-F_h(\theta_2)=F_l(\theta_2)$ by the same arguments as above.
    Hence, 
    \begin{align*}
        F_l(\theta_1) = F_l(\theta_2)+F_h(\theta_2) > F_l\left({F^*}^{-1}\left(\tfrac{1}{2}\right)\right)+F_h\left({F^*}^{-1}\left(\tfrac{1}{2}\right)\right) = 1
    \end{align*}
    where the last equality follows from Equation \eqref{eq:symmetry_F}.
    The overall inequality is impossible.
\end{proof}

\printbibliography

\pagebreak

\section{Online Appendix I: Ambiguous Cursed Equilibria}
\label{app:Ambiguous_Cursed_Equilibria}

This appendix investigates another way of incorporating Knightian uncertainty into the cursed equilibrium.
Definition \ref{def:cursed_knight_nash_equilibrium_given_P*} assumes the existence of a true probability measure $\Prob^*$ in the uncertainty set $\mathcal{P}$ from which the opponents' average strategy is calculated.
The existence of such a distribution in the framework of Knightian uncertainty has been described as an assumption that is ``unverifiable in general and so of a metaphysical nature'' \citep[p. 1064]{marinacci_2015}.\footnote{
The underlying philosophical discussion is related to the debate on how to interpret probabilities.
A \emph{frequentist} view suggests that there exists some true distribution, while this need not hold if one follows a \emph{subjective} or \emph{Bayesian} interpretation of probabilites.
For further reading on this issue, see for example \citet{gillies_2000}.
}
We therefore also provide an alternative to Definition \ref{def:cursed_knight_nash_equilibrium_given_P*} that does not need this assumption.
After that, we will apply the new definition to the trading game investigated in Section \ref{sec:The_Trading_Game_With_Uncertainty}.
We will see that qualitatively, the results remain similar.
However, there are differences with respect to the exact location and the uniqueness of the equilibria.

\begin{definition} \label{def:ambiguous_cursed_knight_nash_equilibrium}
    Let $\Gamma = \left( N, \left(A_k\right)_{k\in N}, \left(\Theta_k\right)_{k\in N}, \left(u_k\right)_{k \in N}, \mathcal{P} \right)$ be an incomplete information game with imprecise probabilistic information.
    For a strategy profile $\sigma_{-k}$ of  player $k$'s opponents, $\sigma_k$ is an \emph{ambiguous maxmin cursed best response} against $\sigma_{-k}$ if for each $\theta_k \in \Theta_k$, and each $a_k^*$ such that $\sigma_k\left(a_k^* \mid \theta_k\right)>0$,
    \begin{align}
        a_k^* &\in \argmax_{a_k \in a_k} \ubar v_k\left(a_k \mid \theta_k, \sigma_{-k}\right), \text{ where} \nonumber \\
        \ubar v_k\left(a_k \mid \theta_k, \sigma_{-k}\right) &\colonequals \inf_{\Prob \in \mathcal{P}} \int_{\Theta_{-k}}
        \sum_{a_{-k} \in A_{-k}} \bar{\sigma}_{-k}\left(a_{-k} \mid \theta_k, \Prob \right) 
        u_i\left(a_k, a_{-k} ; \theta_k, \theta_{-k}\right) \d\Prob\left(\theta_{-k} \mid \theta_k\right). \label{eq:definition_v_uncertainty}
    \end{align} 
    A strategy profile $\sigma^*$ is an \emph{ambiguous cursed Knight-Nash equilibrium} if for each $k$, $\sigma^*_k$ is an ambiguous maxmin cursed best response against $\sigma^*_{-k}$.
\end{definition}

\noindent The notion of ``ambiguous'' in the previous definition refers to the decision makers' fundamental uncertainty about the mechanisms that generate their own type as well as those of the other players.
They know nothing about this type generating process except for the fact that there is some probability distribution $\Prob \in \mathcal{P}$ that generates it.
Or, even less, they \emph{believe} that there is such a distribution, and act accordingly. 
Whether types are really generated according to some distribution from $\mathcal{P}$, or rather by some distribution outside of $\mathcal{P}$, or whether they are actually not random at all\footnote{
For example, there could be an outside experimenter who, unobserved by the players of the game, assigns types to them on a whim, or depending on the weather, or on the performance of her favorite football team in the last league match.
}
does not matter for the subsequent analysis.
In contrast to that, cursed decision makers that behave according to Definition \ref{def:cursed_knight_nash_equilibrium_given_P*} might have an estimate that $\Prob^*$ is the true type-generating distribution, but build their uncertainty set $\mathcal{P}$ around $\Prob^*$ out of caution, as suggested by Assumption \ref{ass:prior_distributions} and the following remarks.

Definition \ref{def:ambiguous_cursed_knight_nash_equilibrium} is somehow detached from the learning foundations of the cursed equilibrium.
The reasoning for this equilibrium concept assumes that players derive the opponents' average strategy from observing previous plays.
Such an argument cannot justify Definition \ref{def:ambiguous_cursed_knight_nash_equilibrium}, as the average strategy $\bar{\sigma}_{-k}\left(\cdot \mid \theta_k, \Prob \right)$ is subject to minimization. 
One might think of a player using Definition \ref{def:ambiguous_cursed_knight_nash_equilibrium} as a Gilboa-Schmeidler decision maker facing an incomplete information game, aiming to find a maxmin best response against their opponents in the sense of the Knight-Nash equilibrium (Definition \ref{def:knight_nash_equilibrium}).
However, due to some cognitive incapacity, they fail to see the connection between their opponents' types and their actions, treating events as stochastically independent that are in fact not.
Such a behavior is also plausible if learning is not possible, for example because agents can actually not observe previous plays of the game, or because the average strategy of the opponents does not converge.
The latter can for example be the case if there exists no true probability measure $\Prob^*$ but if types are determined differently, as outlined above.

We now aim to find ambiguous cursed Knight-Nash equilibria in the trading game described in Subsection \ref{subsec:The_Game}. 
As stated before, the existence of $F^*$ in Assumption \ref{ass:prior_distributions} is not necessary for the analysis of ambiguous equilibria, and can therefore be dropped for this appendix.
As explained in Remark \ref{rmk:true_F_uniform}, we can transform the ambiguity set in a suitable way and obtain that the uncertainty set $\F$ symmetrically encloses the uniform distribution.

\begin{assumption}
\label{ass:prior_distributions_ambiguous}
    There are strictly increasing and continuous functions $F_l$ and $F_h$ on $[0,1]$ that satisfy $F_l(\theta) \leq \theta\leq F_h(\theta)$ for all $\theta \in [0,1]$, as well as
    \begin{align*}
        F_l\left(1-\theta\right) = 1- F_h\left(\theta\right),
    \end{align*}
    such that the ambiguity set is $\mathcal{P}_{\F} = \left\{ \Prob_F^{\otimes 2} \colon F \in \F\right\}$, where $\F$ is the set of all monotonously increasing, continuous cumulative distribution functions $F$ on $[0,1]$ that satisfy $F_l(\theta)\leq F(\theta) \leq F_h(\theta)$ for all $\theta \in [0,1]$.
\end{assumption}

\noindent Again, there is always a trivial equilibrium in which both players almost surely never offer to trade, so we focus on non-trivial equilibria.
By the arguments from Remark \ref{rmk:cursed_player_prefers_non-trivial_equilibrium}, such an equilibrium, if it exists, is always preferred over the trivial one.

While Proposition \ref{prop:cut-offs_ambiguity_in_middle_under_F*} has shown that in the cursed Knight-Nash equilibrium under $F^*$, only cut-off strategies are played, this need not be the case for ambiguous equilibria, as the following remark shows.

\begin{example} \label{ex:equilibrium_not_cut-off}
    Recall the set $A$ and the corresponding strategy $\sigma^A$ from Example \ref{ex:uncursed_equilibrium_not_cut-off}.
    In addition to the properties from that example, let $F_h\left(\bar a\right) < \tfrac{1}{2}$.
    We will show that for $k\in\{1,2\}$, $\sigma_k^A$ is an ambiguous fully cursed best response for player $k$ against $\sigma_{-k}^A$ such that the symmetric strategy profile $\left(\sigma_1^A, \sigma_2^A\right)$ constitutes an ambiguous cursed Knight-Nash equilibrium that does not involve cut-off strategies.
    Without loss of generality, let $k=1$.
    For any type $\theta_1$, player 1's values of not offering to trade and of doing so are given by
    \begin{align}
        \ubar v_1 \left( nt \mid \theta_1, \sigma_2^A \right) &= \inf_{F \in \F} F\left(\theta_1\right) = F_l\left(\theta_1\right) \text{ and} \nonumber \\
        \ubar v_1 \left( tr \mid \theta_1, \sigma_2^A \right) &= \inf_{F \in \F} F \left(\theta_1\right) + \bar \sigma^A_2\left(tr \mid F \right) - 2 F \left(\theta_1\right) \bar \sigma^A_2\left(tr \mid F \right), \label{eq:value(tr)_vs_sigma^A}
    \end{align}
    respectively.
    We will now estimate the value of \eqref{eq:value(tr)_vs_sigma^A} against $F_l\left(\theta_1\right)$ for $\theta_1 \in A$ and $\theta_1 \notin A$ separately.

    If $\theta_1 \in A$, both $1-2F\left(\theta_1\right)$ as well as $\bar \sigma_2 \left(tr \mid F\right)$  are nonnegative for all $F \in \F$ such that
    \begin{align*}
        \ubar v_1 \left( tr \mid \theta_1, \sigma_2^A \right) &= \inf_{F \in \F} \left(1-2F\left(\theta_1\right)\right)\bar \sigma_2 \left(tr \mid F\right) + F\left(\theta_1\right) \\
        &\geq \inf_{F \in \F} F\left(\theta_1\right) = F_l\left(\theta_1\right) = \ubar v_1 \left( 1, \sigma_2^A, nt, \theta_1 \right).
    \end{align*}
    For any $\theta_1 \notin A$, there exists a function $\tilde{F} \in \F$ (whose exact shape depends on $\theta_1$) that satisfies $\tilde{F}\left(\theta_1\right)= F_l\left(\theta_1\right)$ as well as $\tilde{F}(a) = F_h \left(\ubar a\right)= F_l\left(\bar a\right)$ for all $a \in A$.
    In this case, as $\tilde{F}$ is constant on $A$, $\bar \sigma^A_2\left(tr \mid \tilde{F} \right)=0$ such that
    \begin{align*}
        \ubar v_1 \left( tr \mid \theta_1, \sigma_2^A \right) \leq \tilde{F} \left(\theta_1\right) + \bar \sigma^A_2\left(tr \mid \tilde{F} \right) - 2 \tilde{F} \left(\theta_1\right) \bar \sigma^A_2\left(tr \mid \tilde{F} \right) = F_l\left(\theta_1\right) = \ubar v_1 \left( 1, \sigma_2^A, nt, \theta_1 \right).
    \end{align*}
    We have thus shown that player 1 (weakly) prefers to trade their type if $\theta_1 \in A$, and (weakly) prefers not to do so if $\theta_1 \notin A$.
    Hence, $\sigma^A$ is indeed a best response against itself.
\end{example}

\noindent From now on, we will focus on strategy profiles in which both players use cut-off strategies, and characterize all ambiguous fully cursed Knight-Nash equilibria in this class.

We first show that if an ambiguous fully cursed Knight-Nash equilibrium with cut-off strategies exists, both cut-offs must be between $F_h^{-1}\left(\tfrac{1}{2}\right)$ and $F_l^{-1}\left(\tfrac{1}{2}\right)$, therefore proving an analogue to the second part of Proposition \ref{prop:cut-offs_ambiguity_in_middle_under_F*}

\begin{proposition} \label{prop:cut-offs_ambiguity_in_middle}
    Under Assumption \ref{ass:prior_distributions_ambiguous}, in every non-trivial ambiguous fully cursed Knight-Nash equilibrium in cut-off strategies, both thresholds lie in $\left[ F_h^{-1}\left(\tfrac{1}{2}\right), F_l^{-1}\left(\tfrac{1}{2}\right) \right]$.
\end{proposition}

\begin{proof}
    Let us assume that player 2 plays a cut-off strategy with threshold $\hat\theta \in [0,1]$.
    We want to find player 1's ambiguous maxmin cursed best response to that strategy.
    For any type $\theta_1 \in [0,1]$, it holds by Lemma \ref{lem:v(nt)_and_v(tr)} that
    \begin{align}
        \ubar v_1\left( nt \mid \theta_1, \sigma_2^{\hat\theta}\right) &= \inf_{F \in \F} F(\theta_1)=F_l(\theta_1), \nonumber \\
        \ubar v_1\left( tr \mid \theta_1, \sigma_2^{\hat\theta}\right) &= \inf_{F \in \F} F(\theta_1)+F(\hat\theta) - 2 F(\theta_1)F(\hat\theta). \label{eq:ubar_v(tr)} \\
        &\geq \inf_{G,H \in \F} G(\theta_1)+H(\hat\theta) - 2 G(\theta_1)H(\hat\theta) \label{eq:v(tr)_with_G,H} \\
        &= \min_{G,H \in \left\{ F_l, F_h \right\}} G(\theta_1)+H(\hat\theta) - 2 G(\theta_1)H(\hat\theta) \label{eq:v(tr)_with_G,H_2}
    \end{align}
    where the last equality follows from \eqref{eq:v(tr)_with_G,H} being affine linear in $G(\theta_1)$ and $H(\hat\theta)$.
    Hence, it is sufficient to find the value of \eqref{eq:v(tr)_with_G,H_2} and show that there is a function $F \in \F$ such that $F(\theta_1)+F(\hat\theta) - 2 F(\theta_1)F(\hat\theta)$ equals this value.
    
    We divide the set $[0,1]$ into the three disjoint sets $X_1$, $X_2$, and $X_3$, where $X_1 \colonequals \left[0, F_h^{-1}\left(\tfrac{1}{2}\right)\right)$, $X_2 \colonequals \left[ F_h^{-1}\left(\tfrac{1}{2}\right), F_l^{-1}\left(\tfrac{1}{2}\right) \right]$, and $X_3 \colonequals \left( F_l^{-1}\left(\tfrac{1}{2}\right), 1\right]$.
    We start with $\hat\theta \in X_1$.
    For $\theta_1\in X_1\cup X_2$,
    \begin{align*}
        \ubar v_1\left( tr \mid \theta_1, \sigma_2^{\hat\theta}\right) &= \inf_{F\in\F} \left(1-2F(\hat\theta)\right)F(\theta_1)+F(\hat\theta) \\
        &\geq \inf_{F\in\F} \left(1-2F(\hat\theta)\right)F_l(\theta_1)+F(\hat\theta) \\
        &= \inf_{F\in\F} \left(1-2F(\theta_1)\right)F(\hat\theta)+F(\theta_1)
        \geq F_l(\theta_1) = \ubar v_1\left( nt \mid \theta_1, \sigma_2^{\hat\theta}\right),
    \end{align*}
    where in the first inequality, we used that $F(\hat\theta)\leq\frac{1}{2}$ for all $F\in\F$.
    The overall inequality is strict except for $\theta_1=F_l^{-1}\left(\tfrac{1}{2}\right)$ where it holds with equality.
    For $\theta_1\in X_3$,
    \begin{align*}
        \ubar v_1\left( tr \mid \theta_1, \sigma_2^{\hat\theta}\right) = \inf_{F\in\F} \left(1-2F(\hat\theta)\right)F(\theta_1)+F(\hat\theta) < \inf_{F\in\F} F(\theta_1)=F_l(\theta_1) = \ubar v_1\left( nt \mid \theta_1, \sigma_2^{\hat\theta}\right).
    \end{align*}
    It follows that the ambiguous maxmin cursed best response to a cut-off strategy $\sigma_2^{\hat\theta}$ with $\hat\theta \in X_1$ is the cut-off strategy with threshold $F_l^{-1}\left(\tfrac{1}{2}\right) \in X_2$.

    We now turn to the case where $\hat\theta \in X_3$. 
    For $\theta_1 \in X_1$, there exists a function $F\in\F$ that satisfies $F(\theta_1)=F_h(\theta_1)$ and $F(\hat\theta)=F_l(\theta_1)$.
    Hence,
    \begin{align*}
        \ubar v_1\left(tr \mid \theta_1, \sigma_2^{\hat\theta} \right)
        \geq \left( 1-2F_h(\theta_1)\right)F_l(\hat\theta)+F_h(\theta_1)
        > F_h(\theta_1) \geq F_l(\theta_1) = \ubar v_1\left(nt \mid \theta_1, \sigma_2^{\hat\theta} \right).
    \end{align*}
    For $\theta_1 \in X_2 \cup X_3$, the value of \eqref{eq:v(tr)_with_G,H_2} is
    \begin{align*}
        & \quad \min_{H\in \left\{ F_l, F_h \right\}}\min_{G\in \left\{ F_l, F_h \right\}} \left(1-2H(\hat\theta)\right)G(\theta_1) + H(\hat\theta)\\
        &= \min_{H\in \left\{ F_l, F_h \right\}} \left(1-2H(\hat\theta)\right)F_h(\theta_1) + H(\hat\theta) \\
        &= \min_{H\in \left\{ F_l, F_h \right\}} \left(1-2F_h(\theta_1)\right)H(\hat\theta) + F_h(\theta_1) \\
        &= \left(1-2F_h(\theta_1)\right)F_h(\hat\theta) + F_h(\theta_1)
    \end{align*}
    such that
    \begin{align*}
        \ubar v_1\left(tr \mid \theta_1, \sigma_2^{\hat\theta} \right)
        = \left( 1-2F_h(\hat\theta)\right)F_h(\theta_1)+F_h(\hat\theta) = \left( 1-2F_h(\theta_1)\right)F_h(\hat\theta)+F_h(\theta_1).
    \end{align*}
    It follows that for $\hat\theta \in X_3$, the function $\theta_1 \mapsto \ubar v_1\left(tr \mid \theta_1, \sigma_2^{\hat\theta} \right)$ is strictly decreasing on $X_2 \cup X_3$.
    Moreover, 
    \begin{align*}
        \ubar v_1\left(tr \mid F_h^{-1}\left(\tfrac{1}{2}\right), \sigma_2^{\hat\theta} \right)=\tfrac{1}{2}\geq F_l\left(F_h^{-1}\left(\tfrac{1}{2}\right)\right) = \ubar v_1\left(nt \mid F_h^{-1}\left(\tfrac{1}{2}\right), \sigma_2^{\hat\theta} \right)
    \end{align*}
    and 
    \begin{align*}
        \ubar v_1\left(tr \mid F_l^{-1}\left(\tfrac{1}{2}\right), \sigma_2^{\hat\theta} \right)
        &= \left(1-2F_h\left(F_l^{-1}\left(\tfrac{1}{2}\right)\right)\right)F_h(\hat\theta)+F_h\left(F_l^{-1}\left(\tfrac{1}{2}\right)\right) \\
        &< 2\left( F_h\left(F_l^{-1}\left(\tfrac{1}{2}\right)\right) - F_h\left(F_l^{-1}\left(\tfrac{1}{2}\right)\right)^2\right) \\
        &\leq \tfrac{1}{2} \\
        &= F_l\left(F_l^{-1}\left(\tfrac{1}{2}\right)\right) \\
        &= \ubar v_1\left(nt \mid F_l^{-1}\left(\tfrac{1}{2}\right), \sigma_2^{\hat\theta} \right)
    \end{align*}
    where the first inequality follows from $1-2F_h\left(F_l^{-1}\left(\tfrac{1}{2}\right)\right) \leq 0$ and $F_h(\hat\theta)>F_h\left(F_l^{-1}\left(\tfrac{1}{2}\right)\right)$, and the second one from $2\left(x-x^2\right)\leq \tfrac{1}{2}$ for all $x \in \R$.
    Therefore, as $\ubar v_1\left(tr \mid \cdot, \sigma_2^{\hat\theta}\right)$ is strictly decreasing and $\ubar v_1\left(nt \mid \cdot, \sigma_2^{\hat\theta}\right)$ strictly increasing on $X_2 \cup X_3$, there must be a unique $\theta_1^* \in X_2$ such that $\sigma_1^{\theta_1^*}$ is the ambiguous maxmin cursed best response to $\sigma_2^{\hat\theta}$.

    It follows that the ambiguous maxmin cursed best response to a cut-off strategy with threshold in $X_1 \cup X_3$ is a cut-off strategy with threshold in $X_2$.
    Hence, for any $\theta_1, \theta_2 \in X_1 \cup X_3$, $\left(\sigma_1^{\theta_1}, \sigma_2^{\theta_2}\right)$ cannot be an ambiguous cursed Knight-Nash equilibrium.

    Let us now show that it is not possible that one threshold lies in $X_2$ while the other one lies in $X_1 \cup X_3$.
    Assume that one player---without loss of generality, player 2---plays a cut-off strategy with threshold $\hat\theta\in X_2$.
    For any $\theta_1 \in X_1$, similar arguments as above show that
    \begin{align}
        \ubar v_1\left(tr \mid \theta_1, \sigma_2^{\hat\theta}\right) 
        = \left( 1-2F_l(\theta_1)\right)F_l(\hat\theta)+F_l(\theta_1) > F_l(\theta_1) = \ubar v_1\left(nt \mid \theta_1, \sigma_2^{\hat\theta}\right) \label{eq:v1(tr)>v1(nt)_theta1_in_X1}
    \end{align}
    such that player 1 wants to trade for any $\theta_1 \in X_1$, in particular they will not use a cut-off strategy with some threshold in $X_1$.
    For $\theta_1 \in X_3$,
    \begin{align}
        \ubar v_1\left(tr \mid \theta_1, \sigma_2^{\hat\theta}\right) 
        &= F_h(\theta_1)+F_h(\hat\theta) - 2 F_h(\theta_1)F_h(\hat\theta) \nonumber \\
        &= \left( 1-2F_h(\hat\theta)\right)F_h(\theta_1)+F_h(\hat\theta) \nonumber \\
        &\leq \left( 1-2F_h(\hat\theta)\right)F_l(\theta_1)+F_h(\hat\theta) \nonumber \\
        &= \left( 1-2F_l(\theta_1)\right)F_h(\hat\theta)+F_l(\theta_1) \nonumber \\
        &< F_l(\theta_1) = \ubar v_1\left(nt \mid \theta_1, \sigma_2^{\hat\theta}\right) \label{eq:v1(tr)<v1(nt)_theta1_in_X3}
    \end{align}
    such that player 1 will not want to trade for any $\theta_1 \in X_3$, in particular they will not use a cut-off strategy with threshold in $X_3$ against $\sigma_2^{\hat\theta}$.
\end{proof}

\noindent We can now state and prove an analogue to Corollary \ref{cor:fully_cursed_Knight_Nash_equilibrium_certain_median_under_F*} and show that if there is no uncertainty about the median, both players play the cut-off strategy with this unambiguous median in the cursed equilibrium.
In particular, in this case, the two approaches to incorporate Knightian uncertainty into the cursed equilibrium coincide.

\begin{corollary} \label{cor:ambiguous_fully_cursed_Knight_Nash_equilibrium_certain_median}
    In addition to Assumption \ref{ass:prior_distributions_ambiguous}, let $F_h^{-1}\left(\tfrac{1}{2}\right)= F_l^{-1}\left(\tfrac{1}{2}\right)$. 
    Then, there is an ambiguous cursed Knight-Nash equilibrium in which both players play the cut-off strategy with threshold $F_h^{-1}\left(\tfrac{1}{2}\right)= F_l^{-1}\left(\tfrac{1}{2}\right)$, and this equilibrium is the only non-trivial equilibrium in cut-off strategies.
\end{corollary}

\begin{proof}
    By Proposition \ref{prop:cut-offs_ambiguity_in_middle}, the given strategy profile is the only candidate for such an equilibrium. 
    Indeed, if player 2 plays this strategy, we know from the last paragraph of the proof of Proposition \ref{prop:cut-offs_ambiguity_in_middle} that player 1's ambiguous maxmin fully cursed best response involves playing $tr$ for $\theta_1 \in X_1$, i.e., for $\theta_1 < F_h^{-1}\left(\tfrac{1}{2}\right)=F_l^{-1}\left(\tfrac{1}{2}\right)$; and $nt$ for $\theta_1 \in X_3$, i.e., for $\theta_1 > F_h^{-1}\left(\tfrac{1}{2}\right)=F_l^{-1}\left(\tfrac{1}{2}\right)$. 
    Hence, the maxmin cursed best response against a cut-off strategy with threshold $F_h^{-1}\left(\tfrac{1}{2}\right)=F_l^{-1}\left(\tfrac{1}{2}\right)$ is this strategy itself.
\end{proof}

\noindent Before coming to the main result of this appendix, we need two auxiliary lemmas.
The first one characterizes the value of $\ubar v_k\left(tr \mid \theta_k, \sigma_{-k} \right)$ for all ``important'' types $\theta_k$ and strategies $\sigma_{-k}$, therefore providing an analogue to Lemma \ref{lem:v(nt)_and_v(tr)}.
The second lemma shows that in an ambiguous cursed Knight-Nash equilibrium, only symmetric strategies can be played.
This contrasts the equilibrium concept from the main part, where asymmetric equilibria were possible (Remark \ref{rmk:multiple_equilibria}).

\begin{lemma} \label{lem:value_of_minimum_at_cut-off}
    Let Assumption \ref{ass:prior_distributions_ambiguous} be satisfied. 
    For $k \in \{1,2\}$ and $\hat\theta \in \left[ F_h^{-1}\left(\tfrac{1}{2}\right),F_l^{-1}\left(\tfrac{1}{2}\right)\right]$,
    \begin{align} \label{eq:v(tr)_in_X2}
        \ubar v_k\left(tr \mid \theta_k, \sigma_{-k}^{\hat\theta} \right) = 
        \begin{cases}
            F_l(\theta_k)+F_l(\hat\theta) - 2 F_l(\theta_k)F_l(\hat\theta) &\text{ for } \theta_k \in \left[ F_h^{-1}\left(\tfrac{1}{2}\right), 1-\hat\theta \right], \\
            F_h(\theta_k)+F_h(\hat\theta) - 2 F_h(\theta_k)F_h(\hat\theta) &\text{ for } \theta_k \in \left[1-\hat\theta,  F_l^{-1}\left(\tfrac{1}{2}\right) \right],
        \end{cases}
    \end{align}
    and there exists some $b(\hat\theta) \in \left[1-\hat\theta,  F_l^{-1}\left(\tfrac{1}{2}\right) \right]$ such that the unique ambiguous maxmin cursed best response against $\sigma_{-k}^{\hat\theta}$ is $\sigma_k^{b(\hat\theta)}$.
    The function $b$ is continuous and monotonously decreasing in $\hat\theta$.
\end{lemma}

\begin{proof}
    Without loss of generality, let $k=1$.
    We have seen in \eqref{eq:v(tr)_with_G,H_2} that
    \begin{align*}
        \ubar v_1\left(tr \mid \theta_1, \sigma_2^{\hat\theta} \right) \geq \min_{G,H \in \left\{ F_l, F_h \right\}} G(\theta_1)+H(\hat\theta) - 2 G(\theta_1)H(\hat\theta)
    \end{align*}
    for all $\theta_1, \hat\theta \in [0,1]$.
    For $\theta_1, \hat\theta \in X_2=\left[ F_h^{-1}\left(\tfrac{1}{2}\right),F_l^{-1}\left(\tfrac{1}{2}\right)\right]$, 
    \begin{align*}
        F_h(\theta_1)+F_l(\hat\theta) - 2 F_h(\theta_1)F_l(\hat\theta) 
        &= \left( 1-2F_l(\hat\theta)\right)F_h(\theta_1)+F_l(\hat\theta) \\
        &\geq \left( 1-2F_l(\hat\theta)\right)F_l(\theta_1)+F_l(\hat\theta) = F_l(\theta_1)+F_l(\hat\theta) - 2 F_l(\theta_1)F_l(\hat\theta)
    \end{align*}
    and a similar calculation shows that
    \begin{align*}
        F_l(\theta_1)+F_h(\hat\theta) - 2 F_l(\theta_1)F_h(\hat\theta) \geq F_h(\theta_1)+F_h(\hat\theta) - 2 F_h(\theta_1)F_h(\hat\theta).
    \end{align*}
    It follows that $\ubar v_1\left(tr \mid \theta_1, \sigma_2^{\hat\theta} \right) \geq \min_{F \in \left\{ F_l, F_h \right\}} F(\theta_1)+F(\hat\theta) - 2 F(\theta_1)F(\hat\theta)$ and, as the converse inequality directly follows from Equation \eqref{eq:ubar_v(tr)},
    \begin{align*}
        \ubar v_1\left(tr \mid \theta_1, \sigma_2^{\hat\theta} \right) = \min_{F \in \left\{ F_l, F_h \right\}} F(\theta_1)+F(\hat\theta) - 2 F(\theta_1)F(\hat\theta).
    \end{align*}
    As $F_l(\hat\theta)\leq\tfrac{1}{2}$, the function $\theta_1 \mapsto \left( 1-2F_l(\hat\theta)\right)F_l(\theta_1)+F_l(\hat\theta)$ is monotonously increasing in $\theta_1$ and equals $\tfrac{1}{2}$ for $\theta_1=F_l^{-1}\left(\tfrac{1}{2}\right)$.
    As $F_h(\hat\theta)\geq\tfrac{1}{2}$, the function $\theta_1 \mapsto \left( 1-2F_h(\hat\theta)\right)F_h(\theta_1)+F_h(\hat\theta)$ is monotonously decreasing in $\theta_1$ and equals $\tfrac{1}{2}$ for $\theta_1=F_h^{-1}\left(\tfrac{1}{2}\right)$.
    As $F_h^{-1}\left(\tfrac{1}{2}\right)<F_l^{-1}\left(\tfrac{1}{2}\right)$, either $F_l(\hat\theta)<\tfrac{1}{2}$ or $F_h(\hat\theta)>\tfrac{1}{2}$ (or both) such that one of the two functions must be strictly monotonous.
    Hence, they must have a unique intersection in $\left[ F_h^{-1}\left(\tfrac{1}{2}\right),F_l^{-1}\left(\tfrac{1}{2}\right)\right]$.
    It directly follows from Assumption \ref{ass:prior_distributions_ambiguous} that the functions coincide for $\theta_1=1-\hat\theta$, so we obtain Equation \eqref{eq:v(tr)_in_X2}.

    It remains to show the existence and uniqueness of a number $b(\hat\theta) \in \left[1-\hat\theta,  F_l^{-1}\left(\tfrac{1}{2}\right) \right]$ such that $\sigma_1^{b(\hat\theta)}$ is an ambiguous maxmin fully cursed best response against $\sigma_2^{\hat\theta}$.
    While $\ubar v_1 \left(nt \mid \cdot, \sigma_2^{\hat\theta}\right)=F_l$ is strictly increasing, $\ubar v_1 \left(tr \mid \cdot, \sigma_2^{\hat\theta}\right)$ is, by Equation \eqref{eq:v(tr)_in_X2} increasing on $\left[F_h^{-1}\left(\tfrac{1}{2}\right), 1-\hat\theta\right]$ and decreasing on $\left[ 1-\hat\theta, F_l^{-1}\left(\tfrac{1}{2}\right)\right]$.
    Moreover, as $1-2F_h(\hat\theta)\leq 0$,
    \begin{align*}
        \ubar v_1 \left( tr \mid 1-\hat\theta, \sigma_2^{\hat\theta}\right)
        &= \left(1-2F_h(\hat\theta)\right)F_h(1-\hat\theta)+F_h(\hat\theta) \\
        &\geq \left(1-2F_h(\hat\theta)\right)\cdot 1+F_h(\hat\theta) = 1-F_h(\hat\theta) = F_l(1-\hat\theta)= \ubar v_1 \left( nt \mid 1-\hat\theta, \sigma_2^{\hat\theta}\right),
    \end{align*}
    as well as
    \begin{align*}
        \ubar v_1 \left( tr \mid F_l^{-1}\left(\tfrac{1}{2}\right), \sigma_2^{\hat\theta}\right) 
        &= \left(1-2F_h(\hat\theta)\right)F_h\left(F_l^{-1}\left(\tfrac{1}{2}\right)\right)+F_h(\hat\theta) \\
        &\leq \left(1-2F_h(\hat\theta)\right)\tfrac{1}{2}+F_h(\hat\theta) = \tfrac{1}{2} = \ubar v_1 \left( nt \mid F_l^{-1}\left(\tfrac{1}{2}\right), \sigma_2^{\hat\theta}\right).
    \end{align*}
    By continuity, there must be a unique number $b(\hat\theta) \in \left[ 1-\hat\theta, F_l^{-1}\left(\tfrac{1}{2}\right)\right]$ such that
    \begin{align*}
        \ubar v_1 \left(tr \mid \theta_1, \sigma_2^{\hat\theta} \right) &> \ubar v_1 \left(nt \mid \theta_1, \sigma_2^{\hat\theta} \right) &&\text{for } \theta_1 \in \left[1-\hat\theta, b(\hat\theta)\right), \\
        \ubar v_1 \left(tr \mid b(\hat\theta), \sigma_2^{\hat\theta} \right) &= \ubar v_1 \left(nt \mid b(\hat\theta), \sigma_2^{\hat\theta} \right), &&\text{and} \\
        \ubar v_1 \left(tr \mid \theta_1, \sigma_2^{\hat\theta} \right) &< \ubar v_1 \left(nt \mid \theta_1, \sigma_2^{\hat\theta} \right) &&\text{for } \theta_1 \in \left(b(\hat\theta), F_l^{-1}\left(\tfrac{1}{2}\right)\right].
    \end{align*}    
    For $\theta_1 \in \left[ F_h^{-1}\left(\tfrac{1}{2}\right), 1-\hat\theta \right)$, we have $F_l(\theta_1)<\tfrac{1}{2}$ and therefore
    \begin{align*}
        \ubar v_1 \left(tr \mid \theta_1, \sigma_2^{\hat\theta}\right) 
        =\left(1-2F_l(\theta_1)\right) F_l(\hat\theta)+F_l(\theta_1)
        > F_l(\theta_1)= \ubar v_1 \left(nt \mid \theta_1, \sigma_2^{\hat\theta}\right) .
    \end{align*}
    Moreover, we know from Equations \eqref{eq:v1(tr)>v1(nt)_theta1_in_X1} and \eqref{eq:v1(tr)<v1(nt)_theta1_in_X3} that $\ubar v_1 \left(tr \mid \theta_1, \sigma_2^{\hat\theta} \right) > \ubar v_1 \left(nt \mid \theta_1, \sigma_2^{\hat\theta} \right)$ for $\theta_1 \in \left[0, F_h^{-1}\left(\tfrac{1}{2}\right)\right)$ as well as $\ubar v_1 \left(tr \mid \theta_1, \sigma_2^{\hat\theta} \right) < \ubar v_1 \left(nt \mid \theta_1, \sigma_2^{\hat\theta} \right)$ for $\theta_1 \in \left( F_l^{-1}\left(\tfrac{1}{2}\right), 1\right]$. 
    Therefore, $\sigma_1^{b(\hat\theta)}$ is indeed the unique ambiguous maxmin fully cursed best response against $\sigma_2^{\hat\theta}$.

    It remains to show that $b$ is continuous and decreasing in $\hat\theta$.
    To this end, define for each $\hat\theta \in \left[ F_h^{-1}\left(\tfrac{1}{2}\right),F_l^{-1}\left(\tfrac{1}{2}\right)\right]$ the auxiliary function $W_{\hat\theta} \colon \left[ F_h^{-1}\left(\tfrac{1}{2}\right),F_l^{-1}\left(\tfrac{1}{2}\right)\right] \to \R$ via
    \begin{align}
        W_{\hat\theta}(\theta_1) \colonequals F_h(\theta_1)+F_h(\hat\theta)-2F_h(\theta_1)F_h(\hat\theta) - F_l(\theta_1). \label{eq:definition_W}
    \end{align}
    We observe that $(\hat\theta, \theta_1) \mapsto W_{\hat\theta}(\theta_1)$ is continuous and that for each $\hat\theta$, $W_{\hat\theta}$ is strictly decreasing, hence invertible.
    By definition, $b(\hat\theta)=W^{-1}_{\hat\theta}(0) \in \left[ F_h^{-1}\left(\tfrac{1}{2}\right),F_l^{-1}\left(\tfrac{1}{2}\right)\right]$ which is well-defined by the calculations above.
    As all involved functions are continuous, $b$ is continuous.
    Moreover, $\hat\theta \mapsto W_{\hat\theta}(\theta_1)$ is monotonously decreasing for $\theta_1 \in \left[ F_h^{-1}\left(\tfrac{1}{2}\right),F_l^{-1}\left(\tfrac{1}{2}\right)\right]$ as $F_h(\theta_1)\geq \tfrac{1}{2}$ for these $\theta_1$.
    Hence, so is $W^{-1}_{\hat\theta}(0)$.
    Now, let $\hat\theta_k \leq \hat\theta_{ii}$ and observe that $W_{\hat\theta_k}(\theta) \geq W_{\hat\theta_{ii}}(\theta)$ for all $\theta$, in particular $0= W_{\hat\theta_k}\left(W^{-1}_{\hat\theta_k}(0)\right) \geq W_{\hat\theta_{ii}}\left(W^{-1}_{\hat\theta_k}(0)\right)$.
    Applying the decreasing function $W^{-1}_{\hat\theta_{ii}}$ to both sides of the inequality yields that $W^{-1}_{\hat\theta_{ii}}(0) \leq W^{-1}_{\hat\theta_{i}}(0)$.
    Therefore, $b(\hat\theta)=W_{\hat\theta}^{-1}(0)$ is decreasing in $\hat\theta$.
\end{proof}

\begin{lemma} \label{lem:ambiguous_cursed_Knight_Nash_equilibrium_symmetric}
    Let Assumption \ref{ass:prior_distributions_ambiguous} be satisfied.
    In every non-trivial ambiguous cursed Knight-Nash equilibrium in cut-off strategies, both players play the same strategy.
\end{lemma}

\begin{proof}
    Let the equilibrium be such that player $k$ plays the cut-off strategy $\sigma_k^{\theta_k}$.
    We have to show that $\theta_1=\theta_2$.
    By Proposition \ref{prop:cut-offs_ambiguity_in_middle}, $\theta_k \in \left[ F_h^{-1}\left(\tfrac{1}{2}\right), F_l^{-1}\left(\tfrac{1}{2}\right)\right]$ for $k \in \{1,2\}$.
    As $\sigma_1^{\theta_1}$ is an ambiguous fully cursed best response to $\sigma_{2}^{\theta_{2}}$ and vice versa, Lemma \ref{lem:value_of_minimum_at_cut-off} yields that $\theta_1\geq1-\theta_2$ and $\theta_2\geq1-\theta_1$ as well as
    \begin{align*}
        F_l(\theta_1) = \ubar v_1\left(nt \mid \theta_1, \sigma_2^{\theta_2}\right) &= \ubar v_1\left(tr \mid \theta_1, \sigma_2^{\theta_2}\right) \\
        &= F_h(\theta_1)+F_h(\theta_2)-2F_h(\theta_1)F_h(\theta_2) \\
        &= \ubar v_2\left( tr \mid \theta_2, \sigma_1^{\theta_1}\right) 
        = \ubar v_2\left( nt \mid \theta_2, \sigma_1^{\theta_1}\right) = F_l(\theta_2).
    \end{align*}
    As $F_l$ is strictly increasing, it follows that $\theta_1=\theta_2$.
\end{proof}

\noindent We turn to the analogue to Theorem \ref{thm:existence_chracterization_fully_cursed_KNE_under_F*}.
In the framework from the main part, only existence of equilibria is established, while uniqueness may not hold (Remark \ref{rmk:multiple_equilibria}).
We can now show that the ambiguous cursed Knight-Nash equilibrium is unique within the class of cut-off strategies.
However, unlike in the cursed Knight-Nash equilibrium given some true distribution $F^*$, the ambiguous framework also allows for equilibria that are not in cut-off strategies.

\begin{theorem} \label{thm:existence_uniqueness_ambiguous_cursed_KNE}
    Let Assumption \ref{ass:prior_distributions_ambiguous} be satisfied.
    Then, there exists a unique non-trivial ambiguous cursed Knight-Nash equilibrium in cut-off strategies. 
    In this equilibrium, both players play the same threshold strategy\footnote{
    In this appendix, we denote optimal cut-offs by the ``curly'' $\vartheta^*$, in order to avoid confusion with the $\theta^*$ which denoted the optimal cut-offs in Subsection \ref{subsec:How_Uncertainty_Leads_Cursed_Players_To_Trade_Even_More}.
    We will later compare the two values.
    }
    $\sigma^{\vartheta^*}$.
    The threshold $\vartheta^*$ lies in $\left(\tfrac{1}{2}, F_l^{-1}\left(\tfrac{1}{2}\right)\right]$, and satisfies
    \begin{align} \label{eq:threshold_equation_ambigous_CKNE}
        2\left(F_h\left(\vartheta^*\right)-F_h\left(\vartheta^*\right)^2\right) = F_l\left(\vartheta^*\right).
    \end{align}
    Within the class of cut-off strategies, this equilibrium the only non-trivial one.
\end{theorem}

\begin{proof}
    If such an equilibrium exists, both player's thresholds must lie in $\left[ F_h^{-1}\left(\tfrac{1}{2}\right),F_l^{-1}\left(\tfrac{1}{2}\right)\right]$ by Proposition \ref{prop:cut-offs_ambiguity_in_middle}, and by Lemma \ref{lem:ambiguous_cursed_Knight_Nash_equilibrium_symmetric}, they must be identical.
    Now, a symmetric profile of cut-off strategies $\left(\sigma_1^{\vartheta^*}, \sigma_2^{\vartheta^*}\right)$ constitutes an ambiguous cursed Knight-Nash equilibrium if and only if $\vartheta^* \in \left[ F_h^{-1}\left(\tfrac{1}{2}\right),F_l^{-1}\left(\tfrac{1}{2}\right)\right]$ is a fixed point of the best reply function function $b$ from Lemma \ref{lem:value_of_minimum_at_cut-off}, that is, if and only if it satisfies \eqref{eq:threshold_equation_ambigous_CKNE}.
    As $b$ continuously maps $\left[ F_h^{-1}\left(\tfrac{1}{2}\right),F_l^{-1}\left(\tfrac{1}{2}\right)\right]$ onto itself, there must be such a fixed point, and as $b$ monotonously decreases, it is unique.
    As $b(\theta)\geq 1-\theta$ for all $\theta$, the fixed point $\vartheta^*$ cannot be smaller than $\tfrac{1}{2}$.
    Assume by means of contradiction that $\vartheta^*=\tfrac{1}{2}$.
    Then, by the definition of $b$, it follows that $W_{\tfrac{1}{2}}\left(\tfrac{1}{2}\right)=0$---with $W$ being defined in \eqref{eq:definition_W}---, so
    \begin{align*}
        F_l\left(\tfrac{1}{2}\right)=2\left(F_h\left(\tfrac{1}{2}\right)-F_h\left(\tfrac{1}{2}\right)^2\right)=2\left(F_l\left(\tfrac{1}{2}\right)-F_l\left(\tfrac{1}{2}\right)^2\right)<F_l\left(\tfrac{1}{2}\right),
    \end{align*}
    where the second equality follows from choosing $\theta_k=\hat\theta=\tfrac{1}{2}$ in \eqref{eq:v(tr)_in_X2} and the inequality from $F_l\left(\tfrac{1}{2}\right)<\tfrac{1}{2}$.
    The overall inequality is impossible, so $\vartheta^*$ must be greater than $\tfrac{1}{2}$.
\end{proof}

\noindent Theorem \ref{thm:existence_uniqueness_ambiguous_cursed_KNE} parallels Theorem \ref{thm:existence_chracterization_fully_cursed_KNE_under_F*} in showing that a cursed Knight-Nash equilibrium exists, and that in any such equilibrium, more types are willing to engage in trade than if there is no uncertainty:
In the latter case, the only non-trivial cursed equilibrium is where both use the cut-off strategy with threshold $\frac{1}{2}$; if there is uncertainty, the thresholds must be strictly greater.
The underlying reasoning is the same as the one for Theorem \ref{thm:existence_chracterization_fully_cursed_KNE_under_F*}:
while a player who refuses to trade only profits from the opponent having a low type, a cursed agent who engages in trade can seemingly profit from high as well as low types of the opponent, which mitigates the effect of uncertainty.

However, this effect is smaller in the ambiguous cursed Knight-Nash equilibrium than in the cursed Knight-Nash equilibrium given some true distribution $F^*$.

\begin{proposition}
    Let $\theta^*$ and $\vartheta^*$ be the thresholds in the cursed Knight-Nash equilibrium given $Id$ (cf. Theorem \ref{thm:existence_chracterization_fully_cursed_KNE_under_F*}), or the ambiguous cursed Knight-Nash equilibrium (cf. Theorem \ref{thm:existence_uniqueness_ambiguous_cursed_KNE}).
    It then holds that $\theta^*\geq \vartheta^*$, and the inequality is strict if $F_h(\theta)>\theta$ for all $\theta \in (0,1)$.
\end{proposition}

\begin{proof}
    Assume by means of contradiction that $\theta^*< \vartheta^*$.
    Then, as $\vartheta^*$ satisfies \eqref{eq:threshold_equation_ambigous_CKNE}, while $\theta^*$ satisfies \eqref{eq:symmetric_cursed_equilibrium_under_F*}, we obtain that
    \begin{align*}
        F_l\left(\vartheta^*\right) &= 2 \left(F_h\left(\vartheta^*\right) - F_h\left(\vartheta^*\right)^2 \right) \\
        &\leq 2 \left(F_h\left(\theta^*\right) - F_h\left(\theta^*\right)^2 \right) \\
        &= \left(1-2F_h\left(\theta^*\right)\right)F_h\left(\theta^*\right)+F_h\left(\theta^*\right) \\
        &\leq \left(1-2F_h\left(\theta^*\right)\right)\theta^*+F_h\left(\theta^*\right)
        =F_l\left(\theta^*\right).
    \end{align*}
    For the first inequality, we used that $F_h\left(\theta^*\right) < F_h\left(\vartheta^*\right)$ in combination with $x \mapsto 2\left(x-x^2\right)$ being strictly decreasing on $\left(\tfrac{1}{2}, 1\right)$; and for the second inequality, that $1-2F_h\left(\theta^*\right)<0$ and $F_h\left(\theta^*\right)\geq \theta^*$.
    We obtain that $F_l\left(\vartheta^*\right) \leq F_l\left(\theta^*\right)$ which implies $\vartheta^* \leq \theta^*$, thus we arrive at a contradiction.

    The second inequality is strict if we assume that $F_h(\theta)>\theta$ for all $\theta \in (0,1)$, which closes the proof.
\end{proof}

\noindent If the average strategy of the opponent is not derived from a true underlying distribution, but subject to minimization, the expected value of trade becomes smaller, while the expected value of refusing to trade is unaffected.
Therefore, the set of player types that engage in trade is larger in the presence of a true distribution than in the absence of such.

We proceed by revisiting the parametrizations introduced in Example \ref{ex:parametrizations}.

\begin{figure}
\begin{subfigure}[t]{0.5\textwidth}
    \centering
    \includegraphics[width=\textwidth]{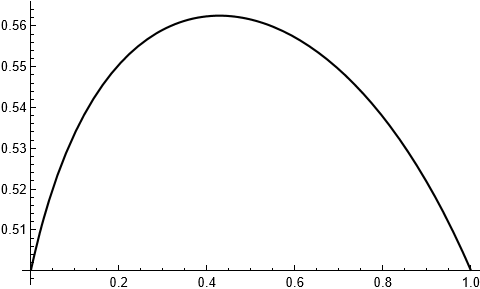}
    \caption{Threshold $\vartheta_{\kappa}^*$}
    \label{subfig:theta^*_kappa_ambiguity}
\end{subfigure}%
\begin{subfigure}[t]{0.5\textwidth}
    \centering
    \includegraphics[width=\textwidth]{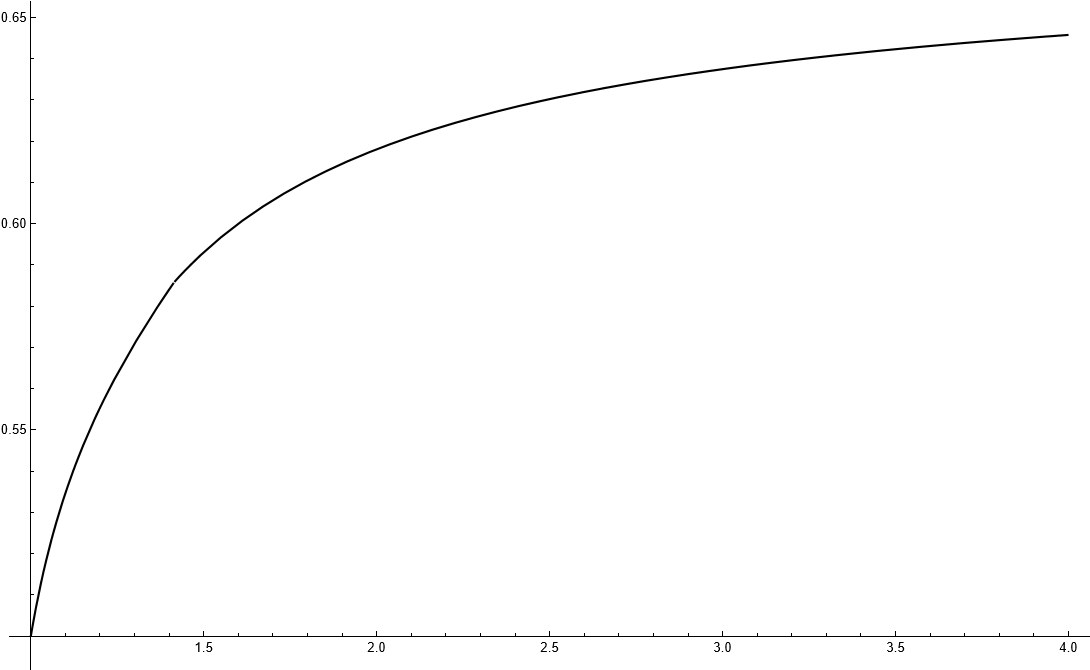}
    \caption{Threshold $\vartheta_{a}^*$}
    \label{subfig:theta^*a_ambiguity}
\end{subfigure}%
\vspace{.8cm}
\centering
\\
\begin{subfigure}[t]{0.5\textwidth}
    \centering
    \includegraphics[width=\textwidth]{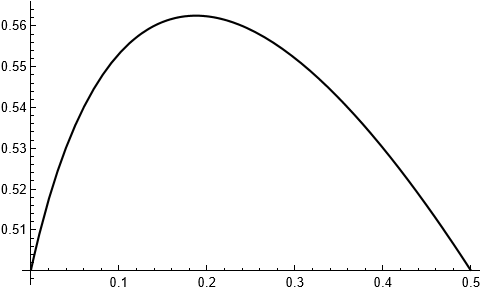}
    \caption{Threshold $\vartheta_{\eps}^*$}
    \label{subfig:theta^*_eps_ambiguity}
\end{subfigure}%
\caption{Thresholds in Example \ref{ex:examples_revisited_II}}
\label{fig:thresholds_ambiguity}
\end{figure}

\begin{example}
\label{ex:examples_revisited_II}
    For $\kappa \in (0,1)$, $a \in (1, \infty)$, and $\eps \in \left(0, \tfrac{1}{2}\right)$, let $\F_{\kappa}$, $\F_a$, and $\F_{\eps}$ be as in Example \ref{ex:parametrizations}.
    We know from Theorem \ref{thm:existence_uniqueness_ambiguous_cursed_KNE} that there exist symmetric cursed Knight-Nash equilibria with thresholds in $\left(\tfrac{1}{2},F_l^{-1}\left(\tfrac{1}{2}\right)\right)$ that can be found by solving Equation \eqref{eq:threshold_equation_ambigous_CKNE}.\footnote{
    Regarding $\F_{\eps}$, the same disclaimer as in Footnote \ref{fn:disclaimer_F_eps} applies.
    }
    In the cases at hand, the thresholds in the equilibria are 
    \begin{align*}
        \vartheta_{\kappa}^* &= \frac{4 \kappa^2 - 7 \kappa +1+\sqrt{- 7 \kappa^2 +10 \kappa+1}}{4 (\kappa^2 - 2 \kappa +1)}, \\
        \vartheta_{a}^* &= \begin{cases}
            \frac{-a^3-2a+4+a \sqrt{a^4+4a^2-4}}{4} &\text{ for } a \leq \sqrt{2}, \\
            \frac{-3a+4+\sqrt{9a^2-8a}}{4} &\text{ for } a \geq \sqrt{2},
        \end{cases}
        \qquad \text{and} \\
        \vartheta_{\eps}^*&=\frac{-4\eps+1+\sqrt{16\eps+1}}{4}.
    \end{align*} 
    Figure \ref{fig:thresholds_ambiguity} depicts these thresholds as functions of their respective parameters.
    One can show that that $\vartheta_{\kappa}$ is first increasing in $\kappa$, attains its maximum at $\kappa=\frac{3}{7}$ and $\vartheta^*_{\kappa}=\frac{9}{16}$, and then decreases again until $\lim_{\kappa \to 1}\vartheta^*_{\kappa}=\tfrac{1}{2}$; that $\vartheta^*_a$ is increasing in $a$ with $\lim_{a\to\infty}\vartheta_a^*=\frac{2}{3}$; and that $\vartheta^*_{\eps}$ first increases, attains its maximum at $\eps=\frac{3}{16}$ and, again, $\vartheta^*_{\eps}=\frac{9}{16}$, and then decreases to $\lim_{\eps \to \tfrac{1}{2}}\vartheta^*_{\eps}=\tfrac{1}{2}$.
\end{example}

\noindent We close this appendix by providing results parallel to Proposition \ref{prop:comparative_statics_max_min_prob_of_trade_under_F*} and Theorem \ref{thm:existence_uniqueness_cursed_uncursed_KNE}:
more uncertainty about the type distribution leads to more uncertainty about the outcome of the game, and the location of the cursed-uncursed equilibrium remains unchanged.

\begin{proposition} \label{prop:comparative_statics_max_min_prob_of_trade_ambiguous}
    Let $\F$ and $\mathcal{G}$ be two uncertainty sets that satisfy Assumption \ref{ass:prior_distributions_ambiguous}, and let $\vartheta^*_{\F}$ and $\vartheta^*_{\mathcal{G}}$ be the thresholds of the cut-off strategies played in the unique symmetric cursed Knight-Nash equilibria.
    If $G_h(\theta)>F_h(\theta)$ and $G_l(\theta)<F_l(\theta)$ for all $\theta \in (0,1)$, then $G_l\left(\vartheta^*_{\mathcal{G}}\right)<F_l\left(\vartheta^*_{\F}\right)$ and $G_h\left(\vartheta^*_{\mathcal{G}}\right)>F_h\left(\vartheta^*_{\F}\right)$.
\end{proposition}

\begin{proof}
    We consider the two cases $\vartheta^*_{\mathcal{G}}\geq \vartheta_{\F}^*$ and $\vartheta^*_{\mathcal{G}}\leq \vartheta_{\F}^*$.\footnote{
    Example \ref{ex:examples_revisited_II} shows that both cases are possible.
    }
    In the first case, as $F_h$ and $G_h$ are increasing, $G_h\left(\vartheta^*_{\mathcal{G}}\right) \geq G_h\left(\vartheta^*_{\F}\right) > F_h\left(\vartheta^*_{\F}\right)$.
    Using that $G_h\left(\vartheta^*_{\mathcal{G}}\right)$ and $F_h\left(\vartheta^*_{\F}\right)$ are both contained in $\left(\frac{1}{2}, 1\right)$ and that the map $x \mapsto 2\left(x-x^2\right)$ is strictly decreasing here, we obtain that
    \begin{align*}
        G_l\left(\vartheta^*_{\mathcal{G}}\right) = 2 \left( G_h\left(\vartheta^*_{\mathcal{G}}\right) - G_h\left(\vartheta^*_{\mathcal{G}}\right)^2\right) 
        < 2 \left( F_h\left(\vartheta^*_{\F}\right) - F_h\left(\vartheta^*_{\F}\right)^2\right) = F_l\left(\vartheta^*_{\F}\right),
    \end{align*}
    where the equalities follow from Equation \eqref{eq:threshold_equation_ambigous_CKNE}.

    In the second case, as $F_l$ and $G_l$ are increasing, $G_l\left(\vartheta^*_{\mathcal{G}}\right) \leq G_l\left(\vartheta^*_{\F}\right) < F_l\left(\vartheta^*_{\F}\right)$.
    Again using \eqref{eq:threshold_equation_ambigous_CKNE}, it holds that
    \begin{align*}
        2 \left( G_h\left(\vartheta^*_{\mathcal{G}}\right) - G_h\left(\vartheta^*_{\mathcal{G}}\right)^2\right) = G_l\left(\vartheta^*_{\mathcal{G}}\right) 
        < F_l\left(\vartheta^*_{\F}\right) = 2 \left( F_h\left(\vartheta^*_{\F}\right) - F_h\left(\vartheta^*_{\F}\right)^2\right),
    \end{align*}
    which, as $x \mapsto 2\left(x-x^2\right)$ is strictly decreasing on $\left(\frac{1}{2}, 1\right)$, completes the proof.
\end{proof}

\begin{theorem}\label{thm:existence_uniqueness_cursed_uncursed_KNE_ambiguous}
    Under Assumption \ref{ass:prior_distributions_ambiguous}, the unique non-trivial ambiguous $\left(\{1\}, \{2\}\right)$-cursed-uncursed Knight-Nash equilibrium in cut-off strategies is $\left(\sigma_1^{F_l^{-1}\left(\frac{1}{2}\right)}, \sigma_2^{\left(F_l+F_h\right)^{-1}\left(\frac{1}{2}\right)}\right)$.
\end{theorem}

\begin{proof}
    From the proof of Proposition \ref{prop:cut-offs_ambiguity_in_middle}, recall the sets $X_1 = \left[0, F_h^{-1}\left(\tfrac{1}{2}\right)\right)$, $X_2 = \left[ F_h^{-1}\left(\tfrac{1}{2}\right), F_l^{-1}\left(\tfrac{1}{2}\right) \right]$, and $X_3 = \left( F_l^{-1}\left(\tfrac{1}{2}\right), 1\right]$.
    The very same arguments as in the proof of Theorem \ref{thm:existence_uniqueness_cursed_uncursed_KNE} show that for the uncursed player 2, $\sigma_2^{\left(F_l+F_h\right)^{-1}\left(\frac{1}{2}\right)}$ is a best response against $\sigma_1^{F_l^{-1}\left(\frac{1}{2}\right)}$.
    As $\left(F_l+F_h\right)^{-1}\left(\frac{1}{2}\right) \in X_1$, it follows from the proof of Proposition \ref{prop:cut-offs_ambiguity_in_middle} that player 1's strategy is an ambiguous maxmin cursed best response against player 2's strategy.

    We turn to uniqueness.
    Assume by means of contradiction that $\left(\sigma_1^{\theta_1}, \sigma_2^{\theta_2}\right)$ is another non-trivial ambiguous cursed-uncursed Knight-Nash equilibrium in cut-off strategies.
    It is straightforward to confirm that if $\theta_1>0$, then $\theta_2=0$ does not define a maxmin best response of player 2; and if $\theta_2>0$, then $\theta_1=0$ does not define an ambiguous maxmin cursed best response of player 1.
    Hence, let both $\theta_1$ and $\theta_2$ be positive.
    By the shape of player 2's interim infimal expected utility functions, it follows that $\theta_2 < \theta_1$.
    Therefore, it is not possible that $\theta_2 \in X_3$, as in this case, by the proof of Proposition \ref{prop:cut-offs_ambiguity_in_middle}, player 1's ambiguous cursed maxmin best reply involves a cut-off in $X_2$.
    For any $\theta_2 \in X_1$, player 1's best reply is the cut-off strategy with threshold $F_l^{-1}\left(\tfrac{1}{2}\right)$, so we arrive at the already established equilibrium $\left(\sigma_1^{F_l^{-1}\left(\tfrac{1}{2}\right)}, \sigma_2^{\left(F_l + F_h\right)^{-1}\left(\tfrac{1}{2}\right)}\right)$.
    The remaining case is where $\theta_2 \in X_2$.
    Again by the proof of Proposition \ref{prop:cut-offs_ambiguity_in_middle}, player 1 will not use a cut-off strategy with threshold in $X_3$, so---as $\theta_1 \in X_1$ is precluded by $\theta_2 < \theta_1$---$\theta_1 \in X_2$ as well. 
    In particular, $\theta_1 \leq F_l^{-1}\left(\tfrac{1}{2}\right)$.
    Using the best reply condition of player 2, $F_h(\theta_2)< F_l(\theta_2)+F_h(\theta_2)=F_l(\theta_1)\leq \tfrac{1}{2}$, so $\theta_2 < F_h^{-1}\left(\tfrac{1}{2}\right)$.
    But this is impossible as $\theta_2 \in X_2$.
\end{proof}

\section{Online Appendix II: Partial Cursedness}
\label{app:Partial_Cursedness}

\citet{eyster_rabin_2005} defined a more general equilibrium concept than the one presented in Definition \ref{def:cursed_equilibrium}, allowing players to be \emph{partially cursed}.
This means that their interim perceived expected utility given their types and their opponent's strategies is a convex combination of the fully rational $w_k$ (Equation \eqref{eq:definition_w}) and the fully cursed $v_k$ (Equation \eqref{eq:definition_v}).

\begin{definition} \label{def:partially_cursed_equilibrium}
    Let  $\Gamma = \left( N, \left(A_k\right)_{k\in N}, \left(\Theta_k\right)_{k\in N}, \left(u_k\right)_{k \in N}, \Prob \right)$ be a Bayesian game as specified at the beginning of Section \ref{sec:Several_Equilibrium_Concepts}, and let $\bar\sigma_k$ be the average strategy of player $k$ as defined in \eqref{eq:average_strategy}.
    For any $\chi \in [0,1]$ and strategy profile $\sigma_{-k}$ of $k$'s opponents, $\sigma_k$ is a \emph{$\chi$-cursed best response} against $\sigma_{-k}$ if for each $\theta_k \in \Theta$, and each $a_k^*$ such that $\sigma_k\left(a_k^* \mid \theta_k\right)>0$,
    \small
    \begin{align}
        a_k^* &\in \argmax_{a_k \in a_k} v_k^{\chi}\left(a_k\mid \theta_k, \sigma_{-k}\right), \text{ where} \nonumber \\
        v_k^{\chi}\left(a_k\mid \theta_k, \sigma_{-k}\right) &\colonequals \int_{\Theta_{-k}}
        \sum_{a_{-k} \in A_{-k}} {\left[ \chi \bar{\sigma}_{-k}\left(a_{-k}\mid \theta_k, \Prob\right) + (1-\chi) \sigma_{-k}\left(a_{-k} \mid \theta_{-k}\right)\right] }
        u_i\left(a_k, a_{-k} ; \theta_k, \theta_{-k}\right) \d\Prob\left(\theta_{-k}\mid \theta_k\right)\label{eq:definition_v^chi}
    \end{align}
    \normalsize
    is the \emph{$\chi$-perceived interim expected utility} of action $a_k$, given the own type $\theta_k$ and the opponents' strategy profile $\sigma_{-k}$.
    A strategy profile $\sigma^*$ is a \emph{$\left(\chi_k\right)_{k \in N}$-cursed equilibrium} if for each $k\in N$, $\sigma^*_k$ is a $\chi_k$-cursed best response against $\sigma^*_{-k}$.
\end{definition}

\noindent It directly follows from the definition that if $\chi_k=0$ for all $k\in N$, we arrive at the Bayesian Nash Equilibrium, and if $\chi_k=1$ for all $k$, the fully cursed equilibrium (Definition \ref{def:cursed_equilibrium}).
Moreover,
\begin{align} \label{eq:partially_cursed_convex_combination}
    v_k^{\chi}\left(a_k \mid \theta_k, \sigma_{-k}\right) &= \chi v_k^1\left(a_k \mid \theta_k, \sigma_{-k}\right) + (1-\chi) v_k^0\left(a_k \mid \theta_k, \sigma_{-k}\right) \nonumber \\
    &= \chi v_k\left(a_k \mid \theta_k, \sigma_{-k}\right) + (1-\chi) w_k\left(a_k \mid \theta_k, \sigma_{-k}\right)
\end{align}
with $v_k$ and $w_k$ having been defined in Equations \eqref{eq:definition_v} and \eqref{eq:definition_w}, respectively.

While the degrees of cursedness can be different among players, they are still fixed. 
This contrasts the concept of \emph{$f$-generalized cursed equilibrium} in \citet{eyster_rabin_2005} where the degrees of cursedness as well as the types of players are drawn from some joint probability distribution $f$.

The partially cursed equilibrium has been found difficult to justify in terms of a learning foundation \citep{eyster_rabin_2005}, but nevertheless used to quantify deviation from full rationality in empirical investigations \citep{eyster_rabin_2005, christensen_2008, szembrot_2018}.

In our two-player game, we have classified $\left(\chi_1, \chi_2\right)$-cursed equilibria for the cases $\left(\chi_1, \chi_2\right)=(0,0)$, $\left(\chi_1, \chi_2\right)=(1,1)$, and $\left(\chi_1, \chi_2\right)=(1,0)$ in the main part (Propositions \ref{prop:BNE_is_nt} and \ref{prop:cNE_is_median_cutoff} and Theorem \ref{thm:existence_uniqueness_cursed_uncursed_KNE}).
In this appendix, we shall find partially cursed equilibria for all other combinations of $\chi_1$ and $\chi_2$.
For tractability, we restrict ourselves to the case where there is no uncertainty about the probability distribution, and follow the assumptions made in Section \ref{sec:The_Trading_Game_Without_Uncertainty}:
types are independent and identically drawn according to some probability distribution $F$ on $[0,1]$ that is strictly increasing and continuous.
As before, we abuse notation by writing $\bar\sigma_{-k}\left(a_{-k}\right)$ instead of $\bar\sigma_{-k}\left(a_{-k} \mid \theta_k, \Prob_F\right)$ 

As a first step to finding the equilibria, we show that we can restrict ourselves to cut-off strategies (Definition \ref{def:cut-off_strategy}).

\begin{lemma}\label{lem:best_response_cut_off}
    Let $k \in \{1,2\}$.
    Whenever $\sigma_{-k}$ is such that player $-k$ sometimes trades ($\Bar{\sigma}_{-k}(tr)>0$) and player $k$ is at least partially cursed ($\chi_k>0$), every $\chi_k$-cursed best response against $\sigma_{-k}$ must be a cut-off strategy.
\end{lemma}
\begin{proof}
    We find that the $\chi_k$-cursed expected difference between not trading and trading at type $\theta_k$, given the opponent's strategy $\sigma_{-k}$, is
    \small
    \begin{align*}
        &d_k^{\chi_k} \left( \theta_k, \sigma_{-k} \right) \\
        \colonequals &v_k^{\chi_k}\left(nt \mid \theta_k, \sigma_{-k}\right) - v_k^{\chi_k}\left(tr \mid \theta_k, \sigma_{-k}\right) \\
        = &\chi_k \Bar{\sigma}_{-k} \left(tr\right) \int_0^1 \mathbbm{1}_{\{\theta_k > \theta_{-k}\}} - \mathbbm{1}_{\{\theta_{-k} > \theta_{k}\}} \d F\left(\theta_{-k}\right) + \left(1-\chi_k\right) \int_0^1 \sigma_{-k}\left(tr \mid \theta_{-k} \right) \left(\mathbbm{1}_{\{\theta_k > \theta_{-k}\}} - \mathbbm{1}_{\{\theta_{-k} > \theta_{k}\}} \right) \d F\left(\theta_{-k}\right) \\
        = &\chi_k \Bar{\sigma}_{-k} \left(tr\right) \left( 2F\left(\theta_k\right)-1\right) + \left(1-\chi_k\right) \int_0^1 \sigma_{-k}\left(tr \mid \theta_{-k} \right) \left(\mathbbm{1}_{\{\theta_k > \theta_{-k}\}} - \mathbbm{1}_{\{\theta_{-k} > \theta_{k}\}} \right) \d F \left(\theta_{-k}\right).
    \end{align*}
    \normalsize
    Both summands are increasing in $\theta_k$, and the first one is strictly increasing. 
    Hence, this also holds for the sum, so there must be a unique threshold $\hat\theta$ such that $d_k^{\chi_k} \left( \theta_k, \sigma_{-k} \right)$ is negative for $\theta \in \left[0,\hat\theta\right)$ and positive for $\theta \in \left(\hat\theta, 1\right]$ (Note that one of these intervals might be empty.), which proves the claim.
\end{proof}

\begin{remark}
    If the opposing player $-k$ (almost surely) never trades, every strategy is a $\chi_k$-cursed best response, as the behavior of player $k$ has no effect on the outcome of the game. 
    If $k$ is not cursed ($\chi_k=0$), then $d_k^0 \left( \theta_k, \sigma_{-k} \right) = \int_0^1 \sigma_{-k}(tr \mid \theta_{-k}) \left(\mathbbm{1}_{\{\theta_k > \theta_{-k}\}} - \mathbbm{1}_{\{\theta_{-k} > \theta_{k}\}} \right) \d F\left(\theta_{-k}\right)$ which, depending on $\sigma_{-k}$, can be constantly 0 on some interval. 
    In this case, there is still a cut-off strategy that is a best response, but also various other strategies.
\end{remark}

\noindent Given that the opponent plays a cut-off strategy with threshold $\hat\theta$, it immediately follows from Lemma \ref{lem:v(nt)_and_v(tr)} in combination with Equation \eqref{eq:partially_cursed_convex_combination} that for any $\chi_k \in [0,1]$, player $k$'s $\chi_k$-perceived interim expected utilities from playing $nt$ or $tr$ are
\begin{align}
    v_k^{\chi_k}\left(nt \mid \theta_k, \sigma_{-k}^{\hat\theta} \right) &= F(\theta_k) \text{ and}\label{eq:v(nt)_partial} \\
    v_k^{\chi_k}\left(tr \mid \theta_k, \sigma_{-k}^{\hat\theta} \right) &= (1-\chi_k) \LV F\left(\theta_k\right)-F(\hat\theta) \RV + \chi_k \left[ F\left(\theta_1\right) + F(\hat\theta) - 2F\left(\theta_k\right)F(\hat\theta)\right]. \label{eq:v(tr)_partial}
\end{align}

\noindent We can now explicitly derive $\chi$-cursed best responses in this game.

\begin{lemma} \label{lem:formula_best_response}
    For player $k$, the unique $\chi$-cursed best response against a cut-off strategy with threshold $\hat\theta_{-k}>0$ is a cut-off strategy with threshold $b_{\chi}^*(\hat\theta_{-k})$, where
    \begin{align*}
        b_{\chi}^*\left(\theta\right)&= F^{-1}\left(\tfrac{1}{2}F(\theta)\right) &&\text{for } \chi=0, \\
         b_{\chi}^*\left(\theta\right) &= F^{-1} \left( \frac{F(\theta)}{2\left(1-\chi+\chi F(\theta)\right)} \right) &&\text{for } \chi \in \left(0, \tfrac{1}{2}\right], \text{ and}\\
         b_{\chi}^*\left(\theta\right) &= \begin{cases}
            F^{-1}\left(\frac{2\chi-1}{2\chi}\right) &\text{ for } \theta \leq F^{-1}\left(\frac{2\chi-1}{2\chi}\right), \\
            F^{-1} \left( \frac{F(\theta)}{2\left(1-\chi+\chi F(\theta)\right)} \right) &\text{ for } \theta > F^{-1}\left(\frac{2\chi-1}{2\chi}\right)
            \end{cases}
            &&\text{for }\chi \in \left(\tfrac{1}{2}, 1\right].
    \end{align*}
\end{lemma}

\begin{proof}
    Without loss of generality, let $k=1$ such that $-k=2$. 
    Assume that player 2 plays the cut-off strategy $\sigma_2^{\hat\theta_2}$ with threshold $\hat\theta_2>0$. 
    Then, for player 1, 
    \begin{align}
        &d_1^{\chi}\left(\theta_1, \sigma_2^{\hat\theta_2}\right) \nonumber\\
        = &v_1^{\chi}\left(nt \mid \theta_1, \sigma_2^{\hat\theta_2} \right) - v_1^{\chi}\left(tr \mid \theta_1, \sigma_2^{\hat\theta_2} \right) \\
        = &\chi F(\hat\theta_2) \left(2F\left(\theta_1\right) -1 \right) + (1-\chi) \int_0^{\hat\theta_2} \mathbbm{1}_{\{\theta_1>\theta_2\}} - \mathbbm{1}_{\{\theta_1<\theta_2\}} \d F\left(\theta_2\right) \nonumber\\
        = &\begin{cases}
        \chi F(\hat\theta_2) \left(2F\left(\theta_1\right) -1 \right) + (1-\chi) \left[ 2F\left(\theta_1\right)-F(\hat\theta_2) \right] &\text{ if } \theta_1< \hat\theta_2,\\
        \chi F(\hat\theta_2) \left(2F\left(\theta_1\right) -1 \right) + (1-\chi)F(\hat\theta_2) &\text{ if } \theta_1\geq\hat\theta_2.            
        \end{cases} \label{eq:u(nt)-u(tr)_formula_best_response}
    \end{align}
    For $\chi=0$, this expression is positive if and only if $2F\left(\theta_1\right)-F(\hat\theta_2)>0$, that is, if and only if $\theta_1> F^{-1}\left(\frac{1}{2}F(\hat\theta_2)\right)=b_{0}^*(\hat\theta_2)$.

    Assume now that $\chi> \frac{1}{2}$. 
    As $d_1^{\chi}\left(\theta_1, \sigma_2^{\hat\theta_2}\right)$ is negative for $\theta_1=0$, positive for $\theta_1=1$, and strictly increasing in $\theta_1$, there is a unique threshold $b_{\chi}^*(\hat\theta_2)$ such that $d_1^{\chi}\left(\theta_1, \sigma_2^{\hat\theta_2}\right)$ is positive if and only if $\theta_1 > b_{\chi}^*(\hat\theta_2)$.
    It is immediate to verify that the first part of \eqref{eq:u(nt)-u(tr)_formula_best_response} is positive if and only if $\theta_1 > F^{-1} \left( \frac{F(\hat\theta_2)}{2\left(1-\chi+\chi F(\hat\theta_2)\right)} \right)$, while the second part is positive if and only if $\theta_1 > F^{-1}\left(\frac{2\chi-1}{2\chi}\right)$.
    As $d_1^{\chi}\left(\theta_1, \sigma_2^{\hat\theta_2}\right)$ is continuous in $\theta_1$, it depends on the sign of $d_1^{\chi}\left(\hat\theta_2, \sigma_2^{\hat\theta_2}\right)$ which of these two values equals $b_{\chi}^*(\hat\theta_2)$.
    If $\hat\theta_2 \leq F^{-1}\left(\frac{2\chi-1}{2\chi}\right)$, it holds that $d_1^{\chi}\left(\hat\theta_2, \sigma_2^{\hat\theta_2}\right)\leq 0$ such that $b_{\chi}^*(\hat\theta_2)$ must be greater than or equal $\hat\theta_2$; hence, $b_{\chi}^*(\hat\theta_2)=F^{-1}\left(\frac{2\chi-1}{2\chi}\right)$.
    Similarly, $\hat\theta_2 > F^{-1}\left(\frac{2\chi-1}{2\chi}\right)$ implies that $d_1^{\chi}\left(\hat\theta_2, \sigma_2^{\hat\theta_2}\right)> 0$, so $\hat\theta_2$ is smaller than $b_{\chi}^*(\hat\theta_2)$ which therefore equals $F^{-1} \left( \frac{F(\hat\theta_2)}{2\left(1-\chi+\chi F(\hat\theta_2)\right)} \right)$.

    Finally, for $\chi \in \left(0, \frac{1}{2}\right]$, the second part of \eqref{eq:u(nt)-u(tr)_formula_best_response} is always nonnegative. 
    Therefore, the threshold for which the utilities from trading and not trading are equal must be at $F^{-1} \left( \frac{F(\hat\theta_2)}{2\left(1-\chi+\chi F(\hat\theta_2)\right)} \right)=b_{\chi}^*(\hat\theta_2)$.
\end{proof}

\begin{corollary} \label{cor:formula_best_response_monotonicity_properties}
    For every $\chi \in [0,1]$, $b^*_{\chi}$ is increasing on $(0,1]$.
    For $\chi \in \left[0, \frac{1}{2}\right]$, the function is strictly increasing on $[0,1]$, while for $\chi \in \left[\frac{1}{2},1\right)$, it is so for $\theta\geq F^{-1} \left( \frac{2\chi-1}{2\chi} \right)$.
    Moreover, for any fixed $\theta \in (0,1)$, $\chi \mapsto b_{\chi}^*\left(\theta\right)$ is strictly increasing on $[0,1]$.
\end{corollary}

\begin{proof}
    Using the definition of $b^*_{\chi}$, one can take the derivative of $F\left(b^*_{\chi}\left(\theta\right) \right)$ with respect to $F(\theta)$, and use that $F$ and $F^{-1}$ are strictly increasing.
\end{proof}

 \begin{figure}
    \centering
    \begin{tikzpicture}[scale=8]
        \draw[->] (0,0) -- (1.1,0) node[right] {$\hat\theta_{-k}$};
        \draw[->] (0,0) -- (0,0.55);
        \draw[scale=1,domain=0:1,variable=\t, black, dotted]  plot ({\t},{0.5*\t}); 
        \draw[scale=1,domain=0:1,variable=\t, black, dashed]  plot ({\t},{(\t)/(2*(1-0.4+0.4*\t)});
        \draw[scale=1,domain=0:0.1667,variable=\t, black, dashdotted]  plot ({\t},{1/6});
        \draw[scale=1,domain=0.1667:1,variable=\t, black, dashdotted]  plot ({\t},{(\t)/(2*(1-0.6+0.6*\t)});
        \draw[scale=1,domain=0:0.375,variable=\t, black]  plot ({\t},{3/8});
        \draw[scale=1,domain=0.375:1,variable=\t, black]  plot ({\t},{(\t)/(2*(1-0.8+0.8*\t)});
        \draw[dotted] (1,0.5) -- (0,0.5) node[left] {$\tfrac{1}{2}$};
        \draw[dotted] (1,0.5) -- (1,0) node[below] {$1$};
    \end{tikzpicture}
    \caption{Thresholds for $\chi$-cursed best responses for $\chi=0$ (dotted), $\chi=0.4$ (dashed), $\chi=0.6$ (dashdotted), and $\chi=0.8$ (solid)}
    \label{fig:different_chi-cursed_best_responses}
\end{figure}
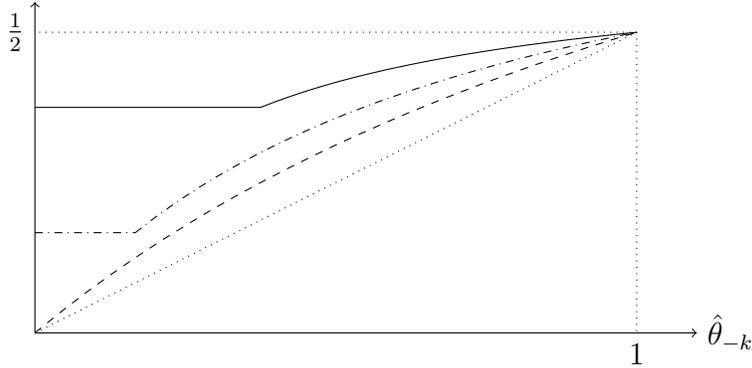

\noindent The monotonicity properties from Corollary \ref{cor:formula_best_response_monotonicity_properties} can be seen in Figure \ref{fig:different_chi-cursed_best_responses} which depicts the functions $b_{\chi}^*$ for $\chi \in \{0, 0.4, 0.6, 0.8\}$, for types being uniformly distributed on $[0,1]$.
In particular, it follows from Lemma \ref{lem:formula_best_response} that partially cursed players ($\chi \in (0,1)$) are always willing to trade more than uncursed players ($\chi=0$).

A fully cursed player ($\chi=1$) does not at all take into account that their opponent actually conditions their behavior on the private type $\theta_{-k}$, and the best response of such a player (i.e., the 1-cursed best response) is independent of the opponent's strategy.\footnote{
Interestingly, this ceases to hold if there is uncertainty about the opponent's type distribution, as in this case, the cursed best response might depend on the opponent's strategy (cf. Proposition \ref{prop:cut-offs_ambiguity_in_middle_under_F*}).
The reason is that in this case, the type distribution that minimizes the interim perceived expected utility is not constant over the opponent's strategies.
}
Such a player offers to trade if and only if their type lies below $F_l^{-1}\left(\tfrac{1}{2}\right)$, for any strategy of the opponent.

Interpreting the degree of cursedness as a measure for the deviation from full rationality, we see that the less rational (or the more cursed) an agent is, the further also their behavior differs from what a rational agent does.
It is remarkable that for sufficiently, but not fully cursed agents ($\chi \in \left(\frac{1}{2},1\right)$), there is an area in which the $\chi$-cursed best response is independent of the cut-off strategy that the opponent plays.
Namely, as long as the opponent plays a cut-off strategy with cut-off below $F^{-1}\left(\frac{2\chi-1}{2\chi}\right)$, a $\chi$-cursed player will respond with a cut-off strategy with cut-off $F^{-1}\left(\frac{2\chi-1}{2\chi}\right)$, no matter the exact value of the opponent's cut-off. 
For cut-offs above $F^{-1}\left(\frac{2\chi-1}{2\chi}\right)$, the $\chi$-cursed best response function $b^*_{\chi}$ strictly increases until 1.
Thus, such a sufficiently cursed agent shows characteristics of both fully cursed (constant best response) and uncursed (strictly increasing best response) players.
In contrast to that, the cut-off of the $\chi$-cursed best response function of a player that is only slightly cursed---i.e., $\chi \in \left(0,\frac{1}{2}\right)$---is strictly increasing on $[0,1]$ in the opponent's cut-off, more resembling the behavior of an uncursed agent.

We are now looking for cursed equilibria in this game. 
For every $\chi_1, \chi_2 \in [0,1]$, it is always a $\left(\chi_1,\chi_2\right)$-cursed equilibrium if both players almost surely never offer to trade.
As before, we call these equilibria \emph{trivial} and look for \emph{non-trivial} cursed equilibria in the subsequent analysis.
In Theorem \ref{thm:cursedness_increases_perceived_gain}, we will see that whenever a non-trivial $\left(\chi_1,\chi_2\right)$-cursed equilibrium exists, both players will strictly prefer that equilibrium to the trivial ones (cf. Remark \ref{rmk:cursed_player_prefers_non-trivial_equilibrium}).

\begin{theorem} \label{thm:cursed_equilibria_different_degrees}
    If $\max\left\{\chi_1,\chi_2\right\} \leq \frac{1}{2}$, there is no non-trivial $\left(\chi_1,\chi_2\right)$-cursed equilibrium.
    If $\max\left\{\chi_1,\chi_2\right\} > \frac{1}{2}$, there is a unique non-trivial $\left(\chi_1,\chi_2\right)$-cursed equilibrium in which the more cursed player plays a cut-off strategy with threshold $F^{-1}\left(\frac{2\max\left\{\chi_1,\chi_2\right\}-1}{2\max\left\{\chi_1,\chi_2\right\}}\right)$, and the less cursed player plays a cut-off strategy with threshold $F^{-1}\left(\frac{2\max\left\{\chi_1,\chi_2\right\}-1}{2\left(2\max\left\{\chi_1,\chi_2\right\}-\min\left\{\chi_1,\chi_2\right\}\right)}\right)$.
\end{theorem}

\begin{proof}
    The case where $\chi_1=\chi_2=0$ has been analyzed in Proposition \ref{prop:BNE_is_nt}.
    We now assume that $\max\{\chi_1,\chi_2\}>0$ and that at least one player sometimes offers to trade, without loss of generality, player 2.
    In this case, $d_1^{\chi_1}(0, \sigma_2)<0$ for every $\chi_1\in[0,1]$.
    As $d_1^{\chi_1}(\cdot, \sigma_2)$ is continuous, any $\chi_1$-cursed best response requires that also $\bar\sigma_1(tr)>0$.
    In a non-trivial $\left(\chi_1,\chi_2\right)$-cursed equilibrium, every player who is at least partially cursed therefore plays a cut-off strategy by Lemma \ref{lem:best_response_cut_off}.
    By Lemma \ref{lem:formula_best_response}, even an uncursed player will react to a cut-off strategy by using a cut-off strategy.
    Hence, in the following analysis, we can restrict ourselves to strategy profiles in which both players play cut-off strategies.
    Let us denote the respective thresholds by $\theta_1^*$ and $\theta_2^*$.
    For $b_{\chi}^*$ as defined in Lemma \ref{lem:formula_best_response}, let $\Tilde{b}_{\chi}$ be given by $\tilde{b}_{\chi}\left(\theta\right) \colonequals F \left( b_{\chi}^* \left( F^{-1}(\theta) \right)\right)$.
    By definition,
    \begin{align*}
        \tilde{b}_{\chi}\left(\theta\right)&=\frac{1}{2}\theta &&\text{for } \chi=0, \\
        \tilde{b}_{\chi}\left(\theta\right) &= \frac{\theta}{2\left(1-\chi+\chi \theta\right)} &&\text{for } \chi \in \left(0, \tfrac{1}{2}\right], \text{ and} \\
        \tilde{b}_{\chi}\left(\theta\right) &= \begin{cases}
            \frac{2\chi-1}{2\chi} &\text{ for } \theta \leq \frac{2\chi-1}{2\chi}, \\
            \frac{\theta}{2\left(1-\chi+\chi \theta\right)} &\text{ for } \theta >\frac{2\chi-1}{2\chi}
        \end{cases}
        &&\text{for } \chi \in \left(0, \tfrac{1}{2}\right].
    \end{align*}
    By Lemma \ref{lem:formula_best_response}, the thresholds must satisfy $\theta_1^*=b^*_{\chi_1}\left(\theta_2^*\right)$ and $\theta_2^*=b^*_{\chi_2}\left(\theta_1^*\right)$ for $\left(\sigma_1^{\theta_1*}, \sigma_2^{\theta_2^*}\right)$ to be a $\left(\chi_1, \chi_2\right)$-cursed equilibrium.
    Then, $b_{\chi_1}^*\left(b_{\chi_2}^*\left(\theta_1^*\right)\right)=\theta_1^*$ which in turn implies that $F\left(\theta_1^*\right) =\Tilde{b}_{\chi_1}\left(\Tilde{b}_{\chi_2}\left(F\left(\theta_1^*\right)\right)\right)$.
    Hence, as a necessary condition for a $\left(\chi_1,\chi_2\right)$-cursed equilibrium, we need to find a fixed point of $\Tilde{b}_{\chi_1}\circ\Tilde{b}_{\chi_2}$ (or equivalently, as $\Tilde{b}_{\chi_1}$ and $\Tilde{b}_{\chi_2}$ commute for every choice of $\chi_1$ and $\chi_2$, a fixed point of $\Tilde{b}_{\chi_2}\circ\Tilde{b}_{\chi_1}$).
    Having found such a fixed point $\theta$ that satisfies $\theta=\tilde{b}_{\chi_1}\left(\tilde{b}_{\chi_2}\left(\theta\right)\right)$, we can set $\theta_1^*\colonequals F^{-1}(\theta)$ and $\theta_2^*\colonequals F^{-1}\left(\Tilde{b}_{\chi_2}(\theta)\right)$.
    Then, $b_{\chi_2}^*\left(\theta_1^*\right)=b_{\chi_2}^*\left(F^{-1}(\theta)\right)=F^{-1}\left(\Tilde{b}_{\chi_2}(\theta)\right)=\theta_2^*$ and $\theta_1^*=b_{\chi_1}^*\left(b_{\chi_2}^*\left(\theta_1^*\right)\right)=b_{\chi_1}^*\left(\theta_2^*\right)$ such that finding a fixed point of $\Tilde{b}_{\chi_1}\circ\Tilde{b}_{\chi_2}$ is not only necessary, but also sufficient in order to find a $\left(\chi_1, \chi_2\right)$-cursed equilibrium.
    
    We first show that for $\chi_1$ and $\chi_2$ being no more than $\frac{1}{2}$, $\Tilde{b}_{\chi_1}\circ\Tilde{b}_{\chi_2}$ does not have a strictly positive fixed point.
    This is obvious for $\chi_1=\chi_2=0$ as then, $(\Tilde{b}_{\chi_1}\circ\Tilde{b}_{\chi_2})\left(\theta\right)=\frac{1}{4}\theta$ for all $\theta$. 
    If $\max\left\{\chi_1,\chi_2\right\}\in \left(0,\frac{1}{2}\right]$, taking derivatives reveals that both $\tilde{b}_{\chi_1}$ and $\tilde{b}_{\chi_2}$ are strictly increasing and concave, while at least one is strictly concave.
    Hence, also $\Tilde{b}_{\chi_1}\circ\Tilde{b}_{\chi_2}$ is strictly increasing and strictly concave.
    Moreover, by a straightforward calculation,
    \begin{align*}
        (\Tilde{b}_{\chi_1}\circ\Tilde{b}_{\chi_2})'\left(0\right) = \Tilde{b}_{\chi_1}'(0)\Tilde{b}_{\chi_2}'(0) = \frac{1}{4\left(1-\chi_1\right)\left(1-\chi_2\right)} \leq 1 \text{ for all } \chi_1, \chi_2 \leq \tfrac{1}{2}.
    \end{align*}
    As $(\Tilde{b}_{\chi_1}\circ\Tilde{b}_{\chi_2})\left(0\right)=0$ and $(\Tilde{b}_{\chi_1}\circ\Tilde{b}_{\chi_2})'$ is strictly decreasing, there cannot be a $\theta>0$ with $(\Tilde{b}_{\chi_1}\circ\Tilde{b}_{\chi_2})\left(\theta\right)=\theta$.
    
    Now, let $\max\left\{\chi_1,\chi_2\right\} > \frac{1}{2}$, and, without loss of generality, $\chi_1 \leq \chi_2$.
    Assume first that $\chi_1>\frac{1}{2}$.
    Then, as $\Tilde{b}_{\chi_2}(\theta)\geq \frac{2\chi_2-1}{2\chi_2}\geq\frac{2\chi_1-1}{2\chi_1}$ for all $\theta\in [0,1]$,
    \begin{align}
        \tilde{b}_{\chi_1}\left(\Tilde{b}_{\chi_2}(\theta)\right) &= \begin{cases}
            \frac{2\chi_1-1}{2\chi_1}\qquad &\text{ for } \Tilde{b}_{\chi_2}(\theta) < \frac{2\chi_1-1}{2\chi_1}, \\
            \frac{\Tilde{b}_{\chi_2}(\theta)}{2\left(1-\chi_1+\chi_1 \Tilde{b}_{\chi_2}(\theta)\right)} \qquad \; &\text{ for } \Tilde{b}_{\chi_2}(\theta) \geq \frac{2\chi_1-1}{2\chi_1}
        \end{cases} \nonumber\\
        &= \frac{\Tilde{b}_{\chi_2}(\theta)}{2\left(1-\chi_1+\chi_1 \Tilde{b}_{\chi_2}(\theta)\right)}\quad \text{ for all } \theta \in [0,1] \label{eq:tilde{b}_{chi_1}(tilde{b}_{chi_2}}.
    \end{align}
    For $\chi_1 \in \left[0, \frac{1}{2}\right]$, \eqref{eq:tilde{b}_{chi_1}(tilde{b}_{chi_2}} holds by definition such that this identity is true whenever $\chi_1\leq\chi_2$ and $\chi_2>\frac{1}{2}$.
    Thus, in this case,
    \begin{align*}
        \tilde{b}_{\chi_1}\left(\tilde{b}_{\chi_2}(\theta)\right) &= \frac{\tilde{b}_{\chi_2}(\theta)}{2\left(1-\chi_1+\chi_1 \tilde{b}_{\chi_2}(\theta)\right)} \qquad \quad\;\;\;\text{for all } \theta \in [0,1] \\
        &=
        \begin{cases}
            \frac{\frac{2\chi_2-1}{2\chi_2}}{2\left(1-\chi_1+\chi_1\frac{2\chi_2-1}{2\chi_2}\right)} = \frac{2\chi_2-1}{2\left(2\chi_2-\chi_1\right)} &\text{ for } \theta \leq \frac{2\chi_2-1}{2\chi_2},\\
            \frac{\frac{\theta}{2\left(1-\chi_2+\chi_2 \theta\right)}}{2\left(1-\chi_1+\chi_1\frac{\theta}{2\left(1-\chi_2+\chi_2 \theta\right)}\right)} &\text{ for } \theta > \frac{2\chi_2-1}{2\chi_2}.
        \end{cases}        
    \end{align*}
    As $\frac{2\chi_2-1}{2\left(2\chi_2-\chi_1\right)} \leq \frac{2\chi_2-1}{2\chi_2}$ and as $\Tilde{b}_{\chi_1}\circ\Tilde{b}_{\chi_2}$ is, for $\theta>\frac{2\chi_2-1}{2\chi_2}$, strictly concave with derivative less than 1 (This can be proven along the same lines as above.), the unique positive fixed point of $\tilde{b}_{\chi_1}\circ\tilde{b}_{\chi_2}$ is given by $\theta=\frac{2\chi_2-1}{2\left(2\chi_2-\chi_1\right)}$.
    As $\Tilde{b}_{\chi_2}\left(\frac{2\chi_2-1}{2\left(2\chi_2-\chi_1\right)}\right) = \frac{2\chi_2-1}{2\chi_2}$, we have indeed shown that the unique positive non-trivial $\left(\chi_1,\chi_2\right)$-cursed equilibrium is given by the cut-off strategies with thresholds $F^{-1}\left(\frac{2\chi_2-1}{2\left(2\chi_2-\chi_1\right)}\right)$ and $F^{-1}\left(\frac{2\chi_2-1}{2\chi_2}\right)$, as claimed.
\end{proof}

\noindent Whether there exists a non-trivial $\left(\chi_1,\chi_2\right)$-cursed equilibrium or not depends only on the degree of cursedness of the more cursed player. 
This player conditions their strategy only on the own degree of cursedness while it is independent of the cursedness of the other player.
In contrast to that, the equilibrium strategy of the less cursed player not only depends on their own degree of cursedness, but also on the degree of the opponent. 
This could be interpreted in such a way that the less cursed player is ``more rational'' and takes the characteristics of the opponent more into account.
As can be seen from the formulas, the thresholds are increasing in the own degree of cursedness, while the threshold of the less cursed player also increases in the cursedness of the opponent.
Interpreting the degree of cursedness as a measure for deviation from rationality, one might say that the amount of trade in the cursed equilibrium is monotone in the rationality of the players:
the less rational (that is, the more cursed) they are, the more they engage in trade.

Inspired by Theorem \ref{thm:cursed_equilibria_different_degrees}, we define, for $k \in \{1,2\}$,
\begin{align} \label{eq:theta^*}
    \theta_k^*\left(\chi_1, \chi_2\right) \colonequals
    \begin{cases}
        0 &\text{ if } \max\{\chi_1,\chi_2\}\leq \frac{1}{2}, \\
        F^{-1}\left(\frac{2\chi_k-1}{2\chi_k}\right) &\text{ if } \max\{\chi_1,\chi_2\}> \frac{1}{2} \text{ and } \chi_k\geq\chi_{-k}, \\
        F^{-1}\left(\frac{2\chi_{-k}-1}{2\left(2\chi_{-k}-\chi_k\right)}\right) &\text{ if } \max\{\chi_1,\chi_2\}> \frac{1}{2} \text{ and } \chi_k<\chi_{-k}.
    \end{cases}
\end{align}
For each $\chi_1, \chi_2 \in [0,1]$, there is a $\left(\chi_1,\chi_2\right)$-cursed equilibrium in which player $k$ plays the cut-off strategy $\sigma_k^{\theta_k^*\left(\chi_1, \chi_2\right)}$ (and, for $\max\{\chi_1,\chi_2\}> \frac{1}{2}$, another one in which both players impose the strategy $\sigma_k^0$).
It is immediate from the definition that the threshold of the more cursed player is greater than that of the less cursed one.

\subsection{Utility Analysis}

We shall now look at the impacts of these $\left(\chi_1,\chi_2\right)$-cursed equilibria on players' utilities and perform comparative statics on the expected gains and losses of both players.
Following Theorem \ref{thm:cursed_equilibria_different_degrees}, players 1 and 2 play cut-off strategies $\sigma_1^{\theta_1^*\left(\chi_1,\chi_2\right)}$ and $\sigma_2^{\theta_2^*\left(\chi_1,\chi_2\right)}$, with $\theta_k^*\left(\chi_1, \chi_2\right)$ as defined in \eqref{eq:theta^*}.

\begin{definition} \label{def:ex-ante_expected_gains}
    In the $\left(\chi_1,\chi_2\right)$-cursed equilibrium, the \emph{ex-ante expected utility} of player $k$, for $k \in \{1,2\}$, is given by
    \begin{align} \label{eq:definition_G}
        U_k\left(\chi_1,\chi_2\right) &\colonequals \mathbb{E}_{\theta_k} \left[ v_k^0\left(\sigma_{k}^{\theta_{k}^*\left(\chi_1,\chi_2\right)}\left(\theta_k\right) \mid \theta_k, \sigma_{-k}^{\theta_{-k}^*\left(\chi_1,\chi_2\right)} \right) \right] \\
        &= \mathbb{E}_{\theta_1,\theta_2}\left[ u_k\left(\sigma_1^{\theta_1^*\left(\chi_1,\chi_2\right)}\left(\theta_1\right),\sigma_2^{\theta_2^*\left(\chi_1,\chi_2\right)}\left(\theta_2\right); \theta_1,\theta_2\right) \right]. \nonumber
        \qedhere
    \end{align}
\end{definition}

\noindent This definition describes the ``real'' expected utilities of the players, as it could be measured by an outstanding (uncursed) observer, and not their cursed perceptions of utilities.
The latter will be investigated later.

\begin{theorem} \label{thm:better_less_cursed}
    Let $\max\left\{\chi_1,\chi_2\right\}>\frac{1}{2}$ such that there is a non-trivial $\left(\chi_1,\chi_2\right)$-cursed equilibrium.
    In this equilibrium, the less cursed player has a higher ex-ante expected utility than the more cursed one.
    Moreover, the ex-ante expected utility of both players is decreasing in their own cursedness, and increasing in the cursedness of the opponent.
\end{theorem}

\begin{proof}
    As $u_1\left(a_1,a_2;\theta_1,\theta_2\right)+u_1\left(a_1,a_2;\theta_1,\theta_2\right)=1$ for all $a_1, a_2, \theta_1, \theta_2$, it suffices to consider only player 1.
    First, we assume that $\chi_1\leq\chi_2$.
    Then, in the non-trivial $\left(\chi_1,\chi_2\right)$-cursed equilibrium, player 1 plays a cut-off strategy with threshold $\theta_1^*\colonequals \theta_1^*\left(\chi_1,\chi_2\right)= F^{-1}\left(\frac{2\chi_2-1}{2\left(2\chi_2-\chi_1\right)}\right)$, and player 2 with $\theta_2^*\colonequals \theta_2^*\left(\chi_1,\chi_2\right)= F^{-1}\left(\frac{2\chi_2-1}{2\chi_2}\right)$.
    Following Equations \eqref{eq:v(nt)} and \eqref{eq:w(tr)}, player 1's interim expected utilities are
    \begin{align*}
        v_1^0\left(\sigma_{1}^{\theta_{1}^*}\left(\theta_1\right) \mid \theta_1, \sigma_{2}^{\theta_{2}^*} \right) = 
        \begin{cases}
            F\left(\theta_2^*\right) - F\left(\theta_1\right) &\text{ if } \theta_1 \leq \theta_1^*, \\
            F\left(\theta_1\right) &\text{ if } \theta_1 > \theta_1^*.
        \end{cases}
    \end{align*}
    Hence, player 1's ex-ante expected utility, given both players' strategies, is\footnote{
    In the third equality, we use the transformation formula for Lebesgue-Stieltjes integrals, see for example Corollary 3 in \citet{winter_1997}.
    }
    \begin{align}
        U_1\left(\chi_1,\chi_2\right) &= \mathbbm{E}_{\theta_1} \left[ v_1^0\left(\sigma_{1}^{\theta_{1}^*}\left(\theta_1\right) \mid \theta_1, \sigma_{2}^{\theta_{2}^*} \right) \right] \nonumber \\
        &= \int_0^{\theta_1^*} F\left(\theta_2^*\right)-F\left(\theta_1\right) \dF\left(\theta_1\right) + \int_{\theta_1^*}^1F\left(\theta_1\right)\dF\left(\theta_1\right) \nonumber\\
        &= F\left(\theta_2^*\right)\int_0^{F\left(\theta_1^*\right)} \d\theta_1 - \int_0^{F\left(\theta_1^*\right)} \theta_1 \d\theta_1 + \int_{F\left(\theta_1^*\right)}^1 \theta_1 \d\theta_1\nonumber\\
        &= F\left(\theta_1^*\right)F\left(\theta_2^*\right)-\tfrac{1}{2}F\left(\theta_1^*\right)^2+\tfrac{1}{2}\left(1-F\left(\theta_1^*\right)^2\right)\nonumber\\
        &= F\left(\theta_1^*\right) \left[F\left(\theta_2^*\right)-F\left(\theta_1^*\right) \right] +\tfrac{1}{2} \label{eq:ex_ante_expected_gain_differently_cursed}
    \end{align}
    which is greater than $\frac{1}{2}$ as $\theta_2^*$ exceeds $\theta_1^*$.
    Therefore, the ex-ante expected utility is higher for player 1 than for player 2, so the less cursed player profits.

    We now want to show that $U_1$ is decreasing in $\chi_1$ and increasing in $\chi_2$ for all $\chi_1$ and $\chi_2$ that satisfy $\max\{\chi_1, \chi_2\}>\frac{1}{2}$.
    By \eqref{eq:ex_ante_expected_gain_differently_cursed} (and the analogous version where $\chi_1$ and $\chi_2$ are permuted), we obtain that $U_1\left(\chi_1,\chi_2\right) > \frac{1}{2}$ for $\chi_1<\chi_2$, $U_1\left(\chi_1,\chi_2\right) = \frac{1}{2}$ for $\chi_1=\chi_2$, and $U_1\left(\chi_1,\chi_2\right) < \frac{1}{2}$ for $\chi_1>\chi_2$.
    Hence, the less cursed player's ex-ante expected utility is indeed higher than the one of the more cursed player.

    We turn to the monotonicity properties.
    As $U_1\left(\chi_1,\chi_2\right)+U_2\left(\chi_1,\chi_2\right)=1$ for all $\chi_1$ and $\chi_2$, we might assume without loss of generality that $\chi_1<\chi_2$.
    In this case, using the definitions of $\theta_1^*$ and $\theta_2^*$, we find that 
    \begin{align*}
        U_1\left(\chi_1,\chi_2\right) &=\frac{2\chi_2-1}{2\left(2\chi_2-\chi_1\right)} \left[\frac{2\chi_2-1}{2\chi_2} -\frac{2\chi_2-1}{2\left(2\chi_2-\chi_1\right)}\right] + \frac{1}{2} \\
        &= \frac{-\chi_1 + \chi_2 + 2 \chi_1 (2 + \chi_1) \chi_2 - 4 (1 + 3 \chi_1) \chi_2^2 + 12 \chi_2^3}{4 \left(\chi_1 - 2 \chi_2\right)^2 \chi_2}
    \end{align*}
    where the last equation follows from straightforward calculations.
    Taking derivatives (and again omitting details of the calculation),
    \begin{align*}
        \frac{\partial}{\partial \chi_1}U_1\left(\chi_1,\chi_2\right)= \frac{\chi_1 (1 - 2 \chi_2)^2}{4 (\chi_1 - 2\chi_2)^3 \chi_2} <0.
    \end{align*}
    
    \noindent It remains to show that the ex-ante expected utility is increasing in the opponent's cursedness.
    To this end, we calculate
    \begin{align*}
        \frac{\partial}{\partial \chi_2}U_1\left(\chi_1,\chi_2\right)&= \frac{(2 \chi_2-1) \left(-6 \chi_1 \chi_2 + 4 \chi_2^2 + \chi_1^2 (1 + 2 \chi_2)\right)}{4 (2 \chi_2-\chi_1)^3 \chi_2^2} \\
        &= \frac{(2\chi_2-1)\chi_2^2}{4 (2 \chi_2-\chi_1)^3 \chi_2^2} \left(-6\frac{\chi_1}{\chi_2}+4+(1+2\chi_2)\left(\frac{\chi_1}{\chi_2}\right)^2\right) \\
        &\geq \frac{(2\chi_2-1)\chi_2^2}{4 (2 \chi_2-\chi_1)^3 \chi_2^2} \left(-6\frac{\chi_1}{\chi_2}+4+2\left(\frac{\chi_1}{\chi_2}\right)^2\right),
    \end{align*}
    where we used that $\chi_2>\frac{1}{2}$.
    For $\chi_1<\chi_2$, this expression is positive as $-6x+4+2x^2>0$ for all $x \in [0,1)$.
\end{proof}

\noindent Theorem \ref{thm:better_less_cursed} shows that from an objective outstanding point of view, a player is better off, the less cursed they are, and the more cursed the opponent is.
When viewing cursedness as a measure of deviation from standard rationality, it is therefore---by definition of ``rationality''---optimal to be fully rational, and utility is monotonously increasing in the own rationality, and decreasing in the opponent's one.
Still, the theorem also shows that players can, in terms of objective expected utility, profit from cursedness, namely if their opponent is even more cursed then they are.

\begin{corollary} \label{cor:less_cursed_better_than_BNE}
    Let $\max\left\{\chi_1,\chi_2\right\}>\frac{1}{2}$ such that there is a non-trivial $\left(\chi_1,\chi_2\right)$-cursed equilibrium.
    Moreover, let $\chi_1 \neq \chi_2$, without loss of generality $\chi_1 < \chi_2$.
    Then, the ex-ante expected utility of the less cursed player 1 is higher in the non-trivial $\left(\chi_1,\chi_2\right)$-cursed equilibrium than in the Bayesian Nash Equilibrium.
    That is,
    \begin{align*}
        U_1 (\chi_1, \chi_2) > U_1 (0, 0).
    \end{align*}
\end{corollary}

\begin{proof}
    By Theorem \ref{thm:better_less_cursed}, $U_1 (\chi_1, \chi_2) > U_2 (\chi_1, \chi_2)$.
    As $U_1 (\chi_1, \chi_2) + U_2 (\chi_1, \chi_2)=1$, it follows that $U_1 (\chi_1, \chi_2) > \frac{1}{2} = U_1(0,0)$.
\end{proof}

\noindent The objective ex-ante expected utility $U_k(\chi_1, \chi_2)$ is of course not what the cursed players themselves try to maximize or even recognize as a relevant measure.
Being cursed, they consider their actions ($tr$ or $nt$) to be optimal, given their type $\theta_k$, as they do not maximize their expectation over $v_k^0$, but instead maximize $v_k^{\chi_k}$ as defined in \eqref{eq:definition_v^chi}.
The following definition takes this into account.
\begin{definition} \label{def:perceived_ex-ante_expected_gain}
    In the $(\chi_1,\chi_2)$-cursed equilibrium, the \emph{$\chi_k$-perceived ex-ante expected utility} of player $k$ is given by
\begin{align} \label{eq:definition_C}
    V_k\left(\chi_1,\chi_2\right) \colonequals \mathbb{E}_{\theta_k} \left[ v_k^{\chi_k}\left( \sigma_{k}^{\theta_{k}^*\left(\chi_1,\chi_2\right)}\left(\theta_k\right) \mid \theta_k, \sigma_{-k}^{\theta_{-k}^*\left(\chi_1,\chi_2\right)} \right) \right].
\end{align}
\end{definition} 

\noindent We can show that the $\chi_k$-perceived ex-ante expected utility $V_k(\chi_1, \chi_2)$ exceeds both the actual ex-ante expected utility $U_k(\chi_1, \chi_2)$ as well as $\tfrac{1}{2}$, which is the ex-ante expected utility in the uncursed Bayesian Nash equilibrium.

\begin{theorem} \label{thm:cursedness_increases_perceived_gain}
    Let $\max\left\{\chi_1,\chi_2\right\}>\frac{1}{2}$ such that there is a non-trivial $\left(\chi_1,\chi_2\right)$-cursed equilibrium.
    Then, 
    \begin{align*}
        V_k(\chi_1, \chi_2) > \max \left\{U_k(\chi_1, \chi_2), \tfrac{1}{2} \right\}.
    \end{align*}
    Moreover, $V_k(\chi_1,\chi_2)$ is increasing in the player's cursedness $\chi_k$.
\end{theorem}

\begin{proof}
    As before, we only consider the utilities of player 1 and write $\theta_k^*$ for $\theta^*_k\left(\chi_1,\chi_2\right)$, for $k \in \{1,2\}$.
    By Equations \eqref{eq:v(nt)_partial} and \eqref{eq:v(tr)_partial},
    \begin{align*}
        v_1^{\chi_1}\left( \sigma_{1}^{\theta_{1}^*}\left(\theta_1\right) \mid \theta_1,\sigma_{2}^{\theta_{2}^*} \right) =
        \begin{cases}
            \chi_1 \left[\Ftt \left(1-F(\theta_1)\right)+\left(1-\Ftt \right)F(\theta_1)\right] & \\
            \qquad \qquad \qquad + (1-\chi_1) \LV F(\theta_2^*)-F(\theta_1) \RV &\text{ for } \theta_1 \leq \theta_1^*, \\
            F(\theta_1) &\text{ for } \theta_1>\theta_1^*.
        \end{cases}
    \end{align*}
    We consider the cases $\chi_1\leq\chi_2$ and $\chi_1>\chi_2$ separately.
    If $\chi_1\leq\chi_2$, it holds that
    \begin{align}
        V_1\left(\chi_1,\chi_2\right) &= \mathbb{E}_{\theta_1} \left[  v_1^{\chi_1}\left( \sigma_{1}^{\theta_{1}^*}\left(\theta_1\right) \mid \theta_1,\sigma_{2}^{\theta_{2}^*} \right) \right] \nonumber \\
        &= \int_0^{\theta_1^*} \chi_1 \left[ \Ftt \left(1-F(\theta_1)\right)+\left(1-\Ftt \right)F(\theta_1) \right] + (1-\chi_1) \left[ \Ftt -F(\theta_1) \right] \mathrm{d}F(\theta_1) \nonumber\\
        & \qquad \qquad + \int_{\theta_1^*}^1 F(\theta_1) \mathrm{d}F(\theta_1) \nonumber \\
        &= \chi_1 \left[ \Ftt \int_0^{F\left(\theta_1^*\right)}1-\theta_1 \mathrm{d}\theta_1 + \left(1-\Ftt\right) \int_0^{\Fto} \theta_1 \mathrm{d}\theta_1 \right] \nonumber\\
        &\qquad \qquad+ (1-\chi_1) \left[ \int_0^{\Fto} \Ftt - \theta_1 \mathrm{d}\theta_1 \right] + \int_{\Fto}^1 \theta_1 \mathrm{d}\theta_1 \nonumber\\
        &= \chi_1 \left[ \Fto\Ftt - \Fto^2\Ftt + \tfrac{1}{2} \Fto^2 \right] + (1-\chi_1) \left[ \Fto \Ftt - \tfrac{1}{2} \Fto^2 \right] \nonumber\\
        &\qquad \qquad + \tfrac{1}{2} \left(1-\Fto^2\right) \nonumber\\
        &= \tfrac{1}{2} - (1-\chi_1) \Fto^2 + \Fto\Ftt - \chi_1\Fto^2\Ftt \label{eq:formula_C.1}\\
        &= U_1(\chi_1,\chi_2) + \chi_1 \Fto^2 \left(1-\Ftt\right) \label{eq:formula_C/G.1} \\
        &> U_1(\chi_1,\chi_2) \nonumber \\
        &\geq \tfrac{1}{2}, \nonumber
    \end{align}
    where the last inequality follows from Theorem \ref{thm:better_less_cursed}, as player 1 is less cursed than player~2.
    If $\chi_1>\chi_2$, we have\footnote{
    In this case, $U_1(\chi_1,\chi_2)=\Ftt\left[\Ftt-\Fto\right]+\frac{1}{2}$, which follows from a calculation similar to \eqref{eq:ex_ante_expected_gain_differently_cursed}.
    }
    \begin{align}
        V_1\left(\chi_1,\chi_2\right) &= \mathbb{E}_{\theta_1} \left[  v_1^{\chi_1}\left( \sigma_{1}^{\theta_{1}^*}\left(\theta_1\right) \mid \theta_1,\sigma_{2}^{\theta_{2}^*} \right) \right]\nonumber \\
        &=\chi_1 \int_0^{\theta_1^*} \Ftt \left(1-F(\theta_1)\right)+\left(1-\Ftt \right)F(\theta_1) \mathrm{d}F(\theta_1) \nonumber\\
        &\qquad \qquad + (1-\chi_1) \left[ \int_0^{\theta_2^*} \Ftt - F(\theta_1) \dF(\theta_1) + \int_{\theta_2^*}^{\theta_1^*} F(\theta_1) - \Ftt \dF(\theta_1) \right] \nonumber\\
        &\qquad \qquad + \int_{\theta_1^*}^1 F(\theta_1) \mathrm{d}F(\theta_1) \nonumber\\
        &= \chi_1 \left[ \Fto\Ftt - \Fto^2\Ftt + \tfrac{1}{2} \Fto^2 \right] \nonumber\\
        &\qquad \qquad + (1-\chi_1) \left[ \tfrac{1}{2} \Fto^2+\Ftt^2 - \Fto\Ftt \right] + \tfrac{1}{2} \left(1-\Fto^2\right) \nonumber\\
        &= \tfrac{1}{2} + (1-\chi_1) \Ftt^2 + (2\chi_1-1) \Fto\Ftt - \chi_1 \Fto^2\Ftt \label{eq:formula_C.2}\\
        &= U_1(\chi_1, \chi_2) + \chi_1 \Ftt \left[ \Fto\left(1-\Fto\right) + \Fto-\Ftt \right] \label{eq:formula_C/G.2}
    \end{align}
    which is greater than $U_1(\chi_1, \chi_2)$ as $\Fto>\Ftt$.
    Rearranging \eqref{eq:formula_C.2}, we also find that
    \begin{align*}
        V_1(\chi_1, \chi_2) &= \tfrac{1}{2} + (1-\chi_1) \Ftt^2 + \Fto \Ftt \left(2\chi_1-1-\chi_1\Fto\right) \\
        &= \tfrac{1}{2} + (1-\chi_1) \Ftt^2 + \Fto \Ftt \left(\chi_1-\tfrac{1}{2}\right) \\
        &>\tfrac{1}{2}
    \end{align*}
    where we used that $\Fto = \tfrac{2\chi_1-1}{2\chi_1}$ and that $\chi_1>\tfrac{1}{2}$.
    
    We see that in either case, $V_1(\chi_1,\chi_2)$ exceeds both $U_1(\chi_1,\chi_2)$ and $\tfrac{1}{2}$.
    
    Next, we investigate how $V_1(\chi_1,\chi_2)$ changes with $\chi_1$, keeping $\chi_2$ constant. 
    From \eqref{eq:formula_C.1}, \eqref{eq:formula_C.2}, and the definitions of $\theta_1^*$ and $\theta_2^*$, one can show that
    \begin{align}
    \label{eq:formula_C.3}
        V_1(\chi_1, \chi_2) = 
        \begin{cases}
            \frac{1 - 4 (1 + \chi_1 - 3 \chi_2) \chi_2}{8 (2 \chi_2 - \chi_1) \chi_2} &\text{ for } \chi_1 \leq \chi_2, \\
            \frac{\chi_2 + 2 \chi_1 \left[\chi_1 (1 + 2 \chi_1)^2 - (3 + 2 \chi_1 + 4 \chi_1^2) \chi_2 + 2 \chi_2^2\right]}{8 \chi_1 (2 \chi_1 - \chi_2)^2} &\text{ for } \chi_1 > \chi_2.
        \end{cases}
    \end{align}
    Inserting $\chi_1=\chi_2$, one finds that $V_1$ is continuous.
    For $\chi_1 \neq \chi_2$, $V_1$ is differentiable in the first argument with partial derivative
    \begin{align*}
        \frac{\partial}{\partial \chi_1} V_1(\chi_1, \chi_2) = 
        \begin{cases}
            \frac{(1 - 2 \chi_2)^2}{8 (\chi_1 - 2 \chi_2)^2 \chi_2} &\text{ for } \chi_1 < \chi_2, \\
            \frac{(2 \chi_1-1) \left[4 \chi_1^3 (1 + 2 \chi_1) - 2 \chi_1 (-3 + \chi_1 (5 + 6 \chi_1)) \chi_2 + (-1 + 2 \chi_1) (1 + 4 \chi_1) \chi_2^2\right]}{8 \chi_1^2 (2 \chi_1 - \chi_2)^3} &\text{ for } \chi_1 > \chi_2.
        \end{cases}
    \end{align*}
    The positivity of the derivative in the case of $\chi_1<\chi_2$ is clear.
    For the second case, we use the following lemma.
    \begin{lemma} \label{lem:function_h_positive}
        Let 
        \begin{align*}
            h_0(\chi_1,\chi_2)\colonequals 4 \chi_1^3 (1 + 2 \chi_1) - 2 \chi_1 (-3 + \chi_1 (5 + 6 \chi_1)) \chi_2 + (-1 + 2 \chi_1) (1 + 4 \chi_1) \chi_2^2.
        \end{align*}
        Then $h_0(\chi_1,\chi_2)>0$ for all $\chi_1,\chi_2 \in [0,1]$ that satisfy $\chi_1>\frac{1}{2}$ and $\chi_1>\chi_2$.
    \end{lemma}
    \begin{proof}[Proof of Lemma \ref{lem:function_h_positive}]
        As $h_0(\chi_1, 0)=8\chi_1^4+4\chi_1^3$, the result is immediate for $\chi_2=0$.
        Let $\chi_2>0$.
        It holds that
        \begin{align*}
            h_0(\chi_1,\chi_2) &= 8\chi_1^4+(12\chi_2+4)\chi_1^3+(8\chi_2^2-10\chi_2)\chi_1^2+(-2\chi_2^2+6\chi_2)\chi_1-\chi_2^2, \\
            h_1(\chi_1,\chi_2)\colonequals\frac{\partial}{\partial \chi_1}h_0(\chi_1,\chi_2) &= 32\chi_1^3+(-36\chi_2+12)\chi_1^2+(16\chi_2^2-20\chi_2)\chi_1+(-2\chi_2^2+6\chi_2), \\
            h_2(\chi_1,\chi_2)\colonequals\frac{\partial^2}{\partial \chi_1^2}h_0(\chi_1,\chi_2) &= 96\chi_1^2+(-72\chi_2+24)\chi_1+(16\chi_2^2-20\chi_2), \\
            h_3(\chi_1,\chi_2)\colonequals\frac{\partial^3}{\partial \chi_1^3}h_0(\chi_1,\chi_2) &= 192\chi_1+(-72\chi_2+24).
        \end{align*}
        Evaluating these functions at $\chi_1=\chi_2=\chi$, we obtain
        \begin{align*}
            h_0(\chi,\chi) &= 4\chi^4-8\chi^3+5\chi^2, \\
            h_1(\chi,\chi) &= 12\chi^3-10\chi^2+6\chi, \\
            h_2(\chi,\chi) &= 40\chi^2+4\chi, \\
            h_3(\chi,\chi) &= 120\chi+24.
        \end{align*}
        One can verify that $h_m(\chi,\chi)>0$ for all $\chi>0$ and $m \in \{0,1,2,3\}$.
        As $h_2(\chi,\chi)>0$ and $h_3(\chi_1,\chi_2)=\frac{\partial}{\partial \chi_1}h_2(\chi_1,\chi_2)>0$ whenever $\chi_1\geq\chi_2$, also $h_2(\chi_1,\chi_2)>0$ whenever $\chi_1\geq\chi_2$.
        Iterating this argument, also $h_1(\chi_1,\chi_2)>0$ whenever $\chi_1\geq\chi_2$, and the same holds for $h_0$ which proves the claim.
    \end{proof}
    
    \noindent As seen before, $\frac{\partial}{\partial \chi_1} V_1(\chi_1, \chi_2) = \frac{(2\chi_1-1)h_0(\chi_1,\chi_2)}{8 \chi_1^2 (2 \chi_1 - \chi_2)^3}$ for $\chi_1>\chi_2$.
    By Lemma \ref{lem:function_h_positive}, this expression is positive whenever $\chi_1>\frac{1}{2}$ and $\chi_1>\chi_2$.
    This closes the proof.
\end{proof}

\noindent While Theorem \ref{thm:better_less_cursed} shows that objectively, it is better for a player to be less cursed, Theorem \ref{thm:cursedness_increases_perceived_gain} provides the complementary result about the average utility a cursed player expects to get.
Despite the fact that the actual ex-ante expected utility $U_k(\chi_1,\chi_2)$ is decreasing in the own cursedness, the $\chi_k$-perceived ex-ante expected utility $V_k(\chi_1,\chi_2)$ is increasing in it.
That is, the more cursed a player is, the more they expect to get from the game, although in reality, it is just the other way round.
To put it a bit bluntly: 
Although they take worse decisions, irrational agents are happier than rational ones.\footnote{
    Note that Theorem \ref{thm:cursedness_increases_perceived_gain} performs comparative statics with respect to one player's cursedness, holding the opponent's degree fixed.
    For some fixed tuple $(\chi_1, \chi_2)$, we still find that the more cursed player has a smaller perceived ex-ante expected utility than the more cursed one:
    for example, Equation \eqref{eq:formula_C.3} yields that $V_1(\chi_1,\chi_2)<V_2(\chi_1,\chi_2)$ if $\chi_1>\chi_2$.
}

More precisely, one can interpret cursedness as the degree to which a player neglects the connection between their opponent's type and their actions.
The more cursed a player is, the more they ignore the information that the action played by the opponent reveals about the opponent's type.
Hence, in light of the results of Theorems \ref{thm:better_less_cursed} and \ref{thm:cursedness_increases_perceived_gain}, one can say that the more a player ignores this information, the higher is their utility at the moment they decide on a strategy themselves, but the lower is the expected actual payoff they will face later.
Thus, ignoring information increases the subjective well-being before a decision, but is detrimental when considering the actual payoffs.
This corresponds to Theorems 6.1 and 6.4 in \citet{preker_karos_2024} where a similar result is obtained in a completely different set-up.

The increase in utility does not only hold for the individual player, but also on the level of collective welfare:
by Theorem \ref{thm:cursedness_increases_perceived_gain},
\begin{align*}
    V_1\left(\chi_1,\chi_2\right)+V_2\left(\chi_1,\chi_2\right) > U_1\left(\chi_1,\chi_2\right)+U_2\left(\chi_1,\chi_2\right)=1
\end{align*}
for all $\chi_1, \chi_2$ with $\max\{\chi_1, \chi_2 \} > \frac{1}{2}$, so cursedness increases the perceived ``size of the cake'' in this constant-sum game.

Moreover, as $V_k\left(\chi_1, \chi_2\right) > \tfrac{1}{2}$, cursed players prefer the outcome of the non-trivial $\left(\chi_1, \chi_2\right)$-cursed equilibrium to the trivial one; the latter coincides with the Bayesian Nash equilibrium in which there is no trade. 
Hence, it is reasonable to assume that cursed players would rather play the non-trivial $\left(\chi_1, \chi_2\right)$-cursed equilibrium than the Bayesian Nash equilibrium, although the latter is also such an equilibrium.

\begin{figure}[t]
     \centering
     \begin{subfigure}[b]{0.32\textwidth}
         \centering
         \includegraphics[width=\textwidth]{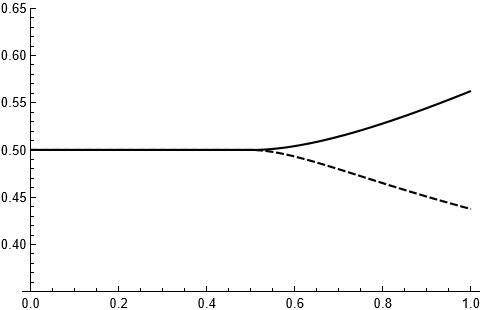}
         \caption{$\chi_{-k}=0$}
         \label{fig:chi2=0}
     \end{subfigure}
     \begin{subfigure}[b]{0.32\textwidth}
         \centering
         \includegraphics[width=\textwidth]{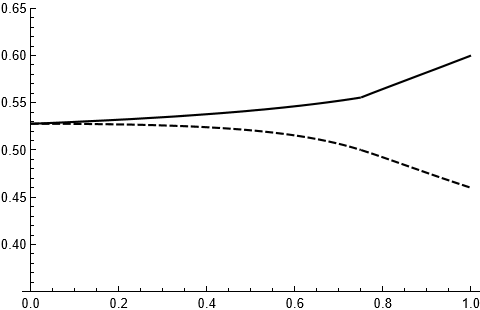}
         \caption{$\chi_{-k}=0.75$}
         \label{fig:chi2=0.75}
     \end{subfigure}
     \begin{subfigure}[b]{0.32\textwidth}
         \centering
         \includegraphics[width=\textwidth]{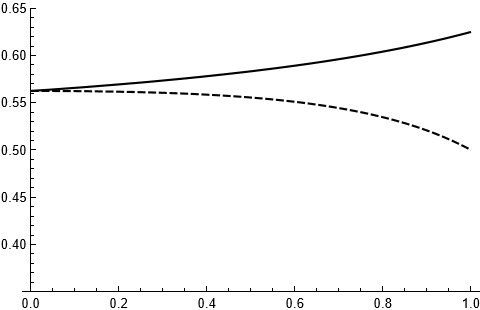}
         \caption{$\chi_{-k}=1$}
         \label{fig:chi2=1}
     \end{subfigure}
         \caption{Actual (dashed) and perceived (solid) ex-ante expected utilities}
         \label{fig:actual_perceived_gains}
\end{figure}

The results of Theorems \ref{thm:better_less_cursed} and \ref{thm:cursedness_increases_perceived_gain} are illustrated in Figure \ref{fig:actual_perceived_gains}. 
The three panels show $U_k(\chi_1,\chi_2)$ (dashed line) and $V_k(\chi_1,\chi_2)$ (solid line) as functions of $\chi_k$, for $\chi_{-k}$ being fixed as 0, 0.75, or 1.
If $\max\{\chi_1,\chi_2\}\leq\frac{1}{2}$ and there is no trade, both the real as well as the perceived ex-ante expected utilities are $\frac{1}{2}$ (as can be seen in Subfigure \ref{fig:chi2=0}). 
In all areas where trade happens, the ex-ante expected utility is strictly decreasing in the own cursedness, while the perceived ex-ante expected utility is increasing in it.

\end{document}